\documentclass[11pt]{article}
\usepackage{latexsym}
\usepackage{color}
\newcommand{\Cpp}{\mbox{C\hspace{-0.12em}\raisebox{0.15em}{\scriptsize $+$\hspace{-0.07em}$+$}}}

\begin{document}
\pagenumbering{roman}

\title{A Fast Generic Sequence Matching Algorithm}

\author{David R. Musser \and Gor V. Nishanov}

\date{Computer Science Department \\
Rensselaer Polytechnic Institute, Troy, NY 12180 \\
\{musser,gorik\}@cs.rpi.edu
~\\
December 1, 1997
}

\maketitle

\begin{abstract} A string matching---and more generally, sequence
matching---algorithm is presented that has a linear worst-case
computing time bound, a low worst-case bound on the number of
comparisons ($2n$), and sublinear average-case behavior that is better
than that of the fastest versions of the Boyer-Moore algorithm. The
algorithm retains its efficiency advantages in a wide variety of
sequence matching problems of practical interest, including
traditional string matching; large-alphabet problems (as in Unicode
strings); and small-alphabet, long-pattern problems (as in DNA
searches).  Since it is expressed as a generic algorithm for searching
in sequences over an arbitrary type $T$, it is well suited for use in
generic software libraries such as the \Cpp\ Standard Template Library.
The algorithm was obtained by adding to the Knuth-Morris-Pratt
algorithm one of the pattern-shifting techniques from the Boyer-Moore
algorithm, with provision for use of hashing in this technique.  In
situations in which a hash function or random access to the sequences
is not available, the algorithm falls back to an optimized version of
the Knuth-Morris-Pratt algorithm.  \end{abstract}

\vspace{0.1in}

\noindent
{\sc key words} ~~String search ~~String matching ~~Pattern matching
~~Sequence matching ~~Generic algorithms ~~Knuth-Morris-Pratt
algorithm ~~Boyer-Moore algorithm ~~DNA pattern matching ~~\Cpp\ 
~~Standard Template Library ~~STL ~~Ada ~~Literate programming

\newpage
\tableofcontents

\newpage
\pagenumbering{arabic}
\section{Introduction}

The traditional string matching problem is to find an
occurrence of a pattern (a string) in a text (another string), or
to decide that none exists. Two of the best known algorithms for the
problem of string matching are the Knuth-Morris-Pratt \cite{KMP} and
Boyer-Moore \cite{BM} algorithms (for short, we will refer to these as
KMP and BM).  Although KMP has a low worst-case bound on number of
comparisons ($2n$, where $n$ is the length of the text), 
it is often considered impractical, since the
number of comparisons it performs in the average case is not
significantly smaller than that of the straightforward (SF) algorithm
\cite{Comp3}, and the overhead for initialization is higher.  On the
other hand, despite the fact that BM has a higher worst-case bound on
the number of comparisons ($\approx 3n$ \cite{Cole}), it has excellent
sublinear behavior in the average case.  This fact often makes BM the
algorithm of choice in practical applications.

In \cite{BM}, Boyer and Moore described both a basic version of their
algorithm and an optimized version based on use of a ``skip loop.''
We will refer to the latter algorithm as Accelerated Boyer-Moore, or
ABM for short.  Unfortunately, this version remained unnoticed by most
researchers despite its much better performance.  For example, ABM
outperforms the Quick Search \cite{Sunday} and Boyer-Moore-Horspool
\cite{Horspool} improvements of the basic BM algorithm.  This state of
affairs was highlighted by Hume and Sunday in 1991 in
\cite{HumeSunday}, in which they introduced two algorithms, LC (least
cost) and TBM (Tuned BM) \cite{HumeSunday}, that perform faster than
ABM in the average case.  These two algorithms use the skip loop of
ABM combined with variants of the straightforward algorithm that use
information on character frequency distribution in the target
text. For traditional string matching LC and TBM have
excellent average case behavior, but in the worst case they behave
like SF, taking time proportional to the product of the 
text and pattern lengths.

Even in the average case, the skip loop as it is used in ABM and other
algorithms performs poorly with small alphabets and long patterns, as
happens, for example, in the problem of DNA pattern matching.  And if
the alphabet is large, as for example with Unicode strings,
initialization overhead and memory requirements for the skip loop
weigh against its use.

This article describes a new linear string-matching algorithm and its
generalization to searching in sequences over an arbitrary type $T$.
The algorithm is based on KMP and has the same low $2n$ worst-case
bound on number of comparisons, but it is better than ABM (comparable
with TBM) in average case performance on English text strings.  It
employs a hash-coded form of the skip loop making it suitable even for
cases with large alphabets or with small alphabets and long patterns.
Since it is expressed as a generic algorithm for searching in
sequences over an arbitrary type $T$, the new algorithm is well suited
for use in generic software libraries such as the \Cpp\ Standard
Template Library (STL).  We present the algorithm in the following
sections by starting with the basic KMP algorithm and transforming it
with optimizations and addition of the skip loop in several
alternative forms. The optimized form of KMP without the skip loop
also serves well in cases in which access to the sequences is
restricted to forward, one-element-at-a-time iteration rather than
random-access.  We also discuss experimental results and some of the
issues in including the new algorithm in a generic algorithm library.

\section{Linear and Accelerated Linear Algorithms}

Let $m \geq a \geq 0$ and $n \geq b \geq 0$ and
suppose $p_a \ldots p_{m-1}$ is a pattern of size $m - a$
to be searched for in the text $t_b \ldots t_{n-1}$ of size
$n - b$.  Characters of the pattern and
the text are drawn from an alphabet $\Sigma$.
The Knuth-Morris-Pratt algorithm can be viewed as an extension of
the straightforward search algorithm. It starts comparing symbols of the
pattern and the text from left to the right. However, when a mismatch
occurs, instead of shifting the pattern by one symbol and repeating
matching from the beginning of the pattern, 
KMP shifts the pattern to the right
in such a way that the scan can be restarted at the point of mismatch
in the text. The amount of shift is determined by precomputed
function \verb|next|, defined by
\[
\mbox{\tt next}(j) = \max_{i < j}\{i | p_a \ldots p_{i-1} = 
p_{a+j-i} \ldots p_{j-1} \wedge p_i \not= p_j\}
\]
(We let $\mbox{\tt next}(j) = a - 1$ if there is no $i$ satisfying the
conditions.)  Here is the basic KMP algorithm as it appeared in
\cite{KMP}, except that we use more general index ranges:\footnote{Although
most authors use pseudocode for expository purposes, we prefer
to be able to check all code with a compiler.  The expository versions
of algorithms in this paper are expressed in Ada 95, which has a syntax
similar to that of most pseudocode languages (at least if one omits the details
of subprogram headers and package declarations, which we include only
in an appendix that deals with actual compilation of the code).
The generic library components developed later in the
paper are written in \Cpp.  Throughout the paper we present expository
and production code in a variant of Knuth's literate programming style
\cite{Knuth:literate}, in which code is presented in ``parts'' numbered
according to the page number on which they appear (with parts on the same 
page distinguished by appending a letter to the number).  This form
of presentation is supported by Briggs' Nuweb tool \cite{Briggs}
(slightly modified, as discussed in a later section),
with which we also generate all code files directly from the 
paper's source file.}

\begin{flushleft} \small
\begin{minipage}{\linewidth} \label{__n:gensearch.w:part1}
$\langle$Basic KMP {\footnotesize 2}$\rangle\equiv$
\vspace{-1ex}
\begin{list}{}{} \item
\mbox{}\verb@pattern_size := m - a; j := a; k := b;@\\
\mbox{}\verb@while j < m and then k < n loop @\\
\mbox{}\verb@  while j >= a and then text(k) /= pattern(j) loop@\\
\mbox{}\verb@    j := next(j); @\\
\mbox{}\verb@  end loop; @\\
\mbox{}\verb@  k := k + 1; j := j + 1; @\\
\mbox{}\verb@end loop; @\\
\mbox{}\verb@if j = m then @\\
\mbox{}\verb@  return k - pattern_size; @\\
\mbox{}\verb@else @\\
\mbox{}\verb@  return n; @\\
\mbox{}\verb@end if;@\\
\mbox{}\verb@@\end{list}
\vspace{-1ex}
\footnotesize\addtolength{\baselineskip}{-1ex}
\begin{list}{}{\setlength{\itemsep}{-\parsep}\setlength{\itemindent}{-\leftmargin}}
\item Used in part 27c.
\end{list}
\end{minipage}\\[4ex]
\end{flushleft}
A return value $i$ between $b$ and $n - \mbox{\tt pattern\_size}$
indicates a match found beginning at position $i$, while a return value of
$n$ means there was no match. Although elegantly short, this
algorithm does redundant operations along the expected execution path.
That is, \verb|text(k)| is usually not equal to \verb|pattern(j)| and
\verb|next(j)| is usually $a - 1$, so the inner loop usually sets
\verb|j| to $a - 1$, redundantly tests it against $a$, and terminates.
\verb|k| and \verb|j| are then both incremented and tested against
their bounds, then \verb|j| is again redundantly compared with $a$.
Knuth, Morris, and Pratt discussed a set of optimizations to the basic
algorithm that required extending the text and pattern with additional
characters, which is possible only under extra assumptions about the
way the inputs are stored.  We must avoid such assumptions when the
goal is a generic algorithm.  Instead, we eliminate the redundant
operations by rewriting the algorithm in the following form, which we
will call Algorithm L (for Linear) in this paper:

\begin{flushleft} \small
\begin{minipage}{\linewidth} \label{__n:gensearch.w:part2}
$\langle$Algorithm L, optimized linear pattern search {\footnotesize 3a}$\rangle\equiv$
\vspace{-1ex}
\begin{list}{}{} \item
\mbox{}\verb@pattern_size := m - a; k := b;@\\
\mbox{}\verb@@\hbox{$\langle$Handle pattern size = 1 as a special case {\footnotesize 3b}$\rangle$}\verb@@\\
\mbox{}\verb@while k <= n - pattern_size loop@\\
\mbox{}\verb@  @\hbox{$\langle$Scan the text for a possible match {\footnotesize 3c}$\rangle$}\verb@@\\
\mbox{}\verb@  @\hbox{$\langle$Verify whether a match is possible at the position found {\footnotesize 4a}$\rangle$}\verb@@\\
\mbox{}\verb@  @\hbox{$\langle$Recover from a mismatch using the next table {\footnotesize 4b}$\rangle$}\verb@@\\
\mbox{}\verb@end loop;@\\
\mbox{}\verb@return n;@\\
\mbox{}\verb@@\end{list}
\vspace{-1ex}
\footnotesize\addtolength{\baselineskip}{-1ex}
\begin{list}{}{\setlength{\itemsep}{-\parsep}\setlength{\itemindent}{-\leftmargin}}
\item Used in part 27c.
\end{list}
\end{minipage}\\[4ex]
\end{flushleft}
The following code allows us to eliminate a test in the main loop:

\begin{flushleft} \small
\begin{minipage}{\linewidth} \label{__n:gensearch.w:part3}
$\langle$Handle pattern size = 1 as a special case {\footnotesize 3b}$\rangle\equiv$
\vspace{-1ex}
\begin{list}{}{} \item
\mbox{}\verb@if pattern_size = 1 then @\\
\mbox{}\verb@  while k /= n and then text(k) /= pattern(a) loop@\\
\mbox{}\verb@    k := k + 1;@\\
\mbox{}\verb@  end loop;@\\
\mbox{}\verb@  return k;@\\
\mbox{}\verb@end if;@\\
\mbox{}\verb@@\end{list}
\vspace{-1ex}
\footnotesize\addtolength{\baselineskip}{-1ex}
\begin{list}{}{\setlength{\itemsep}{-\parsep}\setlength{\itemindent}{-\leftmargin}}
\item Used in parts 3a, 5c, 7b, 12b, 27c.
\end{list}
\end{minipage}\\[4ex]
\end{flushleft}
The three parts of the body of the main loop are defined as follows:

\begin{flushleft} \small
\begin{minipage}{\linewidth} \label{__n:gensearch.w:part4}
$\langle$Scan the text for a possible match {\footnotesize 3c}$\rangle\equiv$
\vspace{-1ex}
\begin{list}{}{} \item
\mbox{}\verb@while text(k) /= pattern(a) loop@\\
\mbox{}\verb@  k := k + 1; @\\
\mbox{}\verb@  if k > n - pattern_size then @\\
\mbox{}\verb@    return n;@\\
\mbox{}\verb@  end if;@\\
\mbox{}\verb@end loop;@\\
\mbox{}\verb@@\end{list}
\vspace{-1ex}
\footnotesize\addtolength{\baselineskip}{-1ex}
\begin{list}{}{\setlength{\itemsep}{-\parsep}\setlength{\itemindent}{-\leftmargin}}
\item Used in parts 3a, 27c.
\end{list}
\end{minipage}\\[4ex]
\end{flushleft}
\begin{flushleft} \small
\begin{minipage}{\linewidth} \label{__n:gensearch.w:part5}
$\langle$Verify whether a match is possible at the position found {\footnotesize 4a}$\rangle\equiv$
\vspace{-1ex}
\begin{list}{}{} \item
\mbox{}\verb@j := a + 1; k := k + 1;@\\
\mbox{}\verb@while text(k) = pattern(j) loop@\\
\mbox{}\verb@  k := k + 1; j := j + 1; @\\
\mbox{}\verb@  if j = m then @\\
\mbox{}\verb@    return k - pattern_size;@\\
\mbox{}\verb@  end if;@\\
\mbox{}\verb@end loop;@\\
\mbox{}\verb@@\end{list}
\vspace{-1ex}
\footnotesize\addtolength{\baselineskip}{-1ex}
\begin{list}{}{\setlength{\itemsep}{-\parsep}\setlength{\itemindent}{-\leftmargin}}
\item Used in parts 3a, 27c.
\end{list}
\end{minipage}\\[4ex]
\end{flushleft}
\begin{flushleft} \small
\begin{minipage}{\linewidth} \label{__n:gensearch.w:part6}
$\langle$Recover from a mismatch using the next table {\footnotesize 4b}$\rangle\equiv$
\vspace{-1ex}
\begin{list}{}{} \item
\mbox{}\verb@loop@\\
\mbox{}\verb@  j := next(j);@\\
\mbox{}\verb@  if j < a then @\\
\mbox{}\verb@     k := k + 1; exit; @\\
\mbox{}\verb@  end if;@\\
\mbox{}\verb@  exit when j = a;@\\
\mbox{}\verb@  while text(k) = pattern(j) loop@\\
\mbox{}\verb@    k := k + 1; j := j + 1; @\\
\mbox{}\verb@    if j = m then @\\
\mbox{}\verb@      return k - pattern_size;@\\
\mbox{}\verb@    end if;@\\
\mbox{}\verb@    if k = n then @\\
\mbox{}\verb@      return n;@\\
\mbox{}\verb@    end if;@\\
\mbox{}\verb@  end loop;@\\
\mbox{}\verb@end loop;@\\
\mbox{}\verb@@\end{list}
\vspace{-1ex}
\footnotesize\addtolength{\baselineskip}{-1ex}
\begin{list}{}{\setlength{\itemsep}{-\parsep}\setlength{\itemindent}{-\leftmargin}}
\item Used in parts 3a, 5c.
\end{list}
\end{minipage}\\[4ex]
\end{flushleft}
This last part guarantees linear worst-case behavior. Notice that if
we simply replace the last part with the code \verb|k := k - (j - a) + 1| we
obtain (an optimized form of) the straightforward algorithm.  

Algorithm L can be further improved by incorporating a skip loop
similar to the one that accounts for the excellent sublinear average
time behavior of ABM.  The idea of this technique is demonstrated in
the following pair of examples:
\begin{verbatim}
Text:        ......uuuuuuuuuua....  ......uuuuuuuuuue....
Before Shift:      bcdabcdabcd            bcdabcdabcd
After Shift:          bcdabcdabcd                    bcdabcdabcd
\end{verbatim}
We inspect the text character $t_j$ that corresponds to the last
character of the pattern, and if $t_j \not= p_{m-1}$ we shift the pattern
by the amount determined by the $\mbox{\tt skip}$ function, which maps any
character of the alphabet to the range $[0, m - a]$ and is
defined as follows:
\[
   \mbox{\tt skip}(x) = \left\{ \begin{array}{ll}
      m - a   & \mbox{if } \forall j: a \le j < m \Rightarrow p_j \not= x \\
        m-1-i & \mbox{otherwise, where}~i = \max \{j: a \le j < m \wedge p_j = x\}
   \end{array} \right.
\]
This is the same function as Boyer and Moore's $\delta_1$ \cite{BM}.
The following code replaces the scan part of Algorithm L:

\begin{flushleft} \small
\begin{minipage}{\linewidth} \label{__n:gensearch.w:part7}
$\langle$Scan the text using the skip loop {\footnotesize 5a}$\rangle\equiv$
\vspace{-1ex}
\begin{list}{}{} \item
\mbox{}\verb@loop@\\
\mbox{}\verb@  d := skip(text(k + pattern_size - 1)); @\\
\mbox{}\verb@  exit when d = 0;@\\
\mbox{}\verb@  k := k + d; @\\
\mbox{}\verb@  if k > n - pattern_size then @\\
\mbox{}\verb@    return n;@\\
\mbox{}\verb@  end if;@\\
\mbox{}\verb@end loop;@\\
\mbox{}\verb@@\end{list}
\vspace{-1ex}
\footnotesize\addtolength{\baselineskip}{-1ex}
\begin{list}{}{\setlength{\itemsep}{-\parsep}\setlength{\itemindent}{-\leftmargin}}
\item Used in part 5c.
\end{list}
\end{minipage}\\[4ex]
\end{flushleft}
If the exit is taken from this loop then 
$\mbox{\tt text(k + pattern\_size - 1)} = \mbox{\tt pattern(m - 1)}$.  
We also change the verifying part of Algorithm L to the following:

\begin{flushleft} \small
\begin{minipage}{\linewidth} \label{__n:gensearch.w:part8}
$\langle$Verify the match for positions a through m - 2 {\footnotesize 5b}$\rangle\equiv$
\vspace{-1ex}
\begin{list}{}{} \item
\mbox{}\verb@j := a; @\\
\mbox{}\verb@while text(k) = pattern(j) loop@\\
\mbox{}\verb@  k := k + 1; j := j + 1; @\\
\mbox{}\verb@  if j = m - 1 then @\\
\mbox{}\verb@    return k - pattern_size + 1;@\\
\mbox{}\verb@  end if;@\\
\mbox{}\verb@end loop;@\\
\mbox{}\verb@@\end{list}
\vspace{-1ex}
\footnotesize\addtolength{\baselineskip}{-1ex}
\begin{list}{}{\setlength{\itemsep}{-\parsep}\setlength{\itemindent}{-\leftmargin}}
\item Used in part 5c.
\end{list}
\end{minipage}\\[4ex]
\end{flushleft}
The algorithm incorporating these changes will be called the
Accelerated Linear algorithm, or AL for short.  In preliminary form,
the algorithm is as follows:

\begin{flushleft} \small
\begin{minipage}{\linewidth} \label{__n:gensearch.w:part9}
$\langle$Accelerated Linear algorithm, preliminary version {\footnotesize 5c}$\rangle\equiv$
\vspace{-1ex}
\begin{list}{}{} \item
\mbox{}\verb@pattern_size := m - a; k := b;@\\
\mbox{}\verb@@\hbox{$\langle$Handle pattern size = 1 as a special case {\footnotesize 3b}$\rangle$}\verb@@\\
\mbox{}\verb@@\hbox{$\langle$Compute next table {\footnotesize 31a}$\rangle$}\verb@@\\
\mbox{}\verb@@\hbox{$\langle$Compute skip table and mismatch shift {\footnotesize 6a}$\rangle$}\verb@@\\
\mbox{}\verb@while k <= n - pattern_size loop@\\
\mbox{}\verb@  @\hbox{$\langle$Scan the text using the skip loop {\footnotesize 5a}$\rangle$}\verb@@\\
\mbox{}\verb@  @\hbox{$\langle$Verify the match for positions a through m - 2 {\footnotesize 5b}$\rangle$}\verb@@\\
\mbox{}\verb@  if mismatch_shift > j - a then@\\
\mbox{}\verb@    k := k + (mismatch_shift - (j - a));@\\
\mbox{}\verb@  else@\\
\mbox{}\verb@    @\hbox{$\langle$Recover from a mismatch using the next table {\footnotesize 4b}$\rangle$}\verb@@\\
\mbox{}\verb@  end if;@\\
\mbox{}\verb@end loop;@\\
\mbox{}\verb@return n;@\\
\mbox{}\verb@@\end{list}
\vspace{-1ex}
\footnotesize\addtolength{\baselineskip}{-1ex}
\begin{list}{}{\setlength{\itemsep}{-\parsep}\setlength{\itemindent}{-\leftmargin}}
\item Used in part 29c.
\end{list}
\end{minipage}\\[4ex]
\end{flushleft}
Following the verification part, we know that the last character of
the pattern and corresponding character of the text are equal, so we
can choose whether to proceed to the recovery part that uses the
\verb|next| table or to shift the pattern by the amount
\verb|mismatch_shift|, predefined as
\[
   \mbox{\tt mismatch\_shift} = \left\{ \begin{array}{ll} m - a & \mbox{if
        } \forall j: a \le j < m-1 \Rightarrow p_j \not= p_{m-1} \\
        m-1-i & \mbox{otherwise, where}~i = \max \{j: a \le j < m-1
        \wedge p_j = p_{m-1}\} \end{array} \right.
\]
This value can be most easily computed if it is done during
the computation of the skip table:

\begin{flushleft} \small
\begin{minipage}{\linewidth} \label{__n:gensearch.w:part10}
$\langle$Compute skip table and mismatch shift {\footnotesize 6a}$\rangle\equiv$
\vspace{-1ex}
\begin{list}{}{} \item
\mbox{}\verb@for i in Character'Range loop@\\
\mbox{}\verb@  skip(i) := pattern_size;@\\
\mbox{}\verb@end loop;@\\
\mbox{}\verb@for j in a .. m - 2 loop@\\
\mbox{}\verb@  skip(pattern(j)) := m - 1 - j;@\\
\mbox{}\verb@end loop;@\\
\mbox{}\verb@mismatch_shift := skip(pattern(m - 1));@\\
\mbox{}\verb@skip(pattern(m - 1)) := 0;@\\
\mbox{}\verb@@\end{list}
\vspace{-1ex}
\footnotesize\addtolength{\baselineskip}{-1ex}
\begin{list}{}{\setlength{\itemsep}{-\parsep}\setlength{\itemindent}{-\leftmargin}}
\item Used in parts 5c, 7b.
\end{list}
\end{minipage}\\[4ex]
\end{flushleft}
The skip loop as described above performs two tests for exit during
each iteration.  As suggested in \cite{BM}, we can eliminate one of
the tests by initializing \verb|skip(pattern(m - 1))| to some value
\verb|large|, chosen large enough to force an exit based on the size
of the index.  Upon exit, we can then perform another test to
distinguish whether a match of a text character with the last pattern
character was found or the pattern was shifted off the end of the text
string. We also add \verb|pattern_size - 1| to \verb|k| outside
the loop and precompute $\mbox{\tt adjustment} = 
\mbox{\tt large} + \mbox{\tt pattern\_size} - 1$.

\begin{flushleft} \small
\begin{minipage}{\linewidth} \label{__n:gensearch.w:part11}
$\langle$Scan the text using a single-test skip loop {\footnotesize 6b}$\rangle\equiv$
\vspace{-1ex}
\begin{list}{}{} \item
\mbox{}\verb@loop@\\
\mbox{}\verb@  k := k + skip(text(k)); @\\
\mbox{}\verb@  exit when k >= n;@\\
\mbox{}\verb@end loop;@\\
\mbox{}\verb@if k < n + pattern_size then@\\
\mbox{}\verb@  return n;@\\
\mbox{}\verb@end if;@\\
\mbox{}\verb@k := k - adjustment;@\\
\mbox{}\verb@@\end{list}
\vspace{-1ex}
\footnotesize\addtolength{\baselineskip}{-1ex}
\begin{list}{}{\setlength{\itemsep}{-\parsep}\setlength{\itemindent}{-\leftmargin}}
\item Not used.
\end{list}
\end{minipage}\\[4ex]
\end{flushleft}
We can further optimize the skip loop by translating \verb|k| by $n$
(by writing \verb|k := k - n| before the main loop), which allows the
exit test to be written as \verb|k >= 0|.  

\begin{flushleft} \small
\begin{minipage}{\linewidth} \label{__n:gensearch.w:part12}
$\langle$Scan the text using a single-test skip loop, with k translated {\footnotesize 7a}$\rangle\equiv$
\vspace{-1ex}
\begin{list}{}{} \item
\mbox{}\verb@loop@\\
\mbox{}\verb@  k := k + skip(text(n + k)); @\\
\mbox{}\verb@  exit when k >= 0;@\\
\mbox{}\verb@end loop;@\\
\mbox{}\verb@if k < pattern_size then@\\
\mbox{}\verb@  return n;@\\
\mbox{}\verb@end if;@\\
\mbox{}\verb@k := k - adjustment;@\\
\mbox{}\verb@@\end{list}
\vspace{-1ex}
\footnotesize\addtolength{\baselineskip}{-1ex}
\begin{list}{}{\setlength{\itemsep}{-\parsep}\setlength{\itemindent}{-\leftmargin}}
\item Used in part 7b.
\end{list}
\end{minipage}\\[4ex]
\end{flushleft}
This saves an instruction over testing \verb|k >= n|, and a good
compiler will compile \verb|text(n + k)| with only one instruction in
the loop since the computation of \verb|text + n| can be moved
outside. (In the \Cpp\ version we make sure of this optimization by
putting it in the source code.)  With this form of the skip loop, some
compilers are able to translate it into only three instructions.

How large is \verb|large|?  At the top of the loop we have 
\[
  \mbox{\tt k} \geq b - n + \mbox{\tt pattern\_size} - 1.
\]
In the case in which \verb|k| is incremented by \verb|large|, we must
have 
\[
  \mbox{\tt large} + b - n + \mbox{\tt pattern\_size} - 1 \geq 
        \mbox{\tt pattern\_size}.
\]
Hence it suffices to choose $\mbox{\tt large} = n - b + 1$.

\begin{flushleft} \small
\begin{minipage}{\linewidth} \label{__n:gensearch.w:part13}
$\langle$Accelerated Linear algorithm {\footnotesize 7b}$\rangle\equiv$
\vspace{-1ex}
\begin{list}{}{} \item
\mbox{}\verb@pattern_size := m - a; text_size := n - b; k := b;@\\
\mbox{}\verb@@\hbox{$\langle$Handle pattern size = 1 as a special case {\footnotesize 3b}$\rangle$}\verb@@\\
\mbox{}\verb@@\hbox{$\langle$Compute next table {\footnotesize 31a}$\rangle$}\verb@@\\
\mbox{}\verb@@\hbox{$\langle$Compute skip table and mismatch shift {\footnotesize 6a}$\rangle$}\verb@@\\
\mbox{}\verb@large := text_size + 1;@\\
\mbox{}\verb@skip(pattern(m - 1)) := large;@\\
\mbox{}\verb@adjustment := large + pattern_size - 1;@\\
\mbox{}\verb@k := k - n;@\\
\mbox{}\verb@loop@\\
\mbox{}\verb@  k := k + pattern_size - 1;@\\
\mbox{}\verb@  exit when k >= 0;@\\
\mbox{}\verb@  @\hbox{$\langle$Scan the text using a single-test skip loop, with k translated {\footnotesize 7a}$\rangle$}\verb@@\\
\mbox{}\verb@  @\hbox{$\langle$Verify match or recover from mismatch {\footnotesize 8a}$\rangle$}\verb@@\\
\mbox{}\verb@end loop;@\\
\mbox{}\verb@return n;@\\
\mbox{}\verb@@\end{list}
\vspace{-1ex}
\footnotesize\addtolength{\baselineskip}{-1ex}
\begin{list}{}{\setlength{\itemsep}{-\parsep}\setlength{\itemindent}{-\leftmargin}}
\item Used in part 27c.
\end{list}
\end{minipage}\\[4ex]
\end{flushleft}
We can also optimize the verification of a match by handling as a
special case the frequently occurring case in which the first
characters do not match.

\begin{flushleft} \small
\begin{minipage}{\linewidth} \label{__n:gensearch.w:part14}
$\langle$Verify match or recover from mismatch {\footnotesize 8a}$\rangle\equiv$
\vspace{-1ex}
\begin{list}{}{} \item
\mbox{}\verb@if text(n + k) /= pattern(a) then@\\
\mbox{}\verb@  k := k + mismatch_shift;@\\
\mbox{}\verb@else @\\
\mbox{}\verb@  @\hbox{$\langle$Verify the match for positions a + 1 through m - 1, with k translated {\footnotesize 8b}$\rangle$}\verb@@\\
\mbox{}\verb@  if mismatch_shift > j - a then@\\
\mbox{}\verb@    k := k + (mismatch_shift - (j - a));@\\
\mbox{}\verb@  else@\\
\mbox{}\verb@    @\hbox{$\langle$Recover from a mismatch using the next table, with k translated {\footnotesize 8c}$\rangle$}\verb@@\\
\mbox{}\verb@  end if;@\\
\mbox{}\verb@end if;@\\
\mbox{}\verb@@\end{list}
\vspace{-1ex}
\footnotesize\addtolength{\baselineskip}{-1ex}
\begin{list}{}{\setlength{\itemsep}{-\parsep}\setlength{\itemindent}{-\leftmargin}}
\item Used in parts 7b, 12b.
\end{list}
\end{minipage}\\[4ex]
\end{flushleft}
The verification loop used here doesn't really need to check position
$m - 1$, but we write it that way in preparation for the hashed
version to be described later.

\begin{flushleft} \small
\begin{minipage}{\linewidth} \label{__n:gensearch.w:part15}
$\langle$Verify the match for positions a + 1 through m - 1, with k translated {\footnotesize 8b}$\rangle\equiv$
\vspace{-1ex}
\begin{list}{}{} \item
\mbox{}\verb@j := a + 1;@\\
\mbox{}\verb@loop@\\
\mbox{}\verb@  k := k + 1;@\\
\mbox{}\verb@  exit when text(n + k) /= pattern(j);@\\
\mbox{}\verb@  j := j + 1; @\\
\mbox{}\verb@  if j = m then @\\
\mbox{}\verb@    return n + k - pattern_size + 1;@\\
\mbox{}\verb@  end if;@\\
\mbox{}\verb@end loop;@\\
\mbox{}\verb@@\end{list}
\vspace{-1ex}
\footnotesize\addtolength{\baselineskip}{-1ex}
\begin{list}{}{\setlength{\itemsep}{-\parsep}\setlength{\itemindent}{-\leftmargin}}
\item Used in part 8a.
\end{list}
\end{minipage}\\[4ex]
\end{flushleft}
\begin{flushleft} \small
\begin{minipage}{\linewidth} \label{__n:gensearch.w:part16}
$\langle$Recover from a mismatch using the next table, with k translated {\footnotesize 8c}$\rangle\equiv$
\vspace{-1ex}
\begin{list}{}{} \item
\mbox{}\verb@loop@\\
\mbox{}\verb@  j := next(j);@\\
\mbox{}\verb@  if j < a then @\\
\mbox{}\verb@     k := k + 1; @\\
\mbox{}\verb@     exit; @\\
\mbox{}\verb@  end if;@\\
\mbox{}\verb@  exit when j = a;@\\
\mbox{}\verb@  while text(n + k) = pattern(j) loop@\\
\mbox{}\verb@    k := k + 1; j := j + 1; @\\
\mbox{}\verb@    if j = m then @\\
\mbox{}\verb@      return n + k - pattern_size;@\\
\mbox{}\verb@    end if;@\\
\mbox{}\verb@    if k = 0 then @\\
\mbox{}\verb@      return n;@\\
\mbox{}\verb@    end if;@\\
\mbox{}\verb@  end loop;@\\
\mbox{}\verb@end loop;@\\
\mbox{}\verb@@\end{list}
\vspace{-1ex}
\footnotesize\addtolength{\baselineskip}{-1ex}
\begin{list}{}{\setlength{\itemsep}{-\parsep}\setlength{\itemindent}{-\leftmargin}}
\item Used in part 8a.
\end{list}
\end{minipage}\\[4ex]
\end{flushleft}
The AL algorithm thus obtained retains the same $2n$ upper
case bound on the number of comparisons as the original KMP algorithm
and acquires sublinear average time behavior equal or
superior to ABM.

\section{Benchmarking with English Texts}

Before generalizing AL by introducing a hash function, let us consider
its use as-is for traditional string matching.  We benchmarked five
algorithms with English text searches: a \Cpp\ version of SF used in the
Hewlett-Packard STL implementation; L and AL in their \Cpp\ versions as
given later in the paper and appendices; and the C versions of ABM
\cite{BM} and TBM as given by Hume and Sunday \cite{HumeSunday}.
(The version of AL actually used is the hashed version, HAL,
discussed in the next section, but using the identity function as the
hash function.)

We searched for patterns of size ranging from 2 to 18 in Lewis
Carroll's \emph{Through the Looking Glass}. The text is composed of
171,556 characters, and the test set included up to 800 different
patterns for each pattern size---400 text strings chosen at evenly
spaced positions in the target text and up to 400 words chosen from
the Unix spell-check dictionary (for longer pattern sizes there were
fewer than 400 words). Table~1 shows search speeds of the five
algorithms with code compiled and executed on three different systems:
\begin{enumerate}
  \item g++ compiler, version 2.7.2.2, 60-Mh Pentium processor;
  \item SGI CC compiler, version 7.10, SGI O$^2$ with MIPS R5000 2.1 processor;
  \item Apogee apCC compiler, version 3.0, 200 MHz UltraSPARC processor.
\end{enumerate}
\begin{table}
\begin{center}
\begin{tabular}{|r|r|r|r|r|}
\hline
Pattern & Algorithm & System 1 & System 2 & System 3\\
Size & & & &\\
\hline
2 & ABM & 8.89665& 24.6946& \bf 32.9261\\
 & HAL & 8.26117& 24.6946& \bf 32.9261\\
 & L & 6.08718& 24.6946& \bf 32.9261\\
 & SF & 4.28357& 9.87784& 24.6946\\
 & TBM & \bf 10.5142& \bf 32.9261& \bf 32.9261\\
\hline
4 & ABM & 20.4425& 46.7838& 68.9446\\
 & HAL & \bf 23.3995& \bf 51.0369& \bf 83.6137\\
 & L & 6.52724& 27.8712& 38.9093\\
 & SF & 4.29622& 9.84923& 23.3919\\
 & TBM & 21.2602& 49.123& 71.4517\\
\hline
6 & ABM & 28.1637& 60.2832& 89.4829\\
 & HAL & \bf 31.2569& \bf 63.6323& \bf 108.055\\
 & L & 6.45279& 27.4015& 37.9265\\
 & SF & 4.28142& 9.84005& 22.1973\\
 & TBM & 29.2294& 62.249& 93.8837\\
\hline
8 & ABM & 33.7463& 69.2828& 106.674\\
 & HAL & \bf 37.0999& \bf 73.0482& \bf 126.801\\
 & L & 6.34086& 26.6684& 36.5241\\
 & SF & 4.23323& 9.78229& 22.0342\\
 & TBM & 35.3437& 72.2627& 112.007\\
\hline
10 & ABM & 39.6329& 76.2308& 117.47\\
 & HAL & \bf 42.5986& \bf 80.5134& \bf 135.202\\
 & L & 6.32525& 26.6383& 36.1904\\
 & SF & 4.22537& 9.74924& 21.9134\\
 & TBM & 41.1973& 78.7439& 125.714\\
\hline
14 & ABM & 47.7986& 89.1214& 129.631\\
 & HAL & \bf 49.8997& \bf 92.9962& \bf 147.511\\
 & L & 6.22037& 25.9262& 33.6837\\
 & SF & 4.189& 9.72233& 21.1774\\
 & TBM & 49.3573& \bf 92.9962& 142.594\\
\hline
18 & ABM & 50.1514& 97.859& 141.352\\
 & HAL & 50.1514& \bf 101.773& \bf 159.021\\
 & L & 5.86185& 24.7023& 31.4115\\
 & SF & 4.05173& 9.63763& 21.0275\\
 & TBM & \bf 51.2912& 97.859& 149.667\\
\hline
\end{tabular}
\end{center}
\caption{Algorithm Speed (Characters Per Microsecond) 
in English Text Searches on Three Systems}
\end{table}
These results show that HAL, ABM, and TBM are quite close in
performance and are substantially better than the SF or L algorithms.
On System~1, TBM is a slightly faster than HAL on the longer
strings, but not enough to outweigh two significant drawbacks:
first, like SF, it takes $\Omega(mn)$ time in the
worst case; and, second, it achieves its slightly better average case
performance though the use of character frequency distribution
information that might need to be changed in applications of the
algorithm other than English text searches.  For both of these
reasons, TBM is not a good candidate for inclusion in a library of
generic algorithms.

For more machine independent performance measures, we show in
a later section the number of operations per character searched, for
various kinds of operations.  

\section{Hashed Accelerated Linear Algorithm}

The skip loop produces a dramatic effect on the algorithm, when we
search for a word or a phrase in an ordinary English text or in the
text in some other natural or programming language with a mid-sized
alphabet (26-256 characters, say).  However, algorithms that use this
technique are dependent on the alphabet size.  In case of a large
alphabet, the result is increased storage requirements and overhead
for initialization of the occurrence table.  Secondary effects are
also possible due to architectural reasons such as cache performance.
Performance of the skip loop is also diminished in cases in which the
pattern size is much greater than the size of the alphabet.  A good
example of this case is searching for DNA patterns, which could be
relatively long, say 250 characters, whereas the alphabet contains
only four characters.  In this section we show how to generalize the
skip loop to handle such adverse cases.

The key idea of the generalization is to apply a hash function
to the current position in the text to obtain an argument for the
skip function.

\begin{flushleft} \small
\begin{minipage}{\linewidth} \label{__n:gensearch.w:part17}
$\langle$Scan the text using a single-test skip loop with hashing {\footnotesize 11}$\rangle\equiv$
\vspace{-1ex}
\begin{list}{}{} \item
\mbox{}\verb@loop@\\
\mbox{}\verb@  k := k + skip(hash(text, n + k)); @\\
\mbox{}\verb@  exit when k >= 0;@\\
\mbox{}\verb@end loop;@\\
\mbox{}\verb@if k < pattern_size then@\\
\mbox{}\verb@  return n;@\\
\mbox{}\verb@end if;@\\
\mbox{}\verb@k := k - adjustment;@\\
\mbox{}\verb@@\end{list}
\vspace{-1ex}
\footnotesize\addtolength{\baselineskip}{-1ex}
\begin{list}{}{\setlength{\itemsep}{-\parsep}\setlength{\itemindent}{-\leftmargin}}
\item Used in part 12b.
\end{list}
\end{minipage}\\[4ex]
\end{flushleft}
We have seen that the skip loop works well when the cardinality of
domain of the skip function is of moderate size, say $\sigma = 256$,
as it is in most conventional string searches. When used
with sequences over a type $T$ with large (even infinite) cardinality,
\verb|hash| can be chosen so that it maps $T$ values to the range $[0,
\sigma)$.  Conversely, if the cardinality of $T$ is smaller than
$\sigma$, we can use more than one element of the text sequence to
compute the hash value in order to obtain $\sigma$ distinct values.  In the
context in which the skip loop appears, we always have available at
least \verb|pattern_size| elements; whenever \verb|pattern_size| 
is too small to yield $\sigma$ different hash values, we can either
make do with fewer values or resort to an algorithm that
does not use a skip loop, such as Algorithm L.  (The skip loop is not very
effective for small pattern lengths anyway.)

Of course, the skip table itself and the mismatch shift value must be
computed using the hash function.  Let \verb|suffix_size| be the
number of sequence elements used in computing the hash function, where
$1 \leq \mbox{\tt suffix\_size} \leq \mbox{\tt pattern\_size}$.

\begin{flushleft} \small
\begin{minipage}{\linewidth} \label{__n:gensearch.w:part18}
$\langle$Compute skip table and mismatch shift using the hash function {\footnotesize 12a}$\rangle\equiv$
\vspace{-1ex}
\begin{list}{}{} \item
\mbox{}\verb@@\\
\mbox{}\verb@for i in hash_range loop@\\
\mbox{}\verb@  skip(i) := pattern_size - suffix_size + 1;@\\
\mbox{}\verb@end loop;@\\
\mbox{}\verb@for j in a + suffix_size - 1 .. m - 2 loop@\\
\mbox{}\verb@  skip(hash(pattern, j)) := m - 1 - j;@\\
\mbox{}\verb@end loop;@\\
\mbox{}\verb@mismatch_shift := skip(hash(pattern, m - 1));@\\
\mbox{}\verb@skip(hash(pattern, m - 1)) := 0;@\\
\mbox{}\verb@@\end{list}
\vspace{-1ex}
\footnotesize\addtolength{\baselineskip}{-1ex}
\begin{list}{}{\setlength{\itemsep}{-\parsep}\setlength{\itemindent}{-\leftmargin}}
\item Used in part 12b.
\end{list}
\end{minipage}\\[4ex]
\end{flushleft}
The remainder of the computation can remain the same, so we have the
following algorithm in which it is assumed that the hash function uses
up to \verb|suffix_size| elements, where $1 \leq \mbox{\tt
suffix\_size} \leq \mbox{\tt pattern\_size}$.

\begin{flushleft} \small
\begin{minipage}{\linewidth} \label{__n:gensearch.w:part19}
$\langle$Hashed Accelerated Linear algorithm {\footnotesize 12b}$\rangle\equiv$
\vspace{-1ex}
\begin{list}{}{} \item
\mbox{}\verb@pattern_size := m - a; text_size := n - b; k := b;@\\
\mbox{}\verb@@\hbox{$\langle$Handle pattern size = 1 as a special case {\footnotesize 3b}$\rangle$}\verb@@\\
\mbox{}\verb@@\hbox{$\langle$Compute next table {\footnotesize 31a}$\rangle$}\verb@@\\
\mbox{}\verb@@\hbox{$\langle$Compute skip table and mismatch shift using the hash function {\footnotesize 12a}$\rangle$}\verb@@\\
\mbox{}\verb@large := text_size + 1;@\\
\mbox{}\verb@skip(hash(pattern, m - 1)) := large;@\\
\mbox{}\verb@adjustment := large + pattern_size - 1;@\\
\mbox{}\verb@k := k - n;@\\
\mbox{}\verb@loop@\\
\mbox{}\verb@  k := k + pattern_size - 1;@\\
\mbox{}\verb@  exit when k >= 0;@\\
\mbox{}\verb@  @\hbox{$\langle$Scan the text using a single-test skip loop with hashing {\footnotesize 11}$\rangle$}\verb@@\\
\mbox{}\verb@  @\hbox{$\langle$Verify match or recover from mismatch {\footnotesize 8a}$\rangle$}\verb@@\\
\mbox{}\verb@end loop;@\\
\mbox{}\verb@return n;@\\
\mbox{}\verb@@\end{list}
\vspace{-1ex}
\footnotesize\addtolength{\baselineskip}{-1ex}
\begin{list}{}{\setlength{\itemsep}{-\parsep}\setlength{\itemindent}{-\leftmargin}}
\item Used in part 29b.
\end{list}
\end{minipage}\\[4ex]
\end{flushleft}
This algorithm will be called HAL.  Note that AL is itself a
special case of HAL, obtained using
\[
\mbox{\tt hash}(\mbox{\tt text}, k) = \mbox{\tt text}(k)~~ 
\mbox{and (hence)}~~\mbox{\tt suffix\_size} = 1,
\]
By inlining \verb|hash|, we can use HAL instead of AL with minimal
performance penalty (none with a good compiler).

It is also noteworthy that in this application of hashing, a ``bad''
hash function causes no great harm, unlike the situation with
associative table searching in which hashing methods usually have
excellent average case performance (constant time) but with a bad hash
function can degrade terribly to linear time. (Thus, in a table with
thousands of elements, searching might take thousands of times longer
than expected.)  Here the worst that can happen---with, say, a hash
function that maps every element to the same value---is that a
sublinear algorithm degrades to linearity.  As a consequence,
in choosing hash functions we can lean toward ease of computation
rather than uniform distribution of the hash values.

There is, however, an essential requirement on the hash function
that must be observed when performing sequence matching in terms of an
equivalence relation $\equiv$ on sequence elements 
that is not an equality relation.  In this
case, we must require that equivalent values hash to the same value:
\[
   x \equiv y ~\supset~ \mbox{\tt hash}(x) = \mbox{\tt hash}(y)
\]
for all $x, y \in T$.  We discuss this requirement further
in a later section on generic library versions of the algorithms.

\section{Searching for DNA Patterns}

As an important example of the use of the HAL algorithm, consider DNA
searches, in which the alphabet has only four characters and patterns
may be very long, say 250 characters.  For this application we
experimented with hash functions $h_{ck}$ that map a string of $c$
characters into the integer range $[0,k)$.  We chose four such
functions, $h_{2,64}$, $h_{3,512}$, $h_{4,256}$, and $h_{5,256}$, all
of which add up the results of various shifts of the characters they
inspect.  For example,\footnote{In the \Cpp\ coding of this computation
we use shifts in place of multiplication and masking in place of
division. The actual \Cpp\ versions are shown in an appendix.}
\[
h_{4,256}(t, k) = (t(k - 3) + 2^2 t(k - 2) + 2^4 t(k - 1) + 2^6 t(k)) \mbox{mod}~ 256.
\]
The algorithms that use these hash functions bear the names HAL2,
HAL3, HAL4, and HAL5, respectively.  The other contestants were SF, L,
ABM and the Giancarlo-Boyer-Moore algorithm (GBM), which was described
in \cite{HumeSunday} and was considered to be the fastest for DNA
pattern matching.

We searched for patterns of size ranging from 20 to 200 in a text of
DNA strings obtained from \cite{DNAsource}.  The text is composed of
997,642 characters, and the test set included up to 80 different
patterns for each pattern size---40 strings chosen at evenly spaced
positions in the target text and up to 40 patterns chosen from another
file from \cite{DNAsource} (for longer pattern sizes there were fewer
than 40 patterns). Table~2 shows search speeds of the five algorithms
with code compiled and executed on the same three systems as in the
English text experiments (the systems described preceding Table~1).
\begin{table}
\begin{center}
\begin{tabular}{|r|r|r|r|r|}
\hline 
Pattern & Algorithm & System 1 & System 2 & System 3 \\
Size & & & & \\
\hline
20 & ABM & 16.0893& 37.3422& 55.6074\\
 & GBM & 25.8827& 62.3888& 138.267\\
 & HAL & 13.1493& 32.5853& 53.8514\\
 & HAL2 & \bf 28.7208& \bf 67.3143& \bf 146.168\\
 & HAL3 & 22.8165& 63.9486& 131.177\\
 & HAL4 & 21.9008& 58.135& 113.686\\
 & L & 4.33091& 16.3971& 18.9477\\
 & SF & 3.28732& 8.31851& 16.94\\
 & TBM & 18.0395& 42.6324& 63.1591\\
\hline
50 & ABM & 19.5708& 44.693& 70.6633\\
 & GBM & 33.6343& 78.046& 193.67\\
 & HAL & 13.6876& 33.736& 60.8033\\
 & HAL2 & \bf 49.5795& \bf 106.716& 275.215\\
 & HAL3 & 43.4625& \bf 106.716& \bf 290.505\\
 & HAL4 & 42.632& 96.8349& 249.004\\
 & L & 4.3519& 16.6003& 19.439\\
 & SF & 3.28906& 8.31333& 16.7599\\
 & TBM & 24.0764& 54.4696& 87.1514\\
\hline
100 & ABM & 21.2655& 49.6781& 73.4371\\
 & GBM & 36.6439& 85.8841& 220.311\\
 & HAL & 12.946& 32.0706& 56.3018\\
 & HAL2 & 70.4997& 163.457& 389.782\\
 & HAL3 & \bf 71.2744& \bf 187.673& \bf 460.651\\
 & HAL4 & 70.4997& 168.905& \bf 460.651\\
 & L & 4.24474& 16.0862& 19.1938\\
 & SF & 3.24623& 8.23929& 16.5054\\
 & TBM & 27.8368& 66.6732& 105.566\\
\hline
150 & ABM & 24.2269& 56.1366& 86.667\\
 & GBM & 37.6383& 86.667& 205.834\\
 & HAL & 14.0205& 34.5456& 60.9879\\
 & HAL2 & 84.3097& 197.601& \bf 548.891\\
 & HAL3 & \bf 91.641& \bf 247.001& \bf 548.891\\
 & HAL4 & 90.3318& 235.239& 494.002\\
 & L & 4.33395& 16.3037& 19.5258\\
 & SF & 3.28992& 8.33056& 17.0935\\
 & TBM & 29.1393& 72.6474& 107.392\\
\hline
200 & ABM & 23.9786& 55.3853& 86.3636\\
 & GBM & 37.0578& 90.9902& 212.31\\
 & HAL & 13.3106& 33.3036& 57.9028\\
 & HAL2 & 89.3449& 221.541& 509.545\\
 & HAL3 & 103.527& \bf 283.081& \bf 636.931\\
 & HAL4 & \bf 105.196& \bf 283.081& 566.161\\
 & L & 4.26565& 16.2275& 19.0841\\
 & SF & 3.25946& 8.28528& 16.8167\\
 & TBM & 28.7321& 73.8471& 113.232\\
\hline
\end{tabular}
\end{center}
\caption{Algorithm Speed (Characters Per Microsecond) 
in DNA Searches on Three Systems}
\end{table}

We see that each of the versions of HAL is significantly faster than
any of the other algorithms, and the speed advantage increases with
longer patterns---for patterns of size 200, HAL5 is over 3.5 to 5.5
times faster than its closest competitor, GBM, depending on the system.
It appears that HAL4 is slightly faster than HAL5 or HAL3, but further
experiments with different hash functions might yield even better
performance.

\section{Large Alphabet Case}

Suppose the alphabet $\Sigma$ has $2^{16} = 65,536$ symbols, as in
Unicode for example. To use the skip loop directly we must
initialize a table with 65,536 entries.  If we are only going to
search, say, for a short pattern in a 10,000 character text, the
initialization overhead dominates the rest of the computation.

One way to eliminate dependency on the alphabet size is to use a large
zero-filled global table $\mbox{\tt skip1}(x)=\mbox{\tt skip}(x)-m$,
so that the algorithm fills at most $m$ positions with
pattern-dependent values at the beginning, performs a search, and then
restores zeroes.  This approach makes the algorithm non-reentrant and
therefore not suitable for multi-threaded applications, but it seems
worth investigating for single-threaded applications.

Another approach is to use HAL with, say, 
\[ 
H(t, k) = t(k)~ \mbox{mod}~ 256
\] 
as the hash function.  In order to compare these two approaches we
implemented a non-hashed version of AL using {\tt skip1}, called NHAL,
and benchmarked it against HAL with $H$ as the hash function.  The
\Cpp\ code for NHAL is shown in an appendix.

We searched for patterns of size ranging from 2 to 18 in randomly
generated texts of size 1,000,000 characters, with each character
being an integer chosen with uniform distribution from 0 to 65,535.
Patterns were chosen from the text at random positions.  The test set
included 500 different patterns for each pattern size.  Table~3
summarizes the timings obtained using the same three systems as
decribed preceding Table~1.  We can see that HAL demonstrates
significantly better performance than NHAL.  On systems~1 and~2 the
ratio of HAL's speed to NHAL's is much higher than on system~3, and we
attribute this disparity to poor optimization abilities of the
compilers we used on systems~1 and~2.

\begin{table}
\begin{center}
\begin{tabular}{|r|r|r|r|r|}
\hline
Pattern & Algorithm & System 1 & System 2 & System 3\\
Size & & & &\\
\hline
2 & HAL & \bf 10.4369& 27.3645& 39.5632\\
 & L & 7.11972& \bf 28.5543& \bf 41.5664\\
 & NHAL & 6.63479& 12.3915& 26.4818\\
 & SF & 4.48334& 9.83157& 23.125\\
\hline
4 & HAL & \bf 18.7455& \bf 44.375& \bf 64.3873\\
 & L & 7.125& 28.3082& 41.5665\\
 & NHAL & 10.9539& 21.0497& 47.5906\\
 & SF & 4.48611& 9.86112& 22.8038\\
\hline
6 & HAL & \bf 25.3206& \bf 54.9243& \bf 86.7225\\
 & L & 7.1179& 28.4091& 41.7146\\
 & NHAL & 14.2358& 28.1663& 63.3741\\
 & SF & 4.505& 9.86663& 23.0451\\
\hline
8 & HAL & \bf 31.1354& \bf 67.1919& \bf 94.0686\\
 & L & 7.12946& 28.6296& 41.676\\
 & NHAL & 16.9606& 33.5959& 80.3025\\
 & SF & 4.49445& 9.88709& 22.7062\\
\hline
10 & HAL & \bf 35.7717& \bf 72.4895& \bf 112.484\\
 & L & 7.09913& 28.3655& 41.2915\\
 & NHAL & 19.1634& 38.3768& 98.8494\\
 & SF & 4.49017& 9.85507& 22.8114\\
\hline
14 & HAL & \bf 42.9195& \bf 78.1701& \bf 149.234\\
 & L & 7.1132& 28.5491& 41.0393\\
 & NHAL & 23.5262& 47.5818& 136.798\\
 & SF & 4.48911& 9.80043& 22.7996\\
\hline
18 & HAL & \bf 47.51862& \bf 96.9324& \bf 173.458\\
 & L & 7.144521& 28.1684& 41.1963\\
 & NHAL & 26.4274& 56.8225& 164.785\\
 & SF & 4.48312& 9.80864& 22.8868\\
\hline
\end{tabular}
\end{center}
\caption{Algorithm Speed (Characters Per Microsecond) in Large
Alphabet Case on Three Systems}
\end{table}

We conclude that the hashing technique presents a viable and efficient
way to eliminate alphabet-size dependency of search algorithms that
use the skip loop.

\section{Generic Search Algorithms}

As we have seen, the HAL algorithm retains its efficiency advantages
in a wide variety of search problems of practical interest, including
traditional string searching with small or large alphabets, and short
or long patterns.  These qualities make it a good candidate for
abstraction to searching in sequences over an arbitrary type $T$, for
inclusion in generic software libraries such as the \Cpp\ Standard
Template Library (STL) \cite{StepanovLee,STLBook}.

By some definitions of genericity, HAL is already a generic algorithm,
since the hash function can be made a parameter and thus the algorithm
can be adapted to work with any type $T$.  In STL, however, another
important issue bearing on the generality of operations on linear
sequences is the kind of access to the sequences assumed---random
access or something weaker, such as forward, single-step advances
only.  STL generic algorithms are specified to access sequences via
iterators, which are generalizations of ordinary C/\Cpp\ pointers. STL
defines five categories of iterators, the most powerful being
random-access iterators, for which computing $i + n$ or $i - n$, for
iterator $i$ and integer $n$, is a constant time operation.  Forward
iterators allow scanning a sequence with only single-step advances and
only in a forward direction.  AL and HAL require random access for
most efficient operation of the skip loop, whereas Algorithm L, with
only minor modifications to the expository version, 
can be made to work efficiently with forward iterators.

The efficiency issue is considered crucial in the STL approach to
genericity.  STL is not a set of specific software components but a
set of requirements which components must satisfy.  By making time
complexity part of the requirements for components, STL ensures that
compliant components not only have the specified interfaces and
semantics but also meet certain computing time bounds.  The
requirements on most components are stated in terms of inputs and
outputs that are linear sequences over some type $T$.  The
requirements are stated as generally as possible, but balanced against
the goal of using efficient algorithms.  In fact, the requirements
were generally chosen based on knowledge of existing efficient
concrete algorithms, by finding the weakest assumptions---about $T$
and about how the sequence elements are accessed---under which those
algorithms could still be used without losing their efficiency.  In
most cases, the computing time requirements are stated as worst-case
bounds, but exceptions are made when the concrete algorithms with the
best worst-case bounds are not as good in the average case as other
algorithms, provided the worst cases occur very infrequently in
practice.

In the case of sequence search algorithms, the concrete algorithms
considered for generalization to include in STL were various
string-search algorithms, including BM, KMP, and SF.  Although KMP has
the lowest worst-case bound, it was stated in the original STL report
\cite{StepanovLee} that SF was superior in the average
case.\footnote{The original STL requirements included the following
statement (which has been dropped in more recent versions of the Draft
\Cpp\ Standard): ``\ldots The Knuth-Morris-Pratt algorithm is not used
here. While the KMP algorithm guarantees linear time, it tends to be
slower in most practical cases than the naive algorithm with
worst-case quadratic behavior \ldots.'' As we have already seen from
Table~1, however, a suitably optimized version of KMP---Algorithm
L---is significantly faster than SF.}  And although BM has excellent
average time behavior, it was evidently ruled out as a generic
algorithm because of its alphabet size dependency. Thus the generic
search algorithm requirements were written with a $O(mn)$ time bound,
to allow its implementation by SF.\footnote{This did not preclude
library implementors from also supplying a specialization of the
search operation for the string search case, implemented with BM.  The
original requirements statement for the search operation noted this
possibility but more recent drafts fail to mention it. We are not
aware of any currently available STL implementations that do provide
such a specialization.}

Thus in the Draft \Cpp\ Standard dated December 1996 \cite{DraftCPP}, two
sequence search functions are required, with the specifications:
\begin{description}
\item[~] 
\begin{verbatim} 
template <typename ForwardIterator1, typename ForwardIterator2> 
ForwardIterator1 search(ForwardIterator1 first1,
                        ForwardIterator1 last1, 
                        ForwardIterator2 first2, 
                        ForwardIterator2 last2);
\end{verbatim}
\item[~]
\begin{verbatim}
template <typename ForwardIterator1, typename ForwardIterator2, 
          typename BinaryPredicate> 
ForwardIterator1 search(ForwardIterator1 first1, 
                        ForwardIterator1 last1, 
                        ForwardIterator2 first2, 
                        ForwardIterator2 last2, 
                        BinaryPredicate pred);
\end{verbatim}
\item[Effects:] Finds a subsequence of equal values in a sequence.

\item[Returns:] The first iterator $i$ in the range $[\mbox{first1},
\mbox{last1} - (\mbox{last2} - \mbox{first2}))$ such that for any
non-negative integer $n$ less than $\mbox{last2} - \mbox{first2}$ the
following corresponding conditions hold: $*(i + n) = *( \mbox{first2}
+ n)$, $\mbox{pred}(*(i + n), *(\mbox{first2} + n)) \not=
\mbox{false}$.  Returns $\mbox{last1}$ if no such iterator is found.

\item[Complexity:] At most $(\mbox{last1} - \mbox{first1}) * (\mbox{last2} -
\mbox{first2})$ applications of the corresponding predicate.
\end{description}
Before going further, we note that the results of the present article
would allow the complexity requirement to be 
replaced with the much stronger requirement that the computing
time be $O((\mbox{last1} - \mbox{first1}) + (\mbox{last2} -
\mbox{first2}))$.

We will base our discussion on the first interface, which assumes
operator== is used for testing sameness of two sequence elements; the
only added issue for the binary predicate case is the requirement
mentioned earlier, that for HAL we must choose a hash function
compatible with the binary predicate, in the sense that any two values
that are equivalent according to the predicate must be mapped to the
same value by the hash function.  Fortunately, for a given predicate
it is usually rather easy to choose a hash function that guarantees
this property. (A heuristic guide is to choose a hash function that uses
less information than the predicate.)

The fact that this standard interface only assumes forward iterators
would seem to preclude HAL, since the skip loop requires random
access. There are however many cases of sequence searching in which we
do have random access, and we do not want to miss the speedup afforded
by the skip loop in those cases.  Fortunately, it is possible to
provide for easy selection of the most appropriate algorithm
under different actual circumstances, including whether random access
or only forward access is available, and whether type $T$ has a small
or large number of distinct values.  For this purpose we use
\emph{traits}, a programming device for compile-time selection of
alternative type and code definitions.  Traits are supported in \Cpp\ by
the ability to give definitions of function templates or class
templates for specializations of their template
parameters.\footnote{Limited forms of the trait device were used in
defining some iterator operations in the first implementations of
STL. More recently the trait device has been adopted more broadly in
other parts of the library, particularly to provide different
definitions of floating point and other parameters used in numeric
algorithms. The most elaborate uses of the device employ the recently
added \Cpp\ feature of \emph{partial specialization}, in which new
definitions can be given with some template parameters specialized
while others are left unspecialized.  Few \Cpp\ compilers currently
support partial specialization, but we do not need it here anyway.}

Algorithm L is not difficult to adapt to work with iterators
instead of array indexing.  The most straightforward translation
would require random access iterators, but with a few adjustments
we can express the algorithm entirely with forward iterator operations,
making it fit the STL \verb|search| function interface.  

\begin{flushleft} \small
\begin{minipage}{\linewidth} \label{__n:gensearch.w:part20}
$\langle$User level search function {\footnotesize 19a}$\rangle\equiv$
\vspace{-1ex}
\begin{list}{}{} \item
\mbox{}\verb@template <typename ForwardIterator1, typename ForwardIterator2>@\\
\mbox{}\verb@inline ForwardIterator1 search(ForwardIterator1 text,@\\
\mbox{}\verb@                               ForwardIterator1 textEnd,@\\
\mbox{}\verb@                               ForwardIterator2 pattern,@\\
\mbox{}\verb@                               ForwardIterator2 patternEnd)@\\
\mbox{}\verb@{@\\
\mbox{}\verb@  typedef iterator_traits<ForwardIterator1> T;@\\
\mbox{}\verb@  return __search(text, textEnd, pattern, patternEnd, T::iterator_category());@\\
\mbox{}\verb@}@\\
\mbox{}\verb@@\end{list}
\vspace{-1ex}
\footnotesize\addtolength{\baselineskip}{-1ex}
\begin{list}{}{\setlength{\itemsep}{-\parsep}\setlength{\itemindent}{-\leftmargin}}
\item Used in part 38b.
\end{list}
\end{minipage}\\[4ex]
\end{flushleft}
When we only have forward iterators, we use Algorithm L.

\begin{flushleft} \small \label{__n:gensearch.w:part21}
$\langle$Forward iterator case {\footnotesize 19b}$\rangle\equiv$
\vspace{-1ex}
\begin{list}{}{} \item
\mbox{}\verb@template <typename ForwardIterator1, typename ForwardIterator2>@\\
\mbox{}\verb@inline ForwardIterator1 __search(ForwardIterator1 text,@\\
\mbox{}\verb@                                 ForwardIterator1 textEnd,@\\
\mbox{}\verb@                                 ForwardIterator2 pattern,@\\
\mbox{}\verb@                                 ForwardIterator2 patternEnd,@\\
\mbox{}\verb@                                 forward_iterator_tag)@\\
\mbox{}\verb@{@\\
\mbox{}\verb@  return __search_L(text, textEnd, pattern, patternEnd);@\\
\mbox{}\verb@}@\\
\mbox{}\verb@@\\
\mbox{}\verb@template <typename ForwardIterator1, typename ForwardIterator2>@\\
\mbox{}\verb@ForwardIterator1 __search_L(ForwardIterator1 text,@\\
\mbox{}\verb@                            ForwardIterator1 textEnd,@\\
\mbox{}\verb@                            ForwardIterator2 pattern,@\\
\mbox{}\verb@                            ForwardIterator2 patternEnd)@\\
\mbox{}\verb@{@\\
\mbox{}\verb@  typedef typename iterator_traits<ForwardIterator2>::difference_type Distance2;@\\
\mbox{}\verb@  ForwardIterator1 advance, hold;@\\
\mbox{}\verb@  ForwardIterator2 p, p1;@\\
\mbox{}\verb@  Distance2 j, m;@\\
\mbox{}\verb@  vector<Distance2> next;@\\
\mbox{}\verb@  vector<ForwardIterator2> pattern_iterator;@\\
\mbox{}\verb@  @\hbox{$\langle$Compute next table (C++ forward) {\footnotesize 20a}$\rangle$}\verb@@\\
\mbox{}\verb@  m = next.size();@\\
\mbox{}\verb@  @\hbox{$\langle$Algorithm L, optimized linear pattern search (C++) {\footnotesize 21a}$\rangle$}\verb@@\\
\mbox{}\verb@}@\\
\mbox{}\verb@@\end{list}
\vspace{-1ex}
\footnotesize\addtolength{\baselineskip}{-1ex}
\begin{list}{}{\setlength{\itemsep}{-\parsep}\setlength{\itemindent}{-\leftmargin}}
\item Used in part 38b.
\end{list}
\end{flushleft}
We store the \verb|next| table in an STL vector, which provides random
access to the integral next values; to be able to get from them 
back to the correct positions in the pattern sequence we also 
store iterators in another vector, \verb|pattern_iterator|.

\begin{flushleft} \small
\begin{minipage}{\linewidth} \label{__n:gensearch.w:part22}
$\langle$Compute next table (C++ forward) {\footnotesize 20a}$\rangle\equiv$
\vspace{-1ex}
\begin{list}{}{} \item
\mbox{}\verb@compute_next(pattern, patternEnd, next, pattern_iterator);@\\
\mbox{}\verb@@\end{list}
\vspace{-1ex}
\footnotesize\addtolength{\baselineskip}{-1ex}
\begin{list}{}{\setlength{\itemsep}{-\parsep}\setlength{\itemindent}{-\leftmargin}}
\item Used in part 19b.
\end{list}
\end{minipage}\\[4ex]
\end{flushleft}
\begin{flushleft} \small \label{__n:gensearch.w:part23}
$\langle$Define procedure to compute next table (C++ forward) {\footnotesize 20b}$\rangle\equiv$
\vspace{-1ex}
\begin{list}{}{} \item
\mbox{}\verb@template <typename ForwardIterator, typename Distance>@\\
\mbox{}\verb@void compute_next(ForwardIterator pattern, @\\
\mbox{}\verb@                  ForwardIterator patternEnd,@\\
\mbox{}\verb@                  vector<Distance>& next, @\\
\mbox{}\verb@                  vector<ForwardIterator>& pattern_iterator)@\\
\mbox{}\verb@{@\\
\mbox{}\verb@  Distance t = -1;@\\
\mbox{}\verb@  next.reserve(32);@\\
\mbox{}\verb@  pattern_iterator.reserve(32);@\\
\mbox{}\verb@  next.push_back(-1);@\\
\mbox{}\verb@  pattern_iterator.push_back(pattern);@\\
\mbox{}\verb@  for (;;) {@\\
\mbox{}\verb@    ForwardIterator advance = pattern;@\\
\mbox{}\verb@    ++advance;@\\
\mbox{}\verb@    if (advance == patternEnd)@\\
\mbox{}\verb@      break;@\\
\mbox{}\verb@    while (t >= 0 && *pattern != *pattern_iterator[t]) @\\
\mbox{}\verb@       t = next[t];@\\
\mbox{}\verb@    ++pattern; ++t;@\\
\mbox{}\verb@    if (*pattern == *pattern_iterator[t]) @\\
\mbox{}\verb@      next.push_back(next[t]);@\\
\mbox{}\verb@    else@\\
\mbox{}\verb@      next.push_back(t);@\\
\mbox{}\verb@    pattern_iterator.push_back(pattern);@\\
\mbox{}\verb@  }@\\
\mbox{}\verb@}@\\
\mbox{}\verb@@\end{list}
\vspace{-1ex}
\footnotesize\addtolength{\baselineskip}{-1ex}
\begin{list}{}{\setlength{\itemsep}{-\parsep}\setlength{\itemindent}{-\leftmargin}}
\item Used in part 38b.
\end{list}
\end{flushleft}
Returning to the search algorithm itself, the details are as follows:

\begin{flushleft} \small
\begin{minipage}{\linewidth} \label{__n:gensearch.w:part24}
$\langle$Algorithm L, optimized linear pattern search (C++) {\footnotesize 21a}$\rangle\equiv$
\vspace{-1ex}
\begin{list}{}{} \item
\mbox{}\verb@@\hbox{$\langle$Handle pattern size = 1 as a special case (C++) {\footnotesize 21b}$\rangle$}\verb@@\\
\mbox{}\verb@p1 = pattern; ++p1;@\\
\mbox{}\verb@while (text != textEnd) {@\\
\mbox{}\verb@  @\hbox{$\langle$Scan the text for a possible match (C++) {\footnotesize 21c}$\rangle$}\verb@@\\
\mbox{}\verb@  @\hbox{$\langle$Verify whether a match is possible at the position found (C++) {\footnotesize 21d}$\rangle$}\verb@@\\
\mbox{}\verb@  @\hbox{$\langle$Recover from a mismatch using the next table (C++ forward) {\footnotesize 22a}$\rangle$}\verb@@\\
\mbox{}\verb@}@\\
\mbox{}\verb@return textEnd;@\\
\mbox{}\verb@@\end{list}
\vspace{-1ex}
\footnotesize\addtolength{\baselineskip}{-1ex}
\begin{list}{}{\setlength{\itemsep}{-\parsep}\setlength{\itemindent}{-\leftmargin}}
\item Used in part 19b.
\end{list}
\end{minipage}\\[4ex]
\end{flushleft}
For the case of pattern size 1, we use the STL generic linear
search algorithm, $\mbox{\tt find}$.

\begin{flushleft} \small
\begin{minipage}{\linewidth} \label{__n:gensearch.w:part25}
$\langle$Handle pattern size = 1 as a special case (C++) {\footnotesize 21b}$\rangle\equiv$
\vspace{-1ex}
\begin{list}{}{} \item
\mbox{}\verb@if (next.size() == 1)@\\
\mbox{}\verb@  return find(text, textEnd, *pattern);@\\
\mbox{}\verb@@\end{list}
\vspace{-1ex}
\footnotesize\addtolength{\baselineskip}{-1ex}
\begin{list}{}{\setlength{\itemsep}{-\parsep}\setlength{\itemindent}{-\leftmargin}}
\item Used in parts 21a, 42a, 46a.
\end{list}
\end{minipage}\\[4ex]
\end{flushleft}
The three parts of the body of the main loop are direct translations
from the Ada versions given earlier, using pointer manipulation
in place of array indexing.

\begin{flushleft} \small
\begin{minipage}{\linewidth} \label{__n:gensearch.w:part26}
$\langle$Scan the text for a possible match (C++) {\footnotesize 21c}$\rangle\equiv$
\vspace{-1ex}
\begin{list}{}{} \item
\mbox{}\verb@while (*text != *pattern) @\\
\mbox{}\verb@  if (++text == textEnd)@\\
\mbox{}\verb@    return textEnd;@\\
\mbox{}\verb@@\end{list}
\vspace{-1ex}
\footnotesize\addtolength{\baselineskip}{-1ex}
\begin{list}{}{\setlength{\itemsep}{-\parsep}\setlength{\itemindent}{-\leftmargin}}
\item Used in part 21a.
\end{list}
\end{minipage}\\[4ex]
\end{flushleft}
\begin{flushleft} \small
\begin{minipage}{\linewidth} \label{__n:gensearch.w:part27}
$\langle$Verify whether a match is possible at the position found (C++) {\footnotesize 21d}$\rangle\equiv$
\vspace{-1ex}
\begin{list}{}{} \item
\mbox{}\verb@p = p1; j = 1;@\\
\mbox{}\verb@hold = text;@\\
\mbox{}\verb@if (++text == textEnd)@\\
\mbox{}\verb@  return textEnd;@\\
\mbox{}\verb@while (*text == *p) {@\\
\mbox{}\verb@  if (++p == patternEnd) @\\
\mbox{}\verb@    return hold;@\\
\mbox{}\verb@  if (++text == textEnd)@\\
\mbox{}\verb@    return textEnd;@\\
\mbox{}\verb@  ++j;@\\
\mbox{}\verb@}@\\
\mbox{}\verb@@\end{list}
\vspace{-1ex}
\footnotesize\addtolength{\baselineskip}{-1ex}
\begin{list}{}{\setlength{\itemsep}{-\parsep}\setlength{\itemindent}{-\leftmargin}}
\item Used in part 21a.
\end{list}
\end{minipage}\\[4ex]
\end{flushleft}
\begin{flushleft} \small
\begin{minipage}{\linewidth} \label{__n:gensearch.w:part28}
$\langle$Recover from a mismatch using the next table (C++ forward) {\footnotesize 22a}$\rangle\equiv$
\vspace{-1ex}
\begin{list}{}{} \item
\mbox{}\verb@for (;;) {@\\
\mbox{}\verb@  j = next[j];@\\
\mbox{}\verb@  if (j < 0) {@\\
\mbox{}\verb@    ++text;@\\
\mbox{}\verb@    break;@\\
\mbox{}\verb@  }@\\
\mbox{}\verb@  if (j == 0)@\\
\mbox{}\verb@    break;@\\
\mbox{}\verb@  p = pattern_iterator[j];@\\
\mbox{}\verb@  while (*text == *p) {@\\
\mbox{}\verb@    ++text; ++p; ++j;@\\
\mbox{}\verb@    if (p == patternEnd) {@\\
\mbox{}\verb@      @\hbox{$\langle$Compute and return position of match {\footnotesize 22b}$\rangle$}\verb@@\\
\mbox{}\verb@    }@\\
\mbox{}\verb@    if (text == textEnd)@\\
\mbox{}\verb@      return textEnd;@\\
\mbox{}\verb@  }@\\
\mbox{}\verb@}@\\
\mbox{}\verb@@\end{list}
\vspace{-1ex}
\footnotesize\addtolength{\baselineskip}{-1ex}
\begin{list}{}{\setlength{\itemsep}{-\parsep}\setlength{\itemindent}{-\leftmargin}}
\item Used in part 21a.
\end{list}
\end{minipage}\\[4ex]
\end{flushleft}
Returning the match position requires use of the \verb|hold|
iterator saved for that purpose.

\begin{flushleft} \small
\begin{minipage}{\linewidth} \label{__n:gensearch.w:part29}
$\langle$Compute and return position of match {\footnotesize 22b}$\rangle\equiv$
\vspace{-1ex}
\begin{list}{}{} \item
\mbox{}\verb@advance = hold;@\\
\mbox{}\verb@for (int i = m; --i >= 0;) @\\
\mbox{}\verb@  ++advance;@\\
\mbox{}\verb@while (advance != text) @\\
\mbox{}\verb@  ++advance, ++hold;@\\
\mbox{}\verb@return hold;@\\
\mbox{}\verb@@\end{list}
\vspace{-1ex}
\footnotesize\addtolength{\baselineskip}{-1ex}
\begin{list}{}{\setlength{\itemsep}{-\parsep}\setlength{\itemindent}{-\leftmargin}}
\item Used in part 22a.
\end{list}
\end{minipage}\\[4ex]
\end{flushleft}
Through the use of traits, we provide for automatic selection of
either the above version of algorithm L in the case of forward or
bidirectional iterators, or the faster HAL algorithm when random
access to the sequences is available.  STL random access iterators
permit the use of either array index notation very similar that in the
expository version of the algorithm, or pointer notation as shown
above for algorithm L, but with additional operations such as $p + k$.
Although it is commonplace to use pointer
notation for efficiency reasons, we avoid it in this case because the
calculation of the \verb|large| value cannot be guaranteed to be valid
in pointer arithmetic.  The advantage of the single-test skip loop
outweighs any disadvantage due to array notation calculations.

\begin{table}[h]
\begin{tabular}{|r|l|r|r|r|r|r|r|}
\hline
Pattern & Algorithm & Comparisons & Other & Big & Other & Distance & Total \\
Size & &  & Accesses & Jumps & Iter Ops & Ops & Ops\\
\hline
 2 & SF & 1.036 & 0.001 & 0.000 & 4.192 & 2.002 & 7.231\\
 & L & 1.028 & 0.001 & 0.000 & 4.095 & 0.177 & 5.301\\
 & HAL & 0.018 & 0.513 & 0.551 & 1.104 & 2.431 & 4.617\\
 & ABM & 0.017 & 0.528 & --- & --- & --- & ---\\
 & TBM & 0.021 & 0.511 & --- & --- & --- & ---\\
\hline
 4 & SF & 1.034 & 0.000 & 0.000 & 4.170 & 2.000 & 7.203\\
 & L & 1.031 & 0.000 & 0.000 & 4.098 & 0.159 & 5.288\\
 & HAL & 0.013 & 0.266 & 0.291 & 0.583 & 0.658 & 1.811\\
 & ABM & 0.013 & 0.277 & --- & --- & --- & ---\\
 & TBM & 0.014 & 0.266 & --- & --- & --- & ---\\
\hline
 6 & SF & 1.042 & 0.000 & 0.000 & 4.211 & 2.000 & 7.254\\
 & L & 1.037 & 0.000 & 0.000 & 4.119 & 0.194 & 5.350\\
 & HAL & 0.011 & 0.189 & 0.211 & 0.422 & 0.482 & 1.315\\
 & ABM & 0.012 & 0.198 & --- & --- & --- & ---\\
 & TBM & 0.012 & 0.189 & --- & --- & --- & ---\\
\hline
 8 & SF & 1.048 & 0.000 & 0.000 & 4.243 & 2.000 & 7.291\\
 & L & 1.042 & 0.000 & 0.000 & 4.135 & 0.220 & 5.396\\
 & HAL & 0.010 & 0.150 & 0.170 & 0.339 & 0.392 & 1.060\\
 & ABM & 0.011 & 0.157 & --- & --- & --- & ---\\
 & TBM & 0.011 & 0.150 & --- & --- & --- & ---\\
\hline
 10 & SF & 1.052 & 0.000 & 0.000 & 4.263 & 2.000 & 7.315\\
 & L & 1.044 & 0.000 & 0.000 & 4.142 & 0.233 & 5.418\\
 & HAL & 0.009 & 0.126 & 0.144 & 0.289 & 0.337 & 0.905\\
 & ABM & 0.010 & 0.132 & --- & --- & --- & ---\\
 & TBM & 0.010 & 0.126 & --- & --- & --- & ---\\
\hline
 14 & SF & 1.077 & 0.000 & 0.000 & 4.384 & 2.000 & 7.460\\
 & L & 1.060 & 0.000 & 0.000 & 4.197 & 0.328 & 5.585\\
 & HAL & 0.010 & 0.105 & 0.125 & 0.250 & 0.305 & 0.796\\
 & ABM & 0.010 & 0.109 & --- & --- & --- & ---\\
 & TBM & 0.011 & 0.105 & --- & --- & --- & ---\\
\hline
 18 & SF & 1.105 & 0.000 & 0.000 & 4.525 & 2.000 & 7.629\\
 & L & 1.077 & 0.000 & 0.000 & 4.257 & 0.436 & 5.770\\
 & HAL & 0.011 & 0.096 & 0.117 & 0.234 & 0.295 & 0.753\\
 & ABM & 0.010 & 0.099 & --- & --- & --- & ---\\
 & TBM & 0.011 & 0.096 & --- & --- & --- & ---\\
\hline
\end{tabular}
\caption{Average Number of Operations Per Character in
English Text Searches}
\label{table:counts}
\end{table}

The trait interface also allows the user to supply the hash function,
but various useful default hash functions can be provided.  The full
details, including complete source code, are shown in an appendix.
The code is available from {\em http://www.cs.rpi.edu/\~{}musser/gp}.
The code supplied includes a set of operation counting components
\cite{MusserCounting} that permit easy gathering of statistics on many
different kinds of operations, including data element accesses and
comparisons, iterator operations, and ``distance operations,'' which
are arithmetic operations on integer results of iterator subtractions.
These counts are obtained without modifying the source code of the
algorithms at all, by specializing their type parameters with classes
whose operations have counters built into them.
Table~\ref{table:counts} shows counts of data comparisons and other
data accesses, iterator ``big jumps'' and other iterator operations,
and distance operations.  In each case the counts are divided by
the number of characters searched.  These statistics come
from searches of the same English text, \emph{Through the Looking
Glass}, with the same selection of patterns, as discussed earlier.
For ABM and TBM, not all operations were counted because the
algorithms are from Hume and Sunday's original C code and therefore
could not be specialized with the counting components.  For these
algorithms a manually instrumented version (supplied as part of the
code distribution \cite{HumeSunday}) kept count of data comparisons and
accesses.

The table shows that HAL, like ABM and TBM, does remarkably few
equality comparison operations on sequence elements---only about 1 per
100 elements for the longer patterns, no more than twice that for the
shorter ones.  They do access the elements substantially more often
than that, in their respective skip loops, but still always
sublinearly.  With string matching, the comparisons and
accesses are inexpensive, but in other applications of sequence
matching they might cost substantially more than iterator or distance
operations.  In such applications the savings in execution time over
SF or L could be even greater.  

For example, an appendix shows one experiment in which the text of
\emph{Through the Looking Glass} was stored as a sequence of words,
each word being a character string, and the patterns were word
sequences of different lengths chosen from evenly spaced positions in
the target word sequence.  In this case, element comparisons were word
comparisons, which could be significantly more costly than iterator or
distance operations.  HAL was again substantially faster than the
other contestants, SF and L. The ABM and TBM algorithms from
\cite{HumeSunday} were not considered because they are only applicable
to string matching, but it was easy to specialize the three generic
algorithms to this case of sequence matching, just by plugging in the
appropriate types and, in the case of HAL, defining a suitable hash
function.  (We used a function that returns the first character of a
word.)

\section{How to Obtain the Appendices and Code}

An expanded version of this paper, including appendices that contain
and document the complete source code for all benchmark experiments
described in the paper, will be maintained indefinitely for public
access on the Internet at \emph{http://www.cs.rpi.edu/\~{}musser/gp/}.
By downloading the Nuweb source file, \emph{gensearch.w}, and using
Briggs' Nuweb tool \cite{Briggs},\footnote{We started with a version
of the Nuweb tool previously modified by Ramsdell and Mengel and made
additional small changes in terminology in the \LaTeX\ file the tool
produces: ``part'' is used in place of ``scrap'' and ``definition'' in
place of ``macro.''  This version, called Nuweb 0.91, is available
from \emph{http://www.cs.rpi.edu/\~{}musser/gp/}. The new version does
not differ from previous versions in the way it produces code files
from Nuweb source files.}  readers can also easily generate all of the
source code described in the paper and appendices.

\section{Conclusion}

When we began this research, our main goal was to develop a generic
sequence search algorithm with a linear worst-case time bound and with
better average case performance than KMP and SF, so that it could be
used in generic software libraries such as the \Cpp\ Standard Template
Library.  We expected that for most of the useful special cases, such
as English text or DNA substring matching, it would probably be better
to provide separate algorithms tailored to those cases.  It was
therefore surprising to discover that for the substring matching
problem itself a new, superior algorithm could be obtained by
combining Boyer and Moore's skip loop with the Knuth-Morris-Pratt
algorithm.  By also developing a hashed version of the skip loop and
providing for selection of different variants of the technique using
traits, we obtained a generic algorithm, HAL, with all of the
attributes we originally sought. Moreover, when specialized to the
usual string matching cases of the most practical interest, such as
English text matching and DNA string matching, the new algorithm beats
most of the existing string matching algorithms.

Since HAL has a linear upper bound on the number of comparisons, it
can be used even in mission-critical applications where the
potential $O(m n)$ behavior of the straightforward algorithm or Hume
and Sunday's TBM algorithm would be a serious concern.  In such
applications, as well as in less-critical applications, HAL's
performance in the average case is not only linear, but sublinear,
beating even the best versions of the Boyer Moore algorithm.  Since we
have provided it in a generic form---in particular, in the
framework of the \Cpp\ Standard Template Library---the new
algorithm is easily reusable in many different contexts.

{\bf Acknowledgement:} This work was partially supported by a grant
from IBM Corporation.

\appendix
\section{Tests of Expository Versions of the Algorithms}

To help ensure against errors in the expository versions of the
algorithms in this paper, we compiled them as part of several Ada test
programs, using both the GNAT Ada 95 compiler, version 3.09, and the
Aonix ObjectAda compiler, Special Edition, version 7.1.  

\subsection{Algorithm Declarations}

We have not attempted to develop Ada generic subprograms based on the
expository versions of the algorithms; instead we encapsulate them
here with non-generic interfaces we can use in simple test programs
based on string (Ada character array) searches.

\begin{flushleft} \small
\begin{minipage}{\linewidth} \label{__n:gensearch.w:part30}
$\langle$Sequence declarations {\footnotesize 27a}$\rangle\equiv$
\vspace{-1ex}
\begin{list}{}{} \item
\mbox{}\verb@type Character_Sequence is array(Integer range <>) of Character;@\\
\mbox{}\verb@type Integer_Sequence is array(Integer range <>) of Integer;@\\
\mbox{}\verb@type Skip_Sequence is array(Character range <>) of Integer;@\\
\mbox{}\verb@@\end{list}
\vspace{-1ex}
\footnotesize\addtolength{\baselineskip}{-1ex}
\begin{list}{}{\setlength{\itemsep}{-\parsep}\setlength{\itemindent}{-\leftmargin}}
\item Used in parts 31b, 35a, 36e.
\end{list}
\end{minipage}\\[4ex]
\end{flushleft}
\begin{flushleft} \small
\begin{minipage}{\linewidth} \label{__n:gensearch.w:part31}
$\langle$Algorithm subprogram declarations {\footnotesize 27b}$\rangle\equiv$
\vspace{-1ex}
\begin{list}{}{} \item
\mbox{}\verb@@\hbox{$\langle$Define procedure to compute next table {\footnotesize 30}$\rangle$}\verb@@\\
\mbox{}\verb@@\hbox{$\langle$Non-hashed algorithms {\footnotesize 27c}$\rangle$}\verb@@\\
\mbox{}\verb@@\hbox{$\langle$Simple hash function declarations {\footnotesize 29a}$\rangle$}\verb@@\\
\mbox{}\verb@@\hbox{$\langle$HAL declaration {\footnotesize 29b}$\rangle$}\verb@@\\
\mbox{}\verb@@\end{list}
\vspace{-1ex}
\footnotesize\addtolength{\baselineskip}{-1ex}
\begin{list}{}{\setlength{\itemsep}{-\parsep}\setlength{\itemindent}{-\leftmargin}}
\item Used in parts 31b, 35a, 36e.
\end{list}
\end{minipage}\\[4ex]
\end{flushleft}
\begin{flushleft} \small \label{__n:gensearch.w:part32}
$\langle$Non-hashed algorithms {\footnotesize 27c}$\rangle\equiv$
\vspace{-1ex}
\begin{list}{}{} \item
\mbox{}\verb@function KMP(text, pattern: Character_Sequence; @\\
\mbox{}\verb@             b, n, a, m: Integer) return Integer is@\\
\mbox{}\verb@  pattern_size, j, k: Integer;@\\
\mbox{}\verb@  next: Integer_Sequence(a .. m - 1);@\\
\mbox{}\verb@begin@\\
\mbox{}\verb@  @\hbox{$\langle$Compute next table {\footnotesize 31a}$\rangle$}\verb@@\\
\mbox{}\verb@  @\hbox{$\langle$Basic KMP {\footnotesize 2}$\rangle$}\verb@@\\
\mbox{}\verb@end KMP;@\\
\mbox{}\verb@@\\
\mbox{}\verb@function L(text, pattern: Character_Sequence; @\\
\mbox{}\verb@              b, n, a, m: Integer) return Integer is@\\
\mbox{}\verb@  pattern_size, j, k: Integer;@\\
\mbox{}\verb@  next: Integer_Sequence(a .. m - 1);@\\
\mbox{}\verb@begin@\\
\mbox{}\verb@  pattern_size := m - a;@\\
\mbox{}\verb@  @\hbox{$\langle$Compute next table {\footnotesize 31a}$\rangle$}\verb@@\\
\mbox{}\verb@  @\hbox{$\langle$Algorithm L, optimized linear pattern search {\footnotesize 3a}$\rangle$}\verb@@\\
\mbox{}\verb@end L;@\\
\mbox{}\verb@@\\
\mbox{}\verb@function SF(text, pattern: Character_Sequence; @\\
\mbox{}\verb@             b, n, a, m: Integer) return Integer is@\\
\mbox{}\verb@  pattern_size, j, k: Integer;@\\
\mbox{}\verb@begin@\\
\mbox{}\verb@  pattern_size := m - a; k := b;@\\
\mbox{}\verb@  @\hbox{$\langle$Handle pattern size = 1 as a special case {\footnotesize 3b}$\rangle$}\verb@@\\
\mbox{}\verb@  while k <= n - pattern_size loop@\\
\mbox{}\verb@    @\hbox{$\langle$Scan the text for a possible match {\footnotesize 3c}$\rangle$}\verb@@\\
\mbox{}\verb@    @\hbox{$\langle$Verify whether a match is possible at the position found {\footnotesize 4a}$\rangle$}\verb@@\\
\mbox{}\verb@    k := k - (j - a) + 1;@\\
\mbox{}\verb@  end loop;@\\
\mbox{}\verb@  return n;@\\
\mbox{}\verb@end SF;@\\
\mbox{}\verb@@\\
\mbox{}\verb@function AL(text, pattern: Character_Sequence; @\\
\mbox{}\verb@            b, n, a, m: Integer) return Integer is@\\
\mbox{}\verb@  pattern_size, text_size, j, k, large, adjustment, mismatch_shift: Integer;@\\
\mbox{}\verb@  next: Integer_Sequence(a .. m - 1);@\\
\mbox{}\verb@  skip: Skip_Sequence(Character'Range);@\\
\mbox{}\verb@begin@\\
\mbox{}\verb@  @\hbox{$\langle$Accelerated Linear algorithm {\footnotesize 7b}$\rangle$}\verb@@\\
\mbox{}\verb@end AL;@\\
\mbox{}\verb@@\end{list}
\vspace{-1ex}
\footnotesize\addtolength{\baselineskip}{-1ex}
\begin{list}{}{\setlength{\itemsep}{-\parsep}\setlength{\itemindent}{-\leftmargin}}
\item Used in part 27b.
\end{list}
\end{flushleft}
The following is a sample hash function definition that 
makes HAL essentially equivalent to AL.

\begin{flushleft} \small
\begin{minipage}{\linewidth} \label{__n:gensearch.w:part33}
$\langle$Simple hash function declarations {\footnotesize 29a}$\rangle\equiv$
\vspace{-1ex}
\begin{list}{}{} \item
\mbox{}\verb@subtype hash_range is Integer range 0..255;@\\
\mbox{}\verb@@\\
\mbox{}\verb@function hash(text: Character_Sequence; k: Integer) return hash_range;@\\
\mbox{}\verb@pragma inline(hash);@\\
\mbox{}\verb@@\\
\mbox{}\verb@function hash(text: Character_Sequence; k: Integer) return hash_range is@\\
\mbox{}\verb@begin@\\
\mbox{}\verb@  return hash_range(character'pos(text(k)));@\\
\mbox{}\verb@end hash;@\\
\mbox{}\verb@@\\
\mbox{}\verb@suffix_size: constant Integer := 1;@\\
\mbox{}\verb@@\end{list}
\vspace{-1ex}
\footnotesize\addtolength{\baselineskip}{-1ex}
\begin{list}{}{\setlength{\itemsep}{-\parsep}\setlength{\itemindent}{-\leftmargin}}
\item Used in part 27b.
\end{list}
\end{minipage}\\[4ex]
\end{flushleft}
\begin{flushleft} \small
\begin{minipage}{\linewidth} \label{__n:gensearch.w:part34}
$\langle$HAL declaration {\footnotesize 29b}$\rangle\equiv$
\vspace{-1ex}
\begin{list}{}{} \item
\mbox{}\verb@function HAL(text, pattern: Character_Sequence; @\\
\mbox{}\verb@               b, n, a, m: Integer) return Integer is@\\
\mbox{}\verb@  pattern_size, text_size, j, k, large, adjustment, mismatch_shift: Integer;@\\
\mbox{}\verb@  next: Integer_Sequence(a .. m - 1);@\\
\mbox{}\verb@  skip: Integer_Sequence(hash_range);@\\
\mbox{}\verb@begin@\\
\mbox{}\verb@  @\hbox{$\langle$Hashed Accelerated Linear algorithm {\footnotesize 12b}$\rangle$}\verb@@\\
\mbox{}\verb@end HAL;@\\
\mbox{}\verb@@\end{list}
\vspace{-1ex}
\footnotesize\addtolength{\baselineskip}{-1ex}
\begin{list}{}{\setlength{\itemsep}{-\parsep}\setlength{\itemindent}{-\leftmargin}}
\item Used in part 27b.
\end{list}
\end{minipage}\\[4ex]
\end{flushleft}
For comparison of HAL with other algorithms we also compose the
following declarations:

\begin{flushleft} \small \label{__n:gensearch.w:part35}
$\langle$Additional algorithms {\footnotesize 29c}$\rangle\equiv$
\vspace{-1ex}
\begin{list}{}{} \item
\mbox{}\verb@function AL0(text, pattern: Character_Sequence; @\\
\mbox{}\verb@               b, n, a, m: Integer) return Integer is@\\
\mbox{}\verb@  pattern_size, j, k, d, mismatch_shift: Integer;@\\
\mbox{}\verb@  next: Integer_Sequence(a .. m - 1);@\\
\mbox{}\verb@  skip: Skip_Sequence(Character'Range);@\\
\mbox{}\verb@begin@\\
\mbox{}\verb@  @\hbox{$\langle$Accelerated Linear algorithm, preliminary version {\footnotesize 5c}$\rangle$}\verb@@\\
\mbox{}\verb@end AL0;@\\
\mbox{}\verb@@\\
\mbox{}\verb@function SF1(text, pattern: Character_Sequence; @\\
\mbox{}\verb@             b, n, a, m: Integer) return Integer is@\\
\mbox{}\verb@  pattern_size, j, k, k0: Integer;@\\
\mbox{}\verb@begin@\\
\mbox{}\verb@  pattern_size := m - a;@\\
\mbox{}\verb@  if n < m then@\\
\mbox{}\verb@    return n;@\\
\mbox{}\verb@  end if;@\\
\mbox{}\verb@  j := a; k := b; k0 := k;@\\
\mbox{}\verb@  while j /= m loop@\\
\mbox{}\verb@    if text(k) /= pattern(j) then@\\
\mbox{}\verb@      if k = n - pattern_size then@\\
\mbox{}\verb@        return n;@\\
\mbox{}\verb@      else@\\
\mbox{}\verb@        k0 := k0 + 1; k := k0; j := a;@\\
\mbox{}\verb@      end if;@\\
\mbox{}\verb@    else@\\
\mbox{}\verb@      k := k + 1; j := j + 1;@\\
\mbox{}\verb@    end if;@\\
\mbox{}\verb@  end loop;@\\
\mbox{}\verb@  return k0;@\\
\mbox{}\verb@end SF1;@\\
\mbox{}\verb@@\\
\mbox{}\verb@function SF2(text, pattern: Character_Sequence; @\\
\mbox{}\verb@             b, n, a, m: Integer) return Integer is@\\
\mbox{}\verb@  pattern_size, j, k, k0, n0: Integer;@\\
\mbox{}\verb@begin@\\
\mbox{}\verb@  pattern_size := m - a;@\\
\mbox{}\verb@  if n - b < pattern_size then@\\
\mbox{}\verb@    return n;@\\
\mbox{}\verb@  end if;@\\
\mbox{}\verb@  j := a; k := b; k0 := k; n0 := n - b;@\\
\mbox{}\verb@  while j /= m loop@\\
\mbox{}\verb@    if text(k) = pattern(j) then@\\
\mbox{}\verb@      k := k + 1; j := j + 1;@\\
\mbox{}\verb@    else@\\
\mbox{}\verb@      if n0 = pattern_size then@\\
\mbox{}\verb@        return n;@\\
\mbox{}\verb@      else@\\
\mbox{}\verb@        k0 := k0 + 1; k := k0; j := a; n0 := n0 - 1;@\\
\mbox{}\verb@      end if;@\\
\mbox{}\verb@    end if;@\\
\mbox{}\verb@  end loop;@\\
\mbox{}\verb@  return k0;@\\
\mbox{}\verb@end SF2;@\\
\mbox{}\verb@@\end{list}
\vspace{-1ex}
\footnotesize\addtolength{\baselineskip}{-1ex}
\begin{list}{}{\setlength{\itemsep}{-\parsep}\setlength{\itemindent}{-\leftmargin}}
\item Used in parts 31b, 35a, 36e.
\end{list}
\end{flushleft}
For computing the KMP next table we provide the following procedure
and calling code:

\begin{flushleft} \small \label{__n:gensearch.w:part36}
$\langle$Define procedure to compute next table {\footnotesize 30}$\rangle\equiv$
\vspace{-1ex}
\begin{list}{}{} \item
\mbox{}\verb@procedure Compute_Next(pattern: Character_Sequence; a, m: Integer; @\\
\mbox{}\verb@                         next: out Integer_Sequence) is@\\
\mbox{}\verb@  j: Integer := a;@\\
\mbox{}\verb@  t: Integer := a - 1;@\\
\mbox{}\verb@begin@\\
\mbox{}\verb@  next(a) := a - 1;@\\
\mbox{}\verb@  while j < m - 1 loop@\\
\mbox{}\verb@    while t >= a and then pattern(j) /= pattern(t) loop@\\
\mbox{}\verb@       t := next(t);@\\
\mbox{}\verb@    end loop;@\\
\mbox{}\verb@    j := j + 1; t := t + 1;@\\
\mbox{}\verb@    if pattern(j) = pattern(t) then@\\
\mbox{}\verb@      next(j) := next(t);@\\
\mbox{}\verb@    else@\\
\mbox{}\verb@      next(j) := t;@\\
\mbox{}\verb@    end if;@\\
\mbox{}\verb@  end loop;@\\
\mbox{}\verb@end Compute_Next;@\\
\mbox{}\verb@@\end{list}
\vspace{-1ex}
\footnotesize\addtolength{\baselineskip}{-1ex}
\begin{list}{}{\setlength{\itemsep}{-\parsep}\setlength{\itemindent}{-\leftmargin}}
\item Used in part 27b.
\end{list}
\end{flushleft}
\begin{flushleft} \small
\begin{minipage}{\linewidth} \label{__n:gensearch.w:part37}
$\langle$Compute next table {\footnotesize 31a}$\rangle\equiv$
\vspace{-1ex}
\begin{list}{}{} \item
\mbox{}\verb@Compute_Next(pattern, a, m, next);@\\
\mbox{}\verb@@\end{list}
\vspace{-1ex}
\footnotesize\addtolength{\baselineskip}{-1ex}
\begin{list}{}{\setlength{\itemsep}{-\parsep}\setlength{\itemindent}{-\leftmargin}}
\item Used in parts 5c, 7b, 12b, 27c.
\end{list}
\end{minipage}\\[4ex]
\end{flushleft}
\subsection{Simple Tests}

The first test program simply reads short test sequences from a file
and reports the results of running the different search algorithms on
them.
\begin{flushleft} \small \label{__n:gensearch.w:part38}
\verb@"Test_Search.adb"@ {\footnotesize 31b }$\equiv$
\vspace{-1ex}
\begin{list}{}{} \item
\mbox{}\verb@with Text_Io; use Text_Io; @\\
\mbox{}\verb@with Ada.Integer_Text_Io; use Ada.Integer_Text_Io; @\\
\mbox{}\verb@with Io_Exceptions;@\\
\mbox{}\verb@procedure Test_Search is@\\
\mbox{}\verb@  @\hbox{$\langle$Sequence declarations {\footnotesize 27a}$\rangle$}\verb@@\\
\mbox{}\verb@  @\hbox{$\langle$Variable declarations {\footnotesize 31c}$\rangle$}\verb@@\\
\mbox{}\verb@  @\hbox{$\langle$Algorithm subprogram declarations {\footnotesize 27b}$\rangle$}\verb@@\\
\mbox{}\verb@  @\hbox{$\langle$Additional algorithms {\footnotesize 29c}$\rangle$}\verb@@\\
\mbox{}\verb@  @\hbox{$\langle$Define procedure to read string into sequence {\footnotesize 33b}$\rangle$}\verb@@\\
\mbox{}\verb@  @\hbox{$\langle$Define procedure to output sequence {\footnotesize 33c}$\rangle$}\verb@@\\
\mbox{}\verb@  @\hbox{$\langle$Define algorithm enumeration type, names, and selector function {\footnotesize 31e}$\rangle$}\verb@@\\
\mbox{}\verb@  @\hbox{$\langle$Define Report procedure {\footnotesize 33e}$\rangle$}\verb@@\\
\mbox{}\verb@begin@\\
\mbox{}\verb@  @\hbox{$\langle$Set file small.txt as input file {\footnotesize 31d}$\rangle$}\verb@@\\
\mbox{}\verb@  loop@\\
\mbox{}\verb@    @\hbox{$\langle$Read test sequences from file {\footnotesize 32}$\rangle$}\verb@@\\
\mbox{}\verb@    @\hbox{$\langle$Run tests and report results {\footnotesize 33d}$\rangle$}\verb@@\\
\mbox{}\verb@  end loop;@\\
\mbox{}\verb@end Test_Search;@\\
\mbox{}\verb@@\end{list}
\vspace{-2ex}
\end{flushleft}
\begin{flushleft} \small
\begin{minipage}{\linewidth} \label{__n:gensearch.w:part39}
$\langle$Variable declarations {\footnotesize 31c}$\rangle\equiv$
\vspace{-1ex}
\begin{list}{}{} \item
\mbox{}\verb@Comment, S1, S2: Character_Sequence(1 .. 100);@\\
\mbox{}\verb@Base_Line, S1_Length, S2_Length, Last: Integer;@\\
\mbox{}\verb@File: Text_Io.File_Type;@\\
\mbox{}\verb@@\end{list}
\vspace{-1ex}
\footnotesize\addtolength{\baselineskip}{-1ex}
\begin{list}{}{\setlength{\itemsep}{-\parsep}\setlength{\itemindent}{-\leftmargin}}
\item Used in part 31b.
\end{list}
\end{minipage}\\[4ex]
\end{flushleft}
\begin{flushleft} \small
\begin{minipage}{\linewidth} \label{__n:gensearch.w:part40}
$\langle$Set file small.txt as input file {\footnotesize 31d}$\rangle\equiv$
\vspace{-1ex}
\begin{list}{}{} \item
\mbox{}\verb@Text_Io.Open(File, Text_IO.In_File, "small.txt");@\\
\mbox{}\verb@Text_Io.Set_Input(File);@\\
\mbox{}\verb@@\end{list}
\vspace{-1ex}
\footnotesize\addtolength{\baselineskip}{-1ex}
\begin{list}{}{\setlength{\itemsep}{-\parsep}\setlength{\itemindent}{-\leftmargin}}
\item Used in part 31b.
\end{list}
\end{minipage}\\[4ex]
\end{flushleft}
\begin{flushleft} \small \label{__n:gensearch.w:part41}
$\langle$Define algorithm enumeration type, names, and selector function {\footnotesize 31e}$\rangle\equiv$
\vspace{-1ex}
\begin{list}{}{} \item
\mbox{}\verb@type Algorithm_Enumeration is (Dummy, SF, SF1, SF2, L, AL, HAL);@\\
\mbox{}\verb@@\\
\mbox{}\verb@  Algorithm_Names: array(Algorithm_Enumeration) of String(1 .. 17) :=@\\
\mbox{}\verb@    ("selection code   ", @\\
\mbox{}\verb@     "SF               ", @\\
\mbox{}\verb@     "HP SF            ", @\\
\mbox{}\verb@     "SGI SF           ", @\\
\mbox{}\verb@     "L                ",@\\
\mbox{}\verb@     "AL               ",@\\
\mbox{}\verb@     "HAL              ");@\\
\mbox{}\verb@  @\\
\mbox{}\verb@function Algorithm(k: Algorithm_Enumeration; @\\
\mbox{}\verb@                   text, pattern: Character_Sequence;@\\
\mbox{}\verb@                   b, n, a, m: Integer) return Integer is@\\
\mbox{}\verb@begin@\\
\mbox{}\verb@  case k is@\\
\mbox{}\verb@    when Dummy => return b;@\\
\mbox{}\verb@    when SF => return SF(text, pattern, b, n, a, m);@\\
\mbox{}\verb@    when SF1 => return SF1(text, pattern, b, n, a, m);@\\
\mbox{}\verb@    when SF2 => return SF2(text, pattern, b, n, a, m);@\\
\mbox{}\verb@    when L => return L(text, pattern, b, n, a, m);@\\
\mbox{}\verb@    when AL => return AL(text, pattern, b, n, a, m);@\\
\mbox{}\verb@    when HAL => return HAL(text, pattern, b, n, a, m);@\\
\mbox{}\verb@  end case;@\\
\mbox{}\verb@end Algorithm;@\\
\mbox{}\verb@@\end{list}
\vspace{-1ex}
\footnotesize\addtolength{\baselineskip}{-1ex}
\begin{list}{}{\setlength{\itemsep}{-\parsep}\setlength{\itemindent}{-\leftmargin}}
\item Used in parts 31b, 35a, 36e.
\end{list}
\end{flushleft}
Test sequences are expected to be found in a file named
\verb|small.txt|.  Each test set is contained on three lines, the
first line being a comment or blank, the second line containing the
text string to be searched, and the third the pattern to search for.

\begin{flushleft} \small
\begin{minipage}{\linewidth} \label{__n:gensearch.w:part42}
$\langle$Read test sequences from file {\footnotesize 32}$\rangle\equiv$
\vspace{-1ex}
\begin{list}{}{} \item
\mbox{}\verb@exit when Text_Io.End_Of_File;@\\
\mbox{}\verb@Get(Comment, Last);@\\
\mbox{}\verb@Put(Comment, Last); New_Line;@\\
\mbox{}\verb@@\hbox{$\langle$Check for unexpected end of file {\footnotesize 33a}$\rangle$}\verb@@\\
\mbox{}\verb@@\\
\mbox{}\verb@Get(S1, Last);@\\
\mbox{}\verb@@\hbox{$\langle$Check for unexpected end of file {\footnotesize 33a}$\rangle$}\verb@@\\
\mbox{}\verb@Put("Text sequence:    "); Put(S1, Last);@\\
\mbox{}\verb@S1_Length := Last;@\\
\mbox{}\verb@@\\
\mbox{}\verb@Get(S2, Last);@\\
\mbox{}\verb@Put("Pattern sequence: "); Put(S2, Last);@\\
\mbox{}\verb@S2_Length := Last;@\\
\mbox{}\verb@@\end{list}
\vspace{-1ex}
\footnotesize\addtolength{\baselineskip}{-1ex}
\begin{list}{}{\setlength{\itemsep}{-\parsep}\setlength{\itemindent}{-\leftmargin}}
\item Used in part 31b.
\end{list}
\end{minipage}\\[4ex]
\end{flushleft}
\begin{flushleft} \small
\begin{minipage}{\linewidth} \label{__n:gensearch.w:part43}
$\langle$Check for unexpected end of file {\footnotesize 33a}$\rangle\equiv$
\vspace{-1ex}
\begin{list}{}{} \item
\mbox{}\verb@if Text_Io.End_Of_File then@\\
\mbox{}\verb@  Put_Line("**** Unexpected end of file."); New_Line;@\\
\mbox{}\verb@  raise Program_Error;@\\
\mbox{}\verb@end if;@\\
\mbox{}\verb@@\end{list}
\vspace{-1ex}
\footnotesize\addtolength{\baselineskip}{-1ex}
\begin{list}{}{\setlength{\itemsep}{-\parsep}\setlength{\itemindent}{-\leftmargin}}
\item Used in part 32.
\end{list}
\end{minipage}\\[4ex]
\end{flushleft}
\begin{flushleft} \small
\begin{minipage}{\linewidth} \label{__n:gensearch.w:part44}
$\langle$Define procedure to read string into sequence {\footnotesize 33b}$\rangle\equiv$
\vspace{-1ex}
\begin{list}{}{} \item
\mbox{}\verb@procedure Get(S: out Character_Sequence; Last: out Integer) is@\\
\mbox{}\verb@  Ch: Character;@\\
\mbox{}\verb@  I : Integer := 0;@\\
\mbox{}\verb@begin@\\
\mbox{}\verb@  while not Text_Io.End_Of_File loop@\\
\mbox{}\verb@    Text_Io.Get_Immediate(Ch);@\\
\mbox{}\verb@    I := I + 1;@\\
\mbox{}\verb@    S(I) := Ch;@\\
\mbox{}\verb@    exit when Text_Io.End_Of_Line;@\\
\mbox{}\verb@  end loop;@\\
\mbox{}\verb@  Last := I;@\\
\mbox{}\verb@  Text_Io.Get_Immediate(Ch);@\\
\mbox{}\verb@end Get;@\\
\mbox{}\verb@@\end{list}
\vspace{-1ex}
\footnotesize\addtolength{\baselineskip}{-1ex}
\begin{list}{}{\setlength{\itemsep}{-\parsep}\setlength{\itemindent}{-\leftmargin}}
\item Used in part 31b.
\end{list}
\end{minipage}\\[4ex]
\end{flushleft}
\begin{flushleft} \small
\begin{minipage}{\linewidth} \label{__n:gensearch.w:part45}
$\langle$Define procedure to output sequence {\footnotesize 33c}$\rangle\equiv$
\vspace{-1ex}
\begin{list}{}{} \item
\mbox{}\verb@procedure Put(S: Character_Sequence; Last: Integer) is@\\
\mbox{}\verb@begin@\\
\mbox{}\verb@  for I in 1 .. Last loop@\\
\mbox{}\verb@    Put(S(I));@\\
\mbox{}\verb@  end loop;@\\
\mbox{}\verb@  New_Line;@\\
\mbox{}\verb@end Put;@\\
\mbox{}\verb@@\end{list}
\vspace{-1ex}
\footnotesize\addtolength{\baselineskip}{-1ex}
\begin{list}{}{\setlength{\itemsep}{-\parsep}\setlength{\itemindent}{-\leftmargin}}
\item Used in part 31b.
\end{list}
\end{minipage}\\[4ex]
\end{flushleft}
\begin{flushleft} \small
\begin{minipage}{\linewidth} \label{__n:gensearch.w:part46}
$\langle$Run tests and report results {\footnotesize 33d}$\rangle\equiv$
\vspace{-1ex}
\begin{list}{}{} \item
\mbox{}\verb@Base_Line := 0;@\\
\mbox{}\verb@for K in Algorithm_Enumeration'Succ(Algorithm_Enumeration'First) .. @\\
\mbox{}\verb@         Algorithm_Enumeration'Last loop@\\
\mbox{}\verb@  Put("  Using "); Put(Algorithm_Names(k)); New_Line;@\\
\mbox{}\verb@  Report(K, S1, S2, 1, S1_Length + 1, 1, S2_Length + 1);@\\
\mbox{}\verb@end loop;@\\
\mbox{}\verb@New_Line;@\\
\mbox{}\verb@@\end{list}
\vspace{-1ex}
\footnotesize\addtolength{\baselineskip}{-1ex}
\begin{list}{}{\setlength{\itemsep}{-\parsep}\setlength{\itemindent}{-\leftmargin}}
\item Used in part 31b.
\end{list}
\end{minipage}\\[4ex]
\end{flushleft}
\begin{flushleft} \small \label{__n:gensearch.w:part47}
$\langle$Define Report procedure {\footnotesize 33e}$\rangle\equiv$
\vspace{-1ex}
\begin{list}{}{} \item
\mbox{}\verb@procedure Report(K: Algorithm_Enumeration;@\\
\mbox{}\verb@                 S1, S2: Character_Sequence; b, n, a, m: Integer) is@\\
\mbox{}\verb@  P: Integer;@\\
\mbox{}\verb@begin@\\
\mbox{}\verb@  P := Algorithm(K, S1, S2, b, n, a, m);@\\
\mbox{}\verb@  Put("    String "); Put('"');@\\
\mbox{}\verb@  @\hbox{$\langle$Output S2 {\footnotesize 34a}$\rangle$}\verb@@\\
\mbox{}\verb@  if P = n then@\\
\mbox{}\verb@    Put(" not found");@\\
\mbox{}\verb@    New_Line;@\\
\mbox{}\verb@  else@\\
\mbox{}\verb@    Put('"'); Put(" found at position ");@\\
\mbox{}\verb@    Put(P);@\\
\mbox{}\verb@    New_Line;@\\
\mbox{}\verb@  end if;@\\
\mbox{}\verb@  if Base_Line = 0 then@\\
\mbox{}\verb@    Base_Line := P - b;@\\
\mbox{}\verb@  else@\\
\mbox{}\verb@    if P - b /= Base_Line then@\\
\mbox{}\verb@      Put("*****Incorrect result!"); New_Line;@\\
\mbox{}\verb@    end if;@\\
\mbox{}\verb@  end if;@\\
\mbox{}\verb@end Report;@\\
\mbox{}\verb@@\end{list}
\vspace{-1ex}
\footnotesize\addtolength{\baselineskip}{-1ex}
\begin{list}{}{\setlength{\itemsep}{-\parsep}\setlength{\itemindent}{-\leftmargin}}
\item Used in parts 31b, 35a.
\end{list}
\end{flushleft}
\begin{flushleft} \small
\begin{minipage}{\linewidth} \label{__n:gensearch.w:part48}
$\langle$Output S2 {\footnotesize 34a}$\rangle\equiv$
\vspace{-1ex}
\begin{list}{}{} \item
\mbox{}\verb@for I in a .. m - 1 loop@\\
\mbox{}\verb@  Put(S2(I));@\\
\mbox{}\verb@end loop;@\\
\mbox{}\verb@@\end{list}
\vspace{-1ex}
\footnotesize\addtolength{\baselineskip}{-1ex}
\begin{list}{}{\setlength{\itemsep}{-\parsep}\setlength{\itemindent}{-\leftmargin}}
\item Used in part 33e.
\end{list}
\end{minipage}\\[4ex]
\end{flushleft}
Here are a few small tests.

\begin{flushleft} \small
\begin{minipage}{\linewidth} \label{__n:gensearch.w:part49}
\verb@"small.txt"@ {\footnotesize 34b }$\equiv$
\vspace{-1ex}
\begin{list}{}{} \item
\mbox{}\verb@#@\\
\mbox{}\verb@Now's the time for all good men and women to come to the aid of their country.@\\
\mbox{}\verb@time@\\
\mbox{}\verb@#@\\
\mbox{}\verb@Now's the time for all good men and women to come to the aid of their country.@\\
\mbox{}\verb@timid@\\
\mbox{}\verb@#@\\
\mbox{}\verb@Now's the time for all good men and women to come to the aid of their country.@\\
\mbox{}\verb@try.@\\
\mbox{}\verb@# The following example is from the KMP paper.@\\
\mbox{}\verb@babcbabcabcaabcabcabcacabc@\\
\mbox{}\verb@abcabcacab@\\
\mbox{}\verb@#@\\
\mbox{}\verb@aaaaaaabcabcadefg@\\
\mbox{}\verb@abcad@\\
\mbox{}\verb@#@\\
\mbox{}\verb@aaaaaaabcabcadefg@\\
\mbox{}\verb@ab@\\
\mbox{}\verb@@\end{list}
\vspace{-2ex}
\end{minipage}\\[4ex]
\end{flushleft}
\subsection{Large Tests}

This Ada test program can read a long character sequence
from a file and run extensive search tests on it.  Patterns
to search for, of a user-specified length, are selected from evenly-spaced
positions in the long sequence.
\begin{flushleft} \small \label{__n:gensearch.w:part50}
\verb@"Test_Long_Search.adb"@ {\footnotesize 35a }$\equiv$
\vspace{-1ex}
\begin{list}{}{} \item
\mbox{}\verb@with Text_Io; use Text_Io; @\\
\mbox{}\verb@with Ada.Integer_Text_Io; use Ada.Integer_Text_Io; @\\
\mbox{}\verb@procedure Test_Long_Search is@\\
\mbox{}\verb@  F: Integer;@\\
\mbox{}\verb@  Number_Of_Tests: Integer;@\\
\mbox{}\verb@  Pattern_Size: Integer;@\\
\mbox{}\verb@  Increment: Integer;@\\
\mbox{}\verb@  @\hbox{$\langle$Sequence declarations {\footnotesize 27a}$\rangle$}\verb@@\\
\mbox{}\verb@  @\hbox{$\langle$Data declarations {\footnotesize 35b}$\rangle$}\verb@@\\
\mbox{}\verb@  @\hbox{$\langle$Algorithm subprogram declarations {\footnotesize 27b}$\rangle$}\verb@@\\
\mbox{}\verb@  @\hbox{$\langle$Additional algorithms {\footnotesize 29c}$\rangle$}\verb@@\\
\mbox{}\verb@  @\hbox{$\langle$Define algorithm enumeration type, names, and selector function {\footnotesize 31e}$\rangle$}\verb@@\\
\mbox{}\verb@  @\hbox{$\langle$Define Report procedure {\footnotesize 33e}$\rangle$}\verb@@\\
\mbox{}\verb@  S2: Character_Sequence(0 .. 100);@\\
\mbox{}\verb@begin@\\
\mbox{}\verb@  @\hbox{$\langle$Read test parameters {\footnotesize 35d}$\rangle$}\verb@@\\
\mbox{}\verb@  @\hbox{$\langle$Set file long.txt as input file {\footnotesize 35c}$\rangle$}\verb@@\\
\mbox{}\verb@  @\hbox{$\langle$Read character sequence from file {\footnotesize 36a}$\rangle$}\verb@@\\
\mbox{}\verb@  Increment := (S1_Length - S2_Length) / Number_Of_Tests;@\\
\mbox{}\verb@  @\hbox{$\langle$Run tests searching for selected subsequences {\footnotesize 36b}$\rangle$}\verb@@\\
\mbox{}\verb@end Test_Long_Search;@\\
\mbox{}\verb@@\end{list}
\vspace{-2ex}
\end{flushleft}
\begin{flushleft} \small
\begin{minipage}{\linewidth} \label{__n:gensearch.w:part51}
$\langle$Data declarations {\footnotesize 35b}$\rangle\equiv$
\vspace{-1ex}
\begin{list}{}{} \item
\mbox{}\verb@Max_Size: constant Integer := 200_000;@\\
\mbox{}\verb@C: Character;@\\
\mbox{}\verb@S1: Character_Sequence(0 .. Max_Size);@\\
\mbox{}\verb@Base_Line, I, S1_Length, S2_Length: Integer;@\\
\mbox{}\verb@File: Text_Io.File_Type;@\\
\mbox{}\verb@@\end{list}
\vspace{-1ex}
\footnotesize\addtolength{\baselineskip}{-1ex}
\begin{list}{}{\setlength{\itemsep}{-\parsep}\setlength{\itemindent}{-\leftmargin}}
\item Used in parts 35a, 36e.
\end{list}
\end{minipage}\\[4ex]
\end{flushleft}
\begin{flushleft} \small
\begin{minipage}{\linewidth} \label{__n:gensearch.w:part52}
$\langle$Set file long.txt as input file {\footnotesize 35c}$\rangle\equiv$
\vspace{-1ex}
\begin{list}{}{} \item
\mbox{}\verb@Text_Io.Open(File, Text_IO.In_File, "long.txt");@\\
\mbox{}\verb@Text_Io.Set_Input(File);@\\
\mbox{}\verb@@\end{list}
\vspace{-1ex}
\footnotesize\addtolength{\baselineskip}{-1ex}
\begin{list}{}{\setlength{\itemsep}{-\parsep}\setlength{\itemindent}{-\leftmargin}}
\item Used in parts 35a, 36e.
\end{list}
\end{minipage}\\[4ex]
\end{flushleft}
\begin{flushleft} \small
\begin{minipage}{\linewidth} \label{__n:gensearch.w:part53}
$\langle$Read test parameters {\footnotesize 35d}$\rangle\equiv$
\vspace{-1ex}
\begin{list}{}{} \item
\mbox{}\verb@Put("Input Number of tests and pattern size: "); Text_Io.Flush;@\\
\mbox{}\verb@Get(Number_Of_Tests); @\\
\mbox{}\verb@Get(Pattern_Size);@\\
\mbox{}\verb@New_Line; Put("Number of tests: "); Put(Number_Of_Tests); New_Line;@\\
\mbox{}\verb@Put("Pattern size: "); Put(Pattern_Size); New_Line;@\\
\mbox{}\verb@S2_Length := Pattern_Size;@\\
\mbox{}\verb@@\end{list}
\vspace{-1ex}
\footnotesize\addtolength{\baselineskip}{-1ex}
\begin{list}{}{\setlength{\itemsep}{-\parsep}\setlength{\itemindent}{-\leftmargin}}
\item Used in parts 35a, 36e.
\end{list}
\end{minipage}\\[4ex]
\end{flushleft}
\begin{flushleft} \small
\begin{minipage}{\linewidth} \label{__n:gensearch.w:part54}
$\langle$Read character sequence from file {\footnotesize 36a}$\rangle\equiv$
\vspace{-1ex}
\begin{list}{}{} \item
\mbox{}\verb@I := 0;@\\
\mbox{}\verb@while not Text_Io.End_Of_File loop@\\
\mbox{}\verb@  Text_Io.Get_Immediate(C);@\\
\mbox{}\verb@  S1(I) := C;@\\
\mbox{}\verb@  I := I + 1;@\\
\mbox{}\verb@end loop;@\\
\mbox{}\verb@S1_Length := I;@\\
\mbox{}\verb@Put(S1_Length); Put(" characters read."); New_Line;@\\
\mbox{}\verb@@\end{list}
\vspace{-1ex}
\footnotesize\addtolength{\baselineskip}{-1ex}
\begin{list}{}{\setlength{\itemsep}{-\parsep}\setlength{\itemindent}{-\leftmargin}}
\item Used in parts 35a, 36e.
\end{list}
\end{minipage}\\[4ex]
\end{flushleft}
\begin{flushleft} \small
\begin{minipage}{\linewidth} \label{__n:gensearch.w:part55}
$\langle$Run tests searching for selected subsequences {\footnotesize 36b}$\rangle\equiv$
\vspace{-1ex}
\begin{list}{}{} \item
\mbox{}\verb@F := 0;@\\
\mbox{}\verb@for K in 1 .. Number_Of_Tests loop@\\
\mbox{}\verb@  @\hbox{$\langle$Select sequence S2 to search for in S1 {\footnotesize 36c}$\rangle$}\verb@@\\
\mbox{}\verb@  @\hbox{$\langle$Run tests {\footnotesize 36d}$\rangle$}\verb@@\\
\mbox{}\verb@end loop;@\\
\mbox{}\verb@@\end{list}
\vspace{-1ex}
\footnotesize\addtolength{\baselineskip}{-1ex}
\begin{list}{}{\setlength{\itemsep}{-\parsep}\setlength{\itemindent}{-\leftmargin}}
\item Used in part 35a.
\end{list}
\end{minipage}\\[4ex]
\end{flushleft}
\begin{flushleft} \small
\begin{minipage}{\linewidth} \label{__n:gensearch.w:part56}
$\langle$Select sequence S2 to search for in S1 {\footnotesize 36c}$\rangle\equiv$
\vspace{-1ex}
\begin{list}{}{} \item
\mbox{}\verb@for I in 0 .. Pattern_Size - 1 loop@\\
\mbox{}\verb@  S2(I) := S1(F + I);@\\
\mbox{}\verb@end loop;@\\
\mbox{}\verb@F := F + Increment;@\\
\mbox{}\verb@@\end{list}
\vspace{-1ex}
\footnotesize\addtolength{\baselineskip}{-1ex}
\begin{list}{}{\setlength{\itemsep}{-\parsep}\setlength{\itemindent}{-\leftmargin}}
\item Used in parts 36b, 37b.
\end{list}
\end{minipage}\\[4ex]
\end{flushleft}
\begin{flushleft} \small
\begin{minipage}{\linewidth} \label{__n:gensearch.w:part57}
$\langle$Run tests {\footnotesize 36d}$\rangle\equiv$
\vspace{-1ex}
\begin{list}{}{} \item
\mbox{}\verb@Base_Line := 0;@\\
\mbox{}\verb@for K in Algorithm_Enumeration'Succ(Algorithm_Enumeration'First) .. @\\
\mbox{}\verb@         Algorithm_Enumeration'Last loop@\\
\mbox{}\verb@  Put("  Using "); Put(Algorithm_Names(k)); New_Line;@\\
\mbox{}\verb@  Report(K, S1, S2, 0, S1_Length, 0, S2_Length);@\\
\mbox{}\verb@end loop;@\\
\mbox{}\verb@New_Line;@\\
\mbox{}\verb@@\end{list}
\vspace{-1ex}
\footnotesize\addtolength{\baselineskip}{-1ex}
\begin{list}{}{\setlength{\itemsep}{-\parsep}\setlength{\itemindent}{-\leftmargin}}
\item Used in part 36b.
\end{list}
\end{minipage}\\[4ex]
\end{flushleft}
\subsection{Timed Tests}

This Ada test program reads a character sequence from a file and times
searches for selected strings.

\begin{flushleft} \small \label{__n:gensearch.w:part58}
\verb@"Time_Long_Search.adb"@ {\footnotesize 36e }$\equiv$
\vspace{-1ex}
\begin{list}{}{} \item
\mbox{}\verb@with Text_Io; use Text_Io; @\\
\mbox{}\verb@with Ada.Integer_Text_Io; use Ada.Integer_Text_Io; @\\
\mbox{}\verb@with Ada.Real_Time;@\\
\mbox{}\verb@procedure Time_Long_Search is@\\
\mbox{}\verb@  use Ada.Real_Time;@\\
\mbox{}\verb@  package My_Float is new Text_IO.Float_IO(Long_Float);@\\
\mbox{}\verb@  Base_Time: Long_Float;@\\
\mbox{}\verb@  Number_Of_Tests, Pattern_Size, Increment: Integer;@\\
\mbox{}\verb@  pragma Suppress(All_Checks);@\\
\mbox{}\verb@  @\hbox{$\langle$Sequence declarations {\footnotesize 27a}$\rangle$}\verb@@\\
\mbox{}\verb@  @\hbox{$\langle$Algorithm subprogram declarations {\footnotesize 27b}$\rangle$}\verb@@\\
\mbox{}\verb@  @\hbox{$\langle$Additional algorithms {\footnotesize 29c}$\rangle$}\verb@@\\
\mbox{}\verb@  @\hbox{$\langle$Define algorithm enumeration type, names, and selector function {\footnotesize 31e}$\rangle$}\verb@@\\
\mbox{}\verb@  @\hbox{$\langle$Data declarations {\footnotesize 35b}$\rangle$}\verb@@\\
\mbox{}\verb@  @\hbox{$\langle$Define run procedure {\footnotesize 37b}$\rangle$}\verb@@\\
\mbox{}\verb@begin@\\
\mbox{}\verb@  @\hbox{$\langle$Read test parameters {\footnotesize 35d}$\rangle$}\verb@@\\
\mbox{}\verb@  @\hbox{$\langle$Set file long.txt as input file {\footnotesize 35c}$\rangle$}\verb@@\\
\mbox{}\verb@  @\hbox{$\langle$Read character sequence from file {\footnotesize 36a}$\rangle$}\verb@@\\
\mbox{}\verb@  Increment := (S1_Length - S2_Length) / Number_Of_Tests;@\\
\mbox{}\verb@  Base_Time := 0.0;@\\
\mbox{}\verb@  @\hbox{$\langle$Run and time tests searching for selected subsequences {\footnotesize 37a}$\rangle$}\verb@@\\
\mbox{}\verb@end Time_Long_Search;@\\
\mbox{}\verb@@\end{list}
\vspace{-2ex}
\end{flushleft}
\begin{flushleft} \small
\begin{minipage}{\linewidth} \label{__n:gensearch.w:part59}
$\langle$Run and time tests searching for selected subsequences {\footnotesize 37a}$\rangle\equiv$
\vspace{-1ex}
\begin{list}{}{} \item
\mbox{}\verb@for K in Algorithm_Enumeration'Range loop@\\
\mbox{}\verb@  Put("Timing "); Put(Algorithm_Names(K)); New_Line;@\\
\mbox{}\verb@  Run(K, S1, S1_Length, S2_Length);@\\
\mbox{}\verb@end loop;@\\
\mbox{}\verb@New_Line;@\\
\mbox{}\verb@@\end{list}
\vspace{-1ex}
\footnotesize\addtolength{\baselineskip}{-1ex}
\begin{list}{}{\setlength{\itemsep}{-\parsep}\setlength{\itemindent}{-\leftmargin}}
\item Used in part 36e.
\end{list}
\end{minipage}\\[4ex]
\end{flushleft}
For a given algorithm, the \verb|Run| procedure conducts a requested number
of searches in sequence \verb|S1| for patterns of a requested size,
selecting the patterns from evenly spaced positions in \verb|S1|.  It
reports the total search length, time taken, and speed (total search
length divided by time taken) of the searches.

\begin{flushleft} \small \label{__n:gensearch.w:part60}
$\langle$Define run procedure {\footnotesize 37b}$\rangle\equiv$
\vspace{-1ex}
\begin{list}{}{} \item
\mbox{}\verb@procedure Run(K: Algorithm_Enumeration;@\\
\mbox{}\verb@              S1: Character_Sequence; Text_Size, Pattern_Size: Integer) is@\\
\mbox{}\verb@  P, F: Integer;@\\
\mbox{}\verb@  Start_Time, Finish_Time: Time;@\\
\mbox{}\verb@  Total_Search: Integer;@\\
\mbox{}\verb@  Time_Taken : Long_Float;@\\
\mbox{}\verb@  S2: Character_Sequence(0 .. Pattern_Size - 1);@\\
\mbox{}\verb@begin@\\
\mbox{}\verb@  F := 0;@\\
\mbox{}\verb@  Total_Search := 0;@\\
\mbox{}\verb@  Start_Time := Clock;@\\
\mbox{}\verb@  for I in 1 .. Number_Of_Tests loop@\\
\mbox{}\verb@    @\hbox{$\langle$Select sequence S2 to search for in S1 {\footnotesize 36c}$\rangle$}\verb@@\\
\mbox{}\verb@    P := Algorithm(K, S1, S2, 0, Text_Size, 0, Pattern_Size);@\\
\mbox{}\verb@    Total_Search := Total_Search + P + Pattern_Size;@\\
\mbox{}\verb@  end loop;@\\
\mbox{}\verb@  Finish_Time := Clock;@\\
\mbox{}\verb@  @\hbox{$\langle$Output statistics {\footnotesize 38a}$\rangle$}\verb@@\\
\mbox{}\verb@end Run;@\\
\mbox{}\verb@@\end{list}
\vspace{-1ex}
\footnotesize\addtolength{\baselineskip}{-1ex}
\begin{list}{}{\setlength{\itemsep}{-\parsep}\setlength{\itemindent}{-\leftmargin}}
\item Used in part 36e.
\end{list}
\end{flushleft}
\begin{flushleft} \small \label{__n:gensearch.w:part61}
$\langle$Output statistics {\footnotesize 38a}$\rangle\equiv$
\vspace{-1ex}
\begin{list}{}{} \item
\mbox{}\verb@Time_Taken := Long_Float((Finish_Time - Start_Time) / Milliseconds(1))@\\
\mbox{}\verb@               / 1000.0 - Base_Time;@\\
\mbox{}\verb@Put("Total search length: ");@\\
\mbox{}\verb@Put(Total_Search); Put(" bytes."); New_Line;@\\
\mbox{}\verb@Put("Time:  "); My_Float.Put(Time_Taken, 5, 4, 0); @\\
\mbox{}\verb@Put(" seconds."); New_Line;@\\
\mbox{}\verb@if K /= Dummy then @\\
\mbox{}\verb@  Put("Speed: ");@\\
\mbox{}\verb@  My_Float.Put(Long_Float(Total_Search) / 1_000_000.0 / Time_Taken, 5, 2, 0); @\\
\mbox{}\verb@  Put(" MBytes/second."); New_Line;@\\
\mbox{}\verb@else@\\
\mbox{}\verb@  Base_Time := Time_Taken;@\\
\mbox{}\verb@end if;@\\
\mbox{}\verb@New_Line;@\\
\mbox{}\verb@@\end{list}
\vspace{-1ex}
\footnotesize\addtolength{\baselineskip}{-1ex}
\begin{list}{}{\setlength{\itemsep}{-\parsep}\setlength{\itemindent}{-\leftmargin}}
\item Used in part 37b.
\end{list}
\end{flushleft}
\section{C++ Library Versions and Test Programs}

The code presented in this section is packaged in files that can be
added to the standard \Cpp\ library and included in user programs with
\verb|#include| directives.  (A few adjustments may be necessary
depending on how well the target compiler conforms to the \Cpp\ 
standard.)  With only minor changes, library maintainers should be
able to incorporate the code into the standard library header files,
replacing whatever \verb|search| implementations they currently
contain.  The only significant work involved would be to construct the
predicate versions of the \verb|search| functions, which are not given
here.

\subsection{Generic Library Interfaces}

\subsubsection{Library Files}

\begin{flushleft} \small
\begin{minipage}{\linewidth} \label{__n:gensearch.w:part62}
\verb@"new_search.h"@ {\footnotesize 38b }$\equiv$
\vspace{-1ex}
\begin{list}{}{} \item
\mbox{}\verb@#ifndef NEW_SEARCH@\\
\mbox{}\verb@#  define NEW_SEARCH@\\
\mbox{}\verb@#  include <vector>@\\
\mbox{}\verb@#  include "search_traits.h"@\\
\mbox{}\verb@#  include <iterator>@\\
\mbox{}\verb@using namespace std;@\\
\mbox{}\verb@@\\
\mbox{}\verb@@\hbox{$\langle$Define procedure to compute next table (C++) {\footnotesize 43d}$\rangle$}\verb@@\\
\mbox{}\verb@@\hbox{$\langle$Define procedure to compute next table (C++ forward) {\footnotesize 20b}$\rangle$}\verb@@\\
\mbox{}\verb@@\hbox{$\langle$User level search function {\footnotesize 19a}$\rangle$}\verb@@\\
\mbox{}\verb@@\hbox{$\langle$Forward iterator case {\footnotesize 19b}$\rangle$}\verb@@\\
\mbox{}\verb@@\hbox{$\langle$Bidirectional iterator case {\footnotesize 40b}$\rangle$}\verb@@\\
\mbox{}\verb@@\hbox{$\langle$HAL with random access iterators, no trait passed {\footnotesize 41a}$\rangle$}\verb@@\\
\mbox{}\verb@@\hbox{$\langle$User level search function with trait argument {\footnotesize 41b}$\rangle$}\verb@@\\
\mbox{}\verb@#endif@\\
\mbox{}\verb@@\end{list}
\vspace{-2ex}
\end{minipage}\\[4ex]
\end{flushleft}
\subsubsection{Search Traits}

\begin{flushleft} \small
\begin{minipage}{\linewidth} \label{__n:gensearch.w:part63}
\verb@"search_traits.h"@ {\footnotesize 39a }$\equiv$
\vspace{-1ex}
\begin{list}{}{} \item
\mbox{}\verb@#ifndef SEARCH_HASH_TRAITS@\\
\mbox{}\verb@#  define SEARCH_HASH_TRAITS@\\
\mbox{}\verb@@\hbox{$\langle$Generic search trait {\footnotesize 39b}$\rangle$}\verb@@\\
\mbox{}\verb@@\hbox{$\langle$Search traits for character sequences {\footnotesize 40a}$\rangle$}\verb@@\\
\mbox{}\verb@#endif@\\
\mbox{}\verb@@\end{list}
\vspace{-2ex}
\end{minipage}\\[4ex]
\end{flushleft}
The generic search trait class is used when there is no search trait
specifically defined, either in the library or by the user, for the
type of values in the sequences being searched, and when no search trait
is explicitly passed to the search function.

\begin{flushleft} \small
\begin{minipage}{\linewidth} \label{__n:gensearch.w:part64}
$\langle$Generic search trait {\footnotesize 39b}$\rangle\equiv$
\vspace{-1ex}
\begin{list}{}{} \item
\mbox{}\verb@template <typename T>@\\
\mbox{}\verb@struct search_trait {@\\
\mbox{}\verb@  enum {hash_range_max = 0};@\\
\mbox{}\verb@  enum {suffix_size = 0};@\\
\mbox{}\verb@  template <typename RandomAccessIterator>@\\
\mbox{}\verb@  inline static unsigned int hash(RandomAccessIterator i) {@\\
\mbox{}\verb@    return 0;              @\\
\mbox{}\verb@  }@\\
\mbox{}\verb@};@\\
\mbox{}\verb@@\end{list}
\vspace{-1ex}
\footnotesize\addtolength{\baselineskip}{-1ex}
\begin{list}{}{\setlength{\itemsep}{-\parsep}\setlength{\itemindent}{-\leftmargin}}
\item Used in part 39a.
\end{list}
\end{minipage}\\[4ex]
\end{flushleft}
The ``hash'' function used in this trait maps everything to 0; it
would be a source of poor performance if it were actually used in the
HAL algorithm.  In fact it is not, because the code in the search
function checks for $\mbox{\tt suffix\_size} = 0$ and uses algorithm L in
that case.  This definition of \verb|hash| permits compilation to
succeed even if the compiler fails to recognize that the code segment
containing the call of \verb|hash| is dead code.

For traditional string searches, the following 
specialized search traits are provided:

\begin{flushleft} \small
\begin{minipage}{\linewidth} \label{__n:gensearch.w:part65}
$\langle$Search traits for character sequences {\footnotesize 40a}$\rangle\equiv$
\vspace{-1ex}
\begin{list}{}{} \item
\mbox{}\verb@template <> struct search_trait<signed char> {@\\
\mbox{}\verb@  enum {hash_range_max = 256};@\\
\mbox{}\verb@  enum {suffix_size = 1};@\\
\mbox{}\verb@  template <typename RandomAccessIterator>@\\
\mbox{}\verb@  inline static unsigned int hash(RandomAccessIterator i) {@\\
\mbox{}\verb@    return *i;              @\\
\mbox{}\verb@  }@\\
\mbox{}\verb@};@\\
\mbox{}\verb@@\\
\mbox{}\verb@typedef unsigned char unsigned_char;@\\
\mbox{}\verb@template <> struct search_trait<unsigned char> {@\\
\mbox{}\verb@  enum {hash_range_max = 256};@\\
\mbox{}\verb@  enum {suffix_size = 1};@\\
\mbox{}\verb@  template <typename RandomAccessIterator>@\\
\mbox{}\verb@  inline static unsigned int hash(RandomAccessIterator i) {@\\
\mbox{}\verb@    return *i;              @\\
\mbox{}\verb@  }@\\
\mbox{}\verb@};@\\
\mbox{}\verb@@\end{list}
\vspace{-1ex}
\footnotesize\addtolength{\baselineskip}{-1ex}
\begin{list}{}{\setlength{\itemsep}{-\parsep}\setlength{\itemindent}{-\leftmargin}}
\item Used in part 39a.
\end{list}
\end{minipage}\\[4ex]
\end{flushleft}
\subsubsection{Search Functions}

The main user-level search function interface and an auxiliary
function \verb|__search_L| for the forward iterator case were given
in the body of the paper.  With bidirectional iterators we again use
the forward iterator version.

\begin{flushleft} \small
\begin{minipage}{\linewidth} \label{__n:gensearch.w:part66}
$\langle$Bidirectional iterator case {\footnotesize 40b}$\rangle\equiv$
\vspace{-1ex}
\begin{list}{}{} \item
\mbox{}\verb@template <typename BidirectionalIterator1, typename BidirectionalIterator2>@\\
\mbox{}\verb@inline BidirectionalIterator1 __search(BidirectionalIterator1 text,@\\
\mbox{}\verb@                                       BidirectionalIterator1 textEnd,@\\
\mbox{}\verb@                                       BidirectionalIterator2 pattern,@\\
\mbox{}\verb@                                       BidirectionalIterator2 patternEnd,@\\
\mbox{}\verb@                                       bidirectional_iterator_tag)@\\
\mbox{}\verb@{@\\
\mbox{}\verb@  return __search_L(text, textEnd, pattern, patternEnd);@\\
\mbox{}\verb@}@\\
\mbox{}\verb@@\end{list}
\vspace{-1ex}
\footnotesize\addtolength{\baselineskip}{-1ex}
\begin{list}{}{\setlength{\itemsep}{-\parsep}\setlength{\itemindent}{-\leftmargin}}
\item Used in part 38b.
\end{list}
\end{minipage}\\[4ex]
\end{flushleft}
When we have random access iterators and no search trait is
passed as an argument, we use a search trait associated with $\mbox{\tt V} =
\mbox{\tt RandomAccessIterator1::value\_type}$ to obtain the hash
function and related parameters.  Then we use the user-level search
function that takes a search trait argument and uses HAL.  If no
search trait has been specifically defined for type \verb|V|,
then the generic \verb|search_hash_trait| is used, causing the 
\verb|search_hashed| algorithm to resort to algorithm L.

\begin{flushleft} \small
\begin{minipage}{\linewidth} \label{__n:gensearch.w:part67}
$\langle$HAL with random access iterators, no trait passed {\footnotesize 41a}$\rangle\equiv$
\vspace{-1ex}
\begin{list}{}{} \item
\mbox{}\verb@template <typename RandomAccessIterator1, typename RandomAccessIterator2>@\\
\mbox{}\verb@inline RandomAccessIterator1 __search(RandomAccessIterator1 text,@\\
\mbox{}\verb@                                      RandomAccessIterator1 textEnd,@\\
\mbox{}\verb@                                      RandomAccessIterator2 pattern,@\\
\mbox{}\verb@                                      RandomAccessIterator2 patternEnd,@\\
\mbox{}\verb@                                      random_access_iterator_tag)@\\
\mbox{}\verb@{@\\
\mbox{}\verb@  typedef iterator_traits<RandomAccessIterator1>::value_type V;@\\
\mbox{}\verb@  typedef search_trait<V> Trait;@\\
\mbox{}\verb@  return search_hashed(text, textEnd, pattern, patternEnd, @\\
\mbox{}\verb@                       static_cast<Trait*>(0)); @\\
\mbox{}\verb@}@\\
\mbox{}\verb@@\end{list}
\vspace{-1ex}
\footnotesize\addtolength{\baselineskip}{-1ex}
\begin{list}{}{\setlength{\itemsep}{-\parsep}\setlength{\itemindent}{-\leftmargin}}
\item Used in part 38b.
\end{list}
\end{minipage}\\[4ex]
\end{flushleft}
Finally, we have a user-level search function for the case of random
access iterators and an explicitly passed search trait.

\begin{flushleft} \small
\begin{minipage}{\linewidth} \label{__n:gensearch.w:part68}
$\langle$User level search function with trait argument {\footnotesize 41b}$\rangle\equiv$
\vspace{-1ex}
\begin{list}{}{} \item
\mbox{}\verb@@\\
\mbox{}\verb@template <typename RandomAccessIterator1, typename RandomAccessIterator2, @\\
\mbox{}\verb@          typename Trait>@\\
\mbox{}\verb@RandomAccessIterator1 search_hashed(RandomAccessIterator1 text,@\\
\mbox{}\verb@                                    RandomAccessIterator1 textEnd,@\\
\mbox{}\verb@                                    RandomAccessIterator2 pattern,@\\
\mbox{}\verb@                                    RandomAccessIterator2 patternEnd,@\\
\mbox{}\verb@                                    Trait*)@\\
\mbox{}\verb@{@\\
\mbox{}\verb@  typedef typename iterator_traits<RandomAccessIterator1>::difference_type @\\
\mbox{}\verb@          Distance1;@\\
\mbox{}\verb@  typedef typename iterator_traits<RandomAccessIterator2>::difference_type @\\
\mbox{}\verb@          Distance2;@\\
\mbox{}\verb@  if (pattern == patternEnd) return text;@\\
\mbox{}\verb@  Distance2 pattern_size, j, m;@\\
\mbox{}\verb@  pattern_size = patternEnd - pattern; @\\
\mbox{}\verb@  if (Trait::suffix_size == 0 || pattern_size < Trait::suffix_size)@\\
\mbox{}\verb@    return __search_L(text, textEnd, pattern, patternEnd);@\\
\mbox{}\verb@  Distance1 i, k, large, adjustment, mismatch_shift, text_size;@\\
\mbox{}\verb@  vector<Distance1> next, skip;@\\
\mbox{}\verb@  @\hbox{$\langle$Hashed Accelerated Linear algorithm (C++) {\footnotesize 42a}$\rangle$}\verb@@\\
\mbox{}\verb@}@\\
\mbox{}\verb@@\end{list}
\vspace{-1ex}
\footnotesize\addtolength{\baselineskip}{-1ex}
\begin{list}{}{\setlength{\itemsep}{-\parsep}\setlength{\itemindent}{-\leftmargin}}
\item Used in part 38b.
\end{list}
\end{minipage}\\[4ex]
\end{flushleft}
The \Cpp\ version of HAL is built from parts corresponding
to those expressed in Ada in the body of the paper.  Note that in
place of \verb|text(n + k)| we can write \verb|textEnd + k| for
the location and \verb|textEnd[k]| for the value at that location.

\begin{flushleft} \small
\begin{minipage}{\linewidth} \label{__n:gensearch.w:part69}
$\langle$Hashed Accelerated Linear algorithm (C++) {\footnotesize 42a}$\rangle\equiv$
\vspace{-1ex}
\begin{list}{}{} \item
\mbox{}\verb@k = 0; @\\
\mbox{}\verb@text_size = textEnd - text;@\\
\mbox{}\verb@@\hbox{$\langle$Compute next table (C++) {\footnotesize 44a}$\rangle$}\verb@@\\
\mbox{}\verb@@\hbox{$\langle$Handle pattern size = 1 as a special case (C++) {\footnotesize 21b}$\rangle$}\verb@@\\
\mbox{}\verb@@\hbox{$\langle$Compute skip table and mismatch shift using the hash function (C++) {\footnotesize 43c}$\rangle$}\verb@@\\
\mbox{}\verb@large = text_size + 1;@\\
\mbox{}\verb@adjustment = large + pattern_size - 1;@\\
\mbox{}\verb@skip[Trait::hash(pattern + pattern_size - 1)] = large;@\\
\mbox{}\verb@k -= text_size;@\\
\mbox{}\verb@for(;;) {@\\
\mbox{}\verb@  k += pattern_size - 1;@\\
\mbox{}\verb@  if (k >= 0) break;@\\
\mbox{}\verb@  @\hbox{$\langle$Scan the text using a single-test skip loop with hashing (C++) {\footnotesize 42b}$\rangle$}\verb@@\\
\mbox{}\verb@  @\hbox{$\langle$Verify match or recover from mismatch (C++) {\footnotesize 42c}$\rangle$}\verb@@\\
\mbox{}\verb@}@\\
\mbox{}\verb@return textEnd;@\\
\mbox{}\verb@@\end{list}
\vspace{-1ex}
\footnotesize\addtolength{\baselineskip}{-1ex}
\begin{list}{}{\setlength{\itemsep}{-\parsep}\setlength{\itemindent}{-\leftmargin}}
\item Used in part 41b.
\end{list}
\end{minipage}\\[4ex]
\end{flushleft}
\begin{flushleft} \small
\begin{minipage}{\linewidth} \label{__n:gensearch.w:part70}
$\langle$Scan the text using a single-test skip loop with hashing (C++) {\footnotesize 42b}$\rangle\equiv$
\vspace{-1ex}
\begin{list}{}{} \item
\mbox{}\verb@do {@\\
\mbox{}\verb@  k += skip[Trait::hash(textEnd + k)]; @\\
\mbox{}\verb@} while (k < 0);@\\
\mbox{}\verb@if (k < pattern_size)@\\
\mbox{}\verb@  return textEnd;@\\
\mbox{}\verb@k -= adjustment;@\\
\mbox{}\verb@@\end{list}
\vspace{-1ex}
\footnotesize\addtolength{\baselineskip}{-1ex}
\begin{list}{}{\setlength{\itemsep}{-\parsep}\setlength{\itemindent}{-\leftmargin}}
\item Used in part 42a.
\end{list}
\end{minipage}\\[4ex]
\end{flushleft}
\begin{flushleft} \small
\begin{minipage}{\linewidth} \label{__n:gensearch.w:part71}
$\langle$Verify match or recover from mismatch (C++) {\footnotesize 42c}$\rangle\equiv$
\vspace{-1ex}
\begin{list}{}{} \item
\mbox{}\verb@if (textEnd[k] != pattern[0])@\\
\mbox{}\verb@  k += mismatch_shift;@\\
\mbox{}\verb@else {@\\
\mbox{}\verb@  @\hbox{$\langle$Verify the match for positions 1 through pattern\_size - 1 (C++) {\footnotesize 43a}$\rangle$}\verb@@\\
\mbox{}\verb@  if (mismatch_shift > j)@\\
\mbox{}\verb@    k += mismatch_shift - j;@\\
\mbox{}\verb@  else@\\
\mbox{}\verb@    @\hbox{$\langle$Recover from a mismatch using the next table (C++) {\footnotesize 43b}$\rangle$}\verb@@\\
\mbox{}\verb@}@\\
\mbox{}\verb@@\end{list}
\vspace{-1ex}
\footnotesize\addtolength{\baselineskip}{-1ex}
\begin{list}{}{\setlength{\itemsep}{-\parsep}\setlength{\itemindent}{-\leftmargin}}
\item Used in parts 42a, 46a.
\end{list}
\end{minipage}\\[4ex]
\end{flushleft}
\begin{flushleft} \small
\begin{minipage}{\linewidth} \label{__n:gensearch.w:part72}
$\langle$Verify the match for positions 1 through pattern\_size - 1 (C++) {\footnotesize 43a}$\rangle\equiv$
\vspace{-1ex}
\begin{list}{}{} \item
\mbox{}\verb@j = 1; @\\
\mbox{}\verb@for (;;) {@\\
\mbox{}\verb@  ++k;@\\
\mbox{}\verb@  if (textEnd[k] != pattern[j])@\\
\mbox{}\verb@    break;@\\
\mbox{}\verb@  ++j;@\\
\mbox{}\verb@  if (j == pattern_size)@\\
\mbox{}\verb@    return textEnd + k - pattern_size + 1;@\\
\mbox{}\verb@}@\\
\mbox{}\verb@@\end{list}
\vspace{-1ex}
\footnotesize\addtolength{\baselineskip}{-1ex}
\begin{list}{}{\setlength{\itemsep}{-\parsep}\setlength{\itemindent}{-\leftmargin}}
\item Used in part 42c.
\end{list}
\end{minipage}\\[4ex]
\end{flushleft}
\begin{flushleft} \small
\begin{minipage}{\linewidth} \label{__n:gensearch.w:part73}
$\langle$Recover from a mismatch using the next table (C++) {\footnotesize 43b}$\rangle\equiv$
\vspace{-1ex}
\begin{list}{}{} \item
\mbox{}\verb@for (;;) {@\\
\mbox{}\verb@  j = next[j];@\\
\mbox{}\verb@  if (j < 0) {@\\
\mbox{}\verb@    ++k;@\\
\mbox{}\verb@    break;@\\
\mbox{}\verb@  }@\\
\mbox{}\verb@  if (j == 0)@\\
\mbox{}\verb@    break;@\\
\mbox{}\verb@  while (textEnd[k] == pattern[j]) {@\\
\mbox{}\verb@    ++k; ++j;@\\
\mbox{}\verb@    if (j == pattern_size) {@\\
\mbox{}\verb@      return textEnd + k - pattern_size;@\\
\mbox{}\verb@    }@\\
\mbox{}\verb@    if (k == 0)@\\
\mbox{}\verb@      return textEnd;@\\
\mbox{}\verb@  }@\\
\mbox{}\verb@}@\\
\mbox{}\verb@@\end{list}
\vspace{-1ex}
\footnotesize\addtolength{\baselineskip}{-1ex}
\begin{list}{}{\setlength{\itemsep}{-\parsep}\setlength{\itemindent}{-\leftmargin}}
\item Used in part 42c.
\end{list}
\end{minipage}\\[4ex]
\end{flushleft}
\subsubsection{Skip Table Computation}

\begin{flushleft} \small
\begin{minipage}{\linewidth} \label{__n:gensearch.w:part74}
$\langle$Compute skip table and mismatch shift using the hash function (C++) {\footnotesize 43c}$\rangle\equiv$
\vspace{-1ex}
\begin{list}{}{} \item
\mbox{}\verb@m = next.size();@\\
\mbox{}\verb@for (i = 0; i < Trait::hash_range_max; ++i)@\\
\mbox{}\verb@  skip.push_back(m - Trait::suffix_size + 1);@\\
\mbox{}\verb@for (j = Trait::suffix_size - 1; j < m - 1; ++j)@\\
\mbox{}\verb@  skip[Trait::hash(pattern + j)] = m - 1 - j;@\\
\mbox{}\verb@mismatch_shift = skip[Trait::hash(pattern + m - 1)];@\\
\mbox{}\verb@skip[Trait::hash(pattern + m - 1)] = 0;@\\
\mbox{}\verb@@\end{list}
\vspace{-1ex}
\footnotesize\addtolength{\baselineskip}{-1ex}
\begin{list}{}{\setlength{\itemsep}{-\parsep}\setlength{\itemindent}{-\leftmargin}}
\item Used in part 42a.
\end{list}
\end{minipage}\\[4ex]
\end{flushleft}
\subsubsection{Next Table Procedure and Call}

When we have random access to the pattern, we take advantage of it in
computing the \verb|next| table (we do not need to create the
\verb|pattern_iterator| table used in the forward iterator version).

\begin{flushleft} \small \label{__n:gensearch.w:part75}
$\langle$Define procedure to compute next table (C++) {\footnotesize 43d}$\rangle\equiv$
\vspace{-1ex}
\begin{list}{}{} \item
\mbox{}\verb@template <typename RandomAccessIterator, typename Distance>@\\
\mbox{}\verb@void compute_next(RandomAccessIterator pattern, @\\
\mbox{}\verb@                  RandomAccessIterator patternEnd,@\\
\mbox{}\verb@                  vector<Distance>& next)@\\
\mbox{}\verb@{@\\
\mbox{}\verb@  Distance pattern_size = patternEnd - pattern, j = 0, t = -1;@\\
\mbox{}\verb@  next.reserve(32);@\\
\mbox{}\verb@  next.push_back(-1);@\\
\mbox{}\verb@  while (j < pattern_size - 1) {@\\
\mbox{}\verb@    while (t >= 0 && pattern[j] != pattern[t]) @\\
\mbox{}\verb@       t = next[t];@\\
\mbox{}\verb@    ++j; ++t;@\\
\mbox{}\verb@    if (pattern[j] == pattern[t]) @\\
\mbox{}\verb@      next.push_back(next[t]);@\\
\mbox{}\verb@    else@\\
\mbox{}\verb@      next.push_back(t);@\\
\mbox{}\verb@  }@\\
\mbox{}\verb@}@\\
\mbox{}\verb@@\end{list}
\vspace{-1ex}
\footnotesize\addtolength{\baselineskip}{-1ex}
\begin{list}{}{\setlength{\itemsep}{-\parsep}\setlength{\itemindent}{-\leftmargin}}
\item Used in part 38b.
\end{list}
\end{flushleft}
\begin{flushleft} \small
\begin{minipage}{\linewidth} \label{__n:gensearch.w:part76}
$\langle$Compute next table (C++) {\footnotesize 44a}$\rangle\equiv$
\vspace{-1ex}
\begin{list}{}{} \item
\mbox{}\verb@compute_next(pattern, patternEnd, next);@\\
\mbox{}\verb@@\end{list}
\vspace{-1ex}
\footnotesize\addtolength{\baselineskip}{-1ex}
\begin{list}{}{\setlength{\itemsep}{-\parsep}\setlength{\itemindent}{-\leftmargin}}
\item Used in parts 42a, 46a.
\end{list}
\end{minipage}\\[4ex]
\end{flushleft}
\subsection{Experimental Version for Large Alphabet Case}

For comparison with HAL in the large alphabet case we also implemented
the experimental version that uses a large skip table and no hashing,
as described in the body of the paper.

\begin{flushleft} \small
\begin{minipage}{\linewidth} \label{__n:gensearch.w:part77}
\verb@"experimental_search.h"@ {\footnotesize 44b }$\equiv$
\vspace{-1ex}
\begin{list}{}{} \item
\mbox{}\verb@@\hbox{$\langle$Experimental search function with skip loop without hashing {\footnotesize 44c}$\rangle$}\verb@@\\
\mbox{}\verb@@\end{list}
\vspace{-2ex}
\end{minipage}\\[4ex]
\end{flushleft}
In our experiments, we assume that the element type is a 2-byte
unsigned short.

\begin{flushleft} \small \label{__n:gensearch.w:part78}
$\langle$Experimental search function with skip loop without hashing {\footnotesize 44c}$\rangle\equiv$
\vspace{-1ex}
\begin{list}{}{} \item
\mbox{}\verb@#include <vector>@\\
\mbox{}\verb@using namespace std;@\\
\mbox{}\verb@@\\
\mbox{}\verb@struct large_alphabet_trait {@\\
\mbox{}\verb@  typedef unsigned short T;@\\
\mbox{}\verb@  enum {suffix_size = 1};@\\
\mbox{}\verb@  enum {hash_range_max = (1u << (sizeof(T) * 8)) - 1};@\\
\mbox{}\verb@};@\\
\mbox{}\verb@@\\
\mbox{}\verb@template <> struct search_trait<unsigned short> {@\\
\mbox{}\verb@  enum {hash_range_max = 256};@\\
\mbox{}\verb@  enum {suffix_size = 1};@\\
\mbox{}\verb@  template <typename RandomAccessIterator>@\\
\mbox{}\verb@  inline static unsigned int hash(RandomAccessIterator i) {@\\
\mbox{}\verb@    return (unsigned char)(*i);@\\
\mbox{}\verb@  }@\\
\mbox{}\verb@};@\\
\mbox{}\verb@@\\
\mbox{}\verb@template <typename T>@\\
\mbox{}\verb@class skewed_value {@\\
\mbox{}\verb@  static T skew;@\\
\mbox{}\verb@  T value;@\\
\mbox{}\verb@public:@\\
\mbox{}\verb@  skewed_value() : value(0) {}@\\
\mbox{}\verb@  skewed_value(T val) : value(val - skew) {}@\\
\mbox{}\verb@  operator T () { return value + skew; }@\\
\mbox{}\verb@  static void setSkew(T askew) { skew = askew; }@\\
\mbox{}\verb@  void clear() { value = 0; }@\\
\mbox{}\verb@};@\\
\mbox{}\verb@@\\
\mbox{}\verb@template <typename T> T skewed_value<T>::skew;@\\
\mbox{}\verb@@\\
\mbox{}\verb@template <typename T, typename RandomAccessIterator, int size>@\\
\mbox{}\verb@class skewed_array {@\\
\mbox{}\verb@  typedef skewed_value<T> value_type;@\\
\mbox{}\verb@  static value_type array[size];@\\
\mbox{}\verb@  RandomAccessIterator pattern, patternEnd;@\\
\mbox{}\verb@public:@\\
\mbox{}\verb@  skewed_array(T skew, RandomAccessIterator pat, RandomAccessIterator patEnd):@\\
\mbox{}\verb@    pattern(pat),patternEnd(patEnd){ value_type::setSkew(skew); }@\\
\mbox{}\verb@  ~skewed_array() {@\\
\mbox{}\verb@    while (pattern != patternEnd) @\\
\mbox{}\verb@      array[*pattern++].clear();@\\
\mbox{}\verb@  }@\\
\mbox{}\verb@  value_type  operator[] (int index) const { return array[index]; }@\\
\mbox{}\verb@  value_type& operator[] (int index)       { return array[index]; }@\\
\mbox{}\verb@};@\\
\mbox{}\verb@@\\
\mbox{}\verb@template <typename T, typename RandomAccessIterator, int size>@\\
\mbox{}\verb@skewed_value<T> skewed_array<T,RandomAccessIterator,size>::array[size];@\\
\mbox{}\verb@@\\
\mbox{}\verb@template <typename RandomAccessIterator1, typename RandomAccessIterator2>@\\
\mbox{}\verb@RandomAccessIterator1 search_no_hashing(RandomAccessIterator1 text,@\\
\mbox{}\verb@                                        RandomAccessIterator1 textEnd,@\\
\mbox{}\verb@                                        RandomAccessIterator2 pattern,@\\
\mbox{}\verb@                                        RandomAccessIterator2 patternEnd)@\\
\mbox{}\verb@{@\\
\mbox{}\verb@  typedef typename iterator_traits<RandomAccessIterator1>::difference_type Distance1;@\\
\mbox{}\verb@  typedef typename iterator_traits<RandomAccessIterator2>::difference_type Distance2;@\\
\mbox{}\verb@  typedef large_alphabet_trait Trait;@\\
\mbox{}\verb@  if (pattern == patternEnd) @\\
\mbox{}\verb@    return text;@\\
\mbox{}\verb@  Distance1 k, text_size, large, adjustment, mismatch_shift;@\\
\mbox{}\verb@  Distance2 j, m, pattern_size;@\\
\mbox{}\verb@  pattern_size = patternEnd - pattern;@\\
\mbox{}\verb@  if (pattern_size < Trait::suffix_size)@\\
\mbox{}\verb@    return __search_L(text, textEnd, pattern, patternEnd);@\\
\mbox{}\verb@  vector<Distance1> next;@\\
\mbox{}\verb@  skewed_array<Distance1, RandomAccessIterator2, Trait::hash_range_max+1>@\\
\mbox{}\verb@    skip(pattern_size - Trait::suffix_size + 1, pattern, patternEnd);@\\
\mbox{}\verb@  @\hbox{$\langle$Accelerated Linear algorithm, no hashing (C++) {\footnotesize 46a}$\rangle$}\verb@@\\
\mbox{}\verb@}@\\
\mbox{}\verb@@\end{list}
\vspace{-1ex}
\footnotesize\addtolength{\baselineskip}{-1ex}
\begin{list}{}{\setlength{\itemsep}{-\parsep}\setlength{\itemindent}{-\leftmargin}}
\item Used in part 44b.
\end{list}
\end{flushleft}
\begin{flushleft} \small
\begin{minipage}{\linewidth} \label{__n:gensearch.w:part79}
$\langle$Accelerated Linear algorithm, no hashing (C++) {\footnotesize 46a}$\rangle\equiv$
\vspace{-1ex}
\begin{list}{}{} \item
\mbox{}\verb@k = 0; @\\
\mbox{}\verb@text_size = textEnd - text;@\\
\mbox{}\verb@@\hbox{$\langle$Compute next table (C++) {\footnotesize 44a}$\rangle$}\verb@@\\
\mbox{}\verb@@\hbox{$\langle$Handle pattern size = 1 as a special case (C++) {\footnotesize 21b}$\rangle$}\verb@@\\
\mbox{}\verb@@\hbox{$\langle$Compute skip table and mismatch shift, no hashing (C++) {\footnotesize 46b}$\rangle$}\verb@@\\
\mbox{}\verb@large = text_size + 1;@\\
\mbox{}\verb@adjustment = large + pattern_size - 1;@\\
\mbox{}\verb@skip[*(pattern + m - 1)] = large;@\\
\mbox{}\verb@@\\
\mbox{}\verb@k -= text_size;@\\
\mbox{}\verb@for (;;) {@\\
\mbox{}\verb@  k += pattern_size - 1;@\\
\mbox{}\verb@  if (k >= 0) break;@\\
\mbox{}\verb@  @\hbox{$\langle$Scan the text using a single-test skip loop, no hashing (C++) {\footnotesize 46c}$\rangle$}\verb@@\\
\mbox{}\verb@  @\hbox{$\langle$Verify match or recover from mismatch (C++) {\footnotesize 42c}$\rangle$}\verb@@\\
\mbox{}\verb@}@\\
\mbox{}\verb@return textEnd;@\\
\mbox{}\verb@@\end{list}
\vspace{-1ex}
\footnotesize\addtolength{\baselineskip}{-1ex}
\begin{list}{}{\setlength{\itemsep}{-\parsep}\setlength{\itemindent}{-\leftmargin}}
\item Used in part 44c.
\end{list}
\end{minipage}\\[4ex]
\end{flushleft}
\begin{flushleft} \small
\begin{minipage}{\linewidth} \label{__n:gensearch.w:part80}
$\langle$Compute skip table and mismatch shift, no hashing (C++) {\footnotesize 46b}$\rangle\equiv$
\vspace{-1ex}
\begin{list}{}{} \item
\mbox{}\verb@m = next.size();@\\
\mbox{}\verb@for (j = Trait::suffix_size - 1; j < m - 1; ++j)@\\
\mbox{}\verb@  skip[*(pattern + j)] = m - 1 - j;@\\
\mbox{}\verb@mismatch_shift = skip[*(pattern + m - 1)];@\\
\mbox{}\verb@skip[*(pattern + m - 1)] = 0;@\\
\mbox{}\verb@@\end{list}
\vspace{-1ex}
\footnotesize\addtolength{\baselineskip}{-1ex}
\begin{list}{}{\setlength{\itemsep}{-\parsep}\setlength{\itemindent}{-\leftmargin}}
\item Used in part 46a.
\end{list}
\end{minipage}\\[4ex]
\end{flushleft}
\begin{flushleft} \small
\begin{minipage}{\linewidth} \label{__n:gensearch.w:part81}
$\langle$Scan the text using a single-test skip loop, no hashing (C++) {\footnotesize 46c}$\rangle\equiv$
\vspace{-1ex}
\begin{list}{}{} \item
\mbox{}\verb@do {@\\
\mbox{}\verb@  k += skip[*(textEnd + k)]; @\\
\mbox{}\verb@} while (k < 0);@\\
\mbox{}\verb@if (k < pattern_size)@\\
\mbox{}\verb@  return textEnd;@\\
\mbox{}\verb@k -= adjustment;@\\
\mbox{}\verb@@\end{list}
\vspace{-1ex}
\footnotesize\addtolength{\baselineskip}{-1ex}
\begin{list}{}{\setlength{\itemsep}{-\parsep}\setlength{\itemindent}{-\leftmargin}}
\item Used in part 46a.
\end{list}
\end{minipage}\\[4ex]
\end{flushleft}
\subsection{DNA Search Functions and Traits}

The following definitions are for use in DNA search experiments.
Four different search functions are defined using 2, 3, 4, or 5
characters as arguments to hash functions.

\begin{flushleft} \small \label{__n:gensearch.w:part82}
\verb@"DNA_search.h"@ {\footnotesize 47a }$\equiv$
\vspace{-1ex}
\begin{list}{}{} \item
\mbox{}\verb@@\hbox{$\langle$Define DNA search traits {\footnotesize 47b}$\rangle$}\verb@@\\
\mbox{}\verb@@\\
\mbox{}\verb@template <typename RandomAccessIterator1, typename RandomAccessIterator2>@\\
\mbox{}\verb@inline RandomAccessIterator1 hal2(RandomAccessIterator1 text, @\\
\mbox{}\verb@                                  RandomAccessIterator1 textEnd,@\\
\mbox{}\verb@                                  RandomAccessIterator2 pattern,@\\
\mbox{}\verb@                                  RandomAccessIterator2 patternEnd)@\\
\mbox{}\verb@{@\\
\mbox{}\verb@  return search_hashed(text, textEnd, pattern, patternEnd,@\\
\mbox{}\verb@                       static_cast<search_trait_dna2*>(0));@\\
\mbox{}\verb@}@\\
\mbox{}\verb@@\\
\mbox{}\verb@template <typename RandomAccessIterator1, typename RandomAccessIterator2>@\\
\mbox{}\verb@inline RandomAccessIterator1 hal3(RandomAccessIterator1 text, @\\
\mbox{}\verb@                                  RandomAccessIterator1 textEnd,@\\
\mbox{}\verb@                                  RandomAccessIterator2 pattern,@\\
\mbox{}\verb@                                  RandomAccessIterator2 patternEnd)@\\
\mbox{}\verb@{@\\
\mbox{}\verb@  return search_hashed(text, textEnd, pattern, patternEnd,@\\
\mbox{}\verb@                       static_cast<search_trait_dna3*>(0));@\\
\mbox{}\verb@}@\\
\mbox{}\verb@    @\\
\mbox{}\verb@template <typename RandomAccessIterator1, typename RandomAccessIterator2>@\\
\mbox{}\verb@inline RandomAccessIterator1 hal4(RandomAccessIterator1 text, @\\
\mbox{}\verb@                                  RandomAccessIterator1 textEnd,@\\
\mbox{}\verb@                                  RandomAccessIterator2 pattern,@\\
\mbox{}\verb@                                  RandomAccessIterator2 patternEnd)@\\
\mbox{}\verb@{@\\
\mbox{}\verb@  return search_hashed(text, textEnd, pattern, patternEnd,@\\
\mbox{}\verb@                       static_cast<search_trait_dna4*>(0));@\\
\mbox{}\verb@} @\\
\mbox{}\verb@@\\
\mbox{}\verb@template <typename RandomAccessIterator1, typename RandomAccessIterator2>@\\
\mbox{}\verb@inline RandomAccessIterator1 hal5(RandomAccessIterator1 text, @\\
\mbox{}\verb@                                  RandomAccessIterator1 textEnd,@\\
\mbox{}\verb@                                  RandomAccessIterator2 pattern,@\\
\mbox{}\verb@                                  RandomAccessIterator2 patternEnd)@\\
\mbox{}\verb@{@\\
\mbox{}\verb@  return search_hashed(text, textEnd, pattern, patternEnd,@\\
\mbox{}\verb@                       static_cast<search_trait_dna5*>(0));@\\
\mbox{}\verb@}@\\
\mbox{}\verb@@\end{list}
\vspace{-2ex}
\end{flushleft}
\begin{flushleft} \small \label{__n:gensearch.w:part83}
$\langle$Define DNA search traits {\footnotesize 47b}$\rangle\equiv$
\vspace{-1ex}
\begin{list}{}{} \item
\mbox{}\verb@struct search_trait_dna2 {@\\
\mbox{}\verb@  enum {hash_range_max = 64};@\\
\mbox{}\verb@  enum {suffix_size = 2};@\\
\mbox{}\verb@  template <typename RAI>@\\
\mbox{}\verb@  inline static unsigned int hash(RAI i) {@\\
\mbox{}\verb@    return (*(i-1) + ((*i) << 3)) & 63;@\\
\mbox{}\verb@  }@\\
\mbox{}\verb@};@\\
\mbox{}\verb@@\\
\mbox{}\verb@struct search_trait_dna3 {@\\
\mbox{}\verb@  enum {hash_range_max = 512};@\\
\mbox{}\verb@  enum {suffix_size = 3};@\\
\mbox{}\verb@  template <typename RAI>@\\
\mbox{}\verb@  inline static unsigned int hash(RAI i) {@\\
\mbox{}\verb@    return (*(i-2) + (*(i-1) << 3) + ((*i) << 6)) & 511;@\\
\mbox{}\verb@  }@\\
\mbox{}\verb@};@\\
\mbox{}\verb@@\\
\mbox{}\verb@struct search_trait_dna4 {@\\
\mbox{}\verb@  enum {hash_range_max = 256};@\\
\mbox{}\verb@  enum {suffix_size = 4};@\\
\mbox{}\verb@  template <typename RAI>@\\
\mbox{}\verb@  inline static unsigned int hash(RAI i) {@\\
\mbox{}\verb@    return (*(i-3) + (*(i-2) << 2) + (*(i-1) << 4)@\\
\mbox{}\verb@           + ((*i) << 6)) & 255;@\\
\mbox{}\verb@  }@\\
\mbox{}\verb@};@\\
\mbox{}\verb@@\\
\mbox{}\verb@struct search_trait_dna5 {@\\
\mbox{}\verb@  enum {hash_range_max = 256};@\\
\mbox{}\verb@  enum {suffix_size = 5};@\\
\mbox{}\verb@  template <typename RAI>@\\
\mbox{}\verb@  inline static unsigned int hash(RAI i) {@\\
\mbox{}\verb@    return (*(i-4) + (*(i-3) << 2) + (*(i-2) << 4)@\\
\mbox{}\verb@           + (*(i-1) << 6) + ((*i) << 8)) & 255;@\\
\mbox{}\verb@  }@\\
\mbox{}\verb@};@\\
\mbox{}\verb@@\end{list}
\vspace{-1ex}
\footnotesize\addtolength{\baselineskip}{-1ex}
\begin{list}{}{\setlength{\itemsep}{-\parsep}\setlength{\itemindent}{-\leftmargin}}
\item Used in part 47a.
\end{list}
\end{flushleft}
\subsection{Simple Tests}

In the test programs we want to compare the new search functions with
the existing search function from an STL algorithm library
implementation, so we rename the existing one.

\begin{flushleft} \small
\begin{minipage}{\linewidth} \label{__n:gensearch.w:part84}
$\langle$Include algorithms header with existing search function renamed {\footnotesize 48}$\rangle\equiv$
\vspace{-1ex}
\begin{list}{}{} \item
\mbox{}\verb@#define search stl_search@\\
\mbox{}\verb@#define __search __stl_search@\\
\mbox{}\verb@#include <algorithm>@\\
\mbox{}\verb@#undef search@\\
\mbox{}\verb@#undef __search@\\
\mbox{}\verb@@\end{list}
\vspace{-1ex}
\footnotesize\addtolength{\baselineskip}{-1ex}
\begin{list}{}{\setlength{\itemsep}{-\parsep}\setlength{\itemindent}{-\leftmargin}}
\item Used in parts 49a, 52, 54f, 57b, 59b, 63b, 65b.
\end{list}
\end{minipage}\\[4ex]
\end{flushleft}
As in the Ada version of the code, the first test program simply reads
short test sequences from a file and reports the results of running
the different search algorithms on them.  


\begin{flushleft} \small
\begin{minipage}{\linewidth} \label{__n:gensearch.w:part85}
\verb@"test_search.cpp"@ {\footnotesize 49a }$\equiv$
\vspace{-1ex}
\begin{list}{}{} \item
\mbox{}\verb@@\hbox{$\langle$Include algorithms header with existing search function renamed {\footnotesize 48}$\rangle$}\verb@@\\
\mbox{}\verb@#include <iostream>@\\
\mbox{}\verb@#include <fstream>@\\
\mbox{}\verb@#include "new_search.h"@\\
\mbox{}\verb@#include "hume.hh"@\\
\mbox{}\verb@#include "DNA_search.h"@\\
\mbox{}\verb@using namespace std;@\\
\mbox{}\verb@int Base_Line;@\\
\mbox{}\verb@@\hbox{$\langle$Define procedure to read string into sequence (C++) {\footnotesize 51b}$\rangle$}\verb@@\\
\mbox{}\verb@typedef unsigned char data;@\\
\mbox{}\verb@@\hbox{$\langle$Define algorithm enumeration type, names, and selector function (C++) {\footnotesize 49b}$\rangle$}\verb@@\\
\mbox{}\verb@@\hbox{$\langle$Define Report procedure (C++) {\footnotesize 51d}$\rangle$}\verb@@\\
\mbox{}\verb@int main() @\\
\mbox{}\verb@{  @\\
\mbox{}\verb@  ostream_iterator<char> out(cout, "");@\\
\mbox{}\verb@  ifstream ifs("small.txt");@\\
\mbox{}\verb@  vector<data> Comment, S1, S2;@\\
\mbox{}\verb@  const char* separator = "";@\\
\mbox{}\verb@  for (;;) {@\\
\mbox{}\verb@    @\hbox{$\langle$Read test sequences from file (C++) {\footnotesize 50}$\rangle$}\verb@@\\
\mbox{}\verb@    @\hbox{$\langle$Run tests and report results (C++) {\footnotesize 51c}$\rangle$}\verb@@\\
\mbox{}\verb@  }@\\
\mbox{}\verb@  return 0;}@\\
\mbox{}\verb@@\end{list}
\vspace{-2ex}
\end{minipage}\\[4ex]
\end{flushleft}
\begin{flushleft} \small \label{__n:gensearch.w:part86}
$\langle$Define algorithm enumeration type, names, and selector function (C++) {\footnotesize 49b}$\rangle\equiv$
\vspace{-1ex}
\begin{list}{}{} \item
\mbox{}\verb@enum algorithm_enumeration {@\\
\mbox{}\verb@     Dummy, SF, L, HAL, ABM, TBM, GBM, HAL2, HAL3, HAL4, HAL5@\\
\mbox{}\verb@};@\\
\mbox{}\verb@const char* algorithm_names[] = {@\\
\mbox{}\verb@     "selection code", "SF", "L", "HAL", "ABM", "TBM", "GBM", @\\
\mbox{}\verb@     "HAL2", "HAL3", "HAL4", "HAL5"@\\
\mbox{}\verb@};@\\
\mbox{}\verb@@\\
\mbox{}\verb@#ifndef DNA_TEST@\\
\mbox{}\verb@  algorithm_enumeration alg[] = {Dummy, SF, L, HAL, ABM, TBM};@\\
\mbox{}\verb@  const char textFileName[] = "long.txt";@\\
\mbox{}\verb@  const char wordFileName[] = "words.txt";@\\
\mbox{}\verb@#else@\\
\mbox{}\verb@  algorithm_enumeration alg[] = {Dummy, SF, L, HAL, ABM, GBM, @\\
\mbox{}\verb@                                 HAL2, HAL3, HAL4, HAL5};@\\
\mbox{}\verb@  const char textFileName[] = "dnatext.txt";@\\
\mbox{}\verb@  const char wordFileName[] = "dnaword.txt";@\\
\mbox{}\verb@#endif@\\
\mbox{}\verb@@\\
\mbox{}\verb@const int number_of_algorithms = sizeof(alg)/sizeof(alg[0]); @\\
\mbox{}\verb@@\\
\mbox{}\verb@template <typename Container, typename Container__const_iterator>@\\
\mbox{}\verb@inline void@\\
\mbox{}\verb@   Algorithm(int k, const Container& x, const Container& y, @\\
\mbox{}\verb@             Container__const_iterator& result)@\\
\mbox{}\verb@{@\\
\mbox{}\verb@  switch (alg[k]) {@\\
\mbox{}\verb@  case Dummy: @\\
\mbox{}\verb@     // does nothing, used for timing overhead of test loop@\\
\mbox{}\verb@     result = x.begin(); return;  @\\
\mbox{}\verb@  case SF: @\\
\mbox{}\verb@     result = stl_search(x.begin(), x.end(), y.begin(), y.end()); return;@\\
\mbox{}\verb@  case L: @\\
\mbox{}\verb@     result =  __search_L(x.begin(), x.end(), y.begin(), y.end()); return;@\\
\mbox{}\verb@  case HAL: @\\
\mbox{}\verb@     result = search(x.begin(), x.end(), y.begin(), y.end()); return;@\\
\mbox{}\verb@  case ABM: @\\
\mbox{}\verb@     result = fbm(x.begin(), x.end(), y.begin(), y.end()); return;@\\
\mbox{}\verb@  case TBM: @\\
\mbox{}\verb@     result = hume(x.begin(), x.end(), y.begin(), y.end()); return;@\\
\mbox{}\verb@  case GBM: @\\
\mbox{}\verb@     result = gdbm(x.begin(), x.end(), y.begin(), y.end()); return;@\\
\mbox{}\verb@  case HAL2: @\\
\mbox{}\verb@     result = hal2(x.begin(), x.end(), y.begin(), y.end()); return;@\\
\mbox{}\verb@  case HAL3: @\\
\mbox{}\verb@     result = hal3(x.begin(), x.end(), y.begin(), y.end()); return;@\\
\mbox{}\verb@  case HAL4: @\\
\mbox{}\verb@     result = hal4(x.begin(), x.end(), y.begin(), y.end()); return;@\\
\mbox{}\verb@  case HAL5: @\\
\mbox{}\verb@     result = hal5(x.begin(), x.end(), y.begin(), y.end()); return;@\\
\mbox{}\verb@  }@\\
\mbox{}\verb@  result = x.begin(); return;@\\
\mbox{}\verb@}@\\
\mbox{}\verb@@\end{list}
\vspace{-1ex}
\footnotesize\addtolength{\baselineskip}{-1ex}
\begin{list}{}{\setlength{\itemsep}{-\parsep}\setlength{\itemindent}{-\leftmargin}}
\item Used in parts 49a, 52, 54f.
\end{list}
\end{flushleft}
\begin{flushleft} \small \label{__n:gensearch.w:part87}
$\langle$Read test sequences from file (C++) {\footnotesize 50}$\rangle\equiv$
\vspace{-1ex}
\begin{list}{}{} \item
\mbox{}\verb@get(ifs, Comment);@\\
\mbox{}\verb@if (ifs.eof())@\\
\mbox{}\verb@  break;@\\
\mbox{}\verb@copy(Comment.begin(), Comment.end(), out); cout << endl;@\\
\mbox{}\verb@@\\
\mbox{}\verb@get(ifs, S1);@\\
\mbox{}\verb@@\hbox{$\langle$Check for unexpected end of file (C++) {\footnotesize 51a}$\rangle$}\verb@@\\
\mbox{}\verb@cout << "Text string:......";@\\
\mbox{}\verb@copy(S1.begin(), S1.end(), out);@\\
\mbox{}\verb@cout << endl;@\\
\mbox{}\verb@@\\
\mbox{}\verb@get(ifs, S2);@\\
\mbox{}\verb@@\hbox{$\langle$Check for unexpected end of file (C++) {\footnotesize 51a}$\rangle$}\verb@@\\
\mbox{}\verb@cout << "Pattern string:...";@\\
\mbox{}\verb@copy(S2.begin(), S2.end(), out); cout << endl;@\\
\mbox{}\verb@@\end{list}
\vspace{-1ex}
\footnotesize\addtolength{\baselineskip}{-1ex}
\begin{list}{}{\setlength{\itemsep}{-\parsep}\setlength{\itemindent}{-\leftmargin}}
\item Used in part 49a.
\end{list}
\end{flushleft}
\begin{flushleft} \small
\begin{minipage}{\linewidth} \label{__n:gensearch.w:part88}
$\langle$Check for unexpected end of file (C++) {\footnotesize 51a}$\rangle\equiv$
\vspace{-1ex}
\begin{list}{}{} \item
\mbox{}\verb@if (ifs.eof()) {@\\
\mbox{}\verb@  cout << "**** Unexpected end of file." << endl;@\\
\mbox{}\verb@  exit(1);@\\
\mbox{}\verb@}@\\
\mbox{}\verb@@\end{list}
\vspace{-1ex}
\footnotesize\addtolength{\baselineskip}{-1ex}
\begin{list}{}{\setlength{\itemsep}{-\parsep}\setlength{\itemindent}{-\leftmargin}}
\item Used in part 50.
\end{list}
\end{minipage}\\[4ex]
\end{flushleft}
\begin{flushleft} \small
\begin{minipage}{\linewidth} \label{__n:gensearch.w:part89}
$\langle$Define procedure to read string into sequence (C++) {\footnotesize 51b}$\rangle\equiv$
\vspace{-1ex}
\begin{list}{}{} \item
\mbox{}\verb@template <typename Container>@\\
\mbox{}\verb@void get(istream& is, Container& S) {@\\
\mbox{}\verb@  S.erase(S.begin(), S.end());@\\
\mbox{}\verb@  char ch;@\\
\mbox{}\verb@  while (is.get(ch)) {@\\
\mbox{}\verb@    if (ch == '\n')@\\
\mbox{}\verb@      break;@\\
\mbox{}\verb@    S.push_back(ch);@\\
\mbox{}\verb@  }@\\
\mbox{}\verb@}@\\
\mbox{}\verb@@\end{list}
\vspace{-1ex}
\footnotesize\addtolength{\baselineskip}{-1ex}
\begin{list}{}{\setlength{\itemsep}{-\parsep}\setlength{\itemindent}{-\leftmargin}}
\item Used in part 49a.
\end{list}
\end{minipage}\\[4ex]
\end{flushleft}
\begin{flushleft} \small
\begin{minipage}{\linewidth} \label{__n:gensearch.w:part90}
$\langle$Run tests and report results (C++) {\footnotesize 51c}$\rangle\equiv$
\vspace{-1ex}
\begin{list}{}{} \item
\mbox{}\verb@Base_Line = 0;@\\
\mbox{}\verb@for (int k = 1; k < number_of_algorithms; ++k) {@\\
\mbox{}\verb@  cout << "Using " << algorithm_names[k] << ":" << endl;@\\
\mbox{}\verb@  Report(algorithm_enumeration(k), S1, S2, separator);@\\
\mbox{}\verb@}@\\
\mbox{}\verb@cout << endl;@\\
\mbox{}\verb@@\end{list}
\vspace{-1ex}
\footnotesize\addtolength{\baselineskip}{-1ex}
\begin{list}{}{\setlength{\itemsep}{-\parsep}\setlength{\itemindent}{-\leftmargin}}
\item Used in parts 49a, 54ce.
\end{list}
\end{minipage}\\[4ex]
\end{flushleft}
\begin{flushleft} \small \label{__n:gensearch.w:part91}
$\langle$Define Report procedure (C++) {\footnotesize 51d}$\rangle\equiv$
\vspace{-1ex}
\begin{list}{}{} \item
\mbox{}\verb@template <typename Container>@\\
\mbox{}\verb@void Report(algorithm_enumeration k, const Container& S1, @\\
\mbox{}\verb@            const Container& S2, const char* separator)@\\
\mbox{}\verb@{@\\
\mbox{}\verb@  typename Container::const_iterator P;@\\
\mbox{}\verb@  typedef typename Container::value_type value_t;@\\
\mbox{}\verb@  Algorithm(k, S1, S2, P);@\\
\mbox{}\verb@  cout << "  String " << '"';@\\
\mbox{}\verb@  copy(S2.begin(), S2.end(), ostream_iterator<value_t>(cout, separator));@\\
\mbox{}\verb@  if (P == S1.end())@\\
\mbox{}\verb@    cout << '"' << " not found" << endl;@\\
\mbox{}\verb@  else@\\
\mbox{}\verb@    cout << '"' << " found at position " << P - S1.begin() << endl;@\\
\mbox{}\verb@  if (Base_Line == 0)@\\
\mbox{}\verb@    Base_Line = P - S1.begin();@\\
\mbox{}\verb@  else@\\
\mbox{}\verb@    if (P - S1.begin() != Base_Line)@\\
\mbox{}\verb@      cout << "*****Incorrect result!" << endl;@\\
\mbox{}\verb@}@\\
\mbox{}\verb@@\end{list}
\vspace{-1ex}
\footnotesize\addtolength{\baselineskip}{-1ex}
\begin{list}{}{\setlength{\itemsep}{-\parsep}\setlength{\itemindent}{-\leftmargin}}
\item Used in parts 49a, 52, 63b.
\end{list}
\end{flushleft}
\subsection{Large Tests}

The following program for conducting tests on a long text sequence performs
the same tests as the Ada version, plus searches for words of the requested
pattern size selected from a given dictionary (which is read from a file).

\begin{flushleft} \small \label{__n:gensearch.w:part92}
\verb@"test_long_search.cpp"@ {\footnotesize 52 }$\equiv$
\vspace{-1ex}
\begin{list}{}{} \item
\mbox{}\verb@@\hbox{$\langle$Include algorithms header with existing search function renamed {\footnotesize 48}$\rangle$}\verb@@\\
\mbox{}\verb@#include  "new_search.h"@\\
\mbox{}\verb@#include "hume.hh"@\\
\mbox{}\verb@#include "DNA_search.h"@\\
\mbox{}\verb@#include <iterator>@\\
\mbox{}\verb@#include <vector>@\\
\mbox{}\verb@#include <map>@\\
\mbox{}\verb@#include <iostream>@\\
\mbox{}\verb@#include <fstream>@\\
\mbox{}\verb@#include <string>@\\
\mbox{}\verb@using namespace std;@\\
\mbox{}\verb@@\\
\mbox{}\verb@typedef unsigned char data;@\\
\mbox{}\verb@typedef vector<data> sequence;@\\
\mbox{}\verb@sequence S1, S2;@\\
\mbox{}\verb@@\\
\mbox{}\verb@int Base_Line;@\\
\mbox{}\verb@unsigned int Number_Of_Tests, Number_Of_Pattern_Sizes, Increment;@\\
\mbox{}\verb@@\hbox{$\langle$Define algorithm enumeration type, names, and selector function (C++) {\footnotesize 49b}$\rangle$}\verb@@\\
\mbox{}\verb@@\hbox{$\langle$Define Report procedure (C++) {\footnotesize 51d}$\rangle$}\verb@@\\
\mbox{}\verb@int main() @\\
\mbox{}\verb@{  @\\
\mbox{}\verb@  unsigned int F, K, j;@\\
\mbox{}\verb@  @\hbox{$\langle$Read test parameters (C++) {\footnotesize 53a}$\rangle$}\verb@@\\
\mbox{}\verb@  @\hbox{$\langle$Read dictionary from file, placing words of size j in dictionary[j] {\footnotesize 53b}$\rangle$}\verb@@\\
\mbox{}\verb@  @\hbox{$\langle$Read character sequence from file (C++) {\footnotesize 54a}$\rangle$}\verb@@\\
\mbox{}\verb@  for (j = 0; j < Number_Of_Pattern_Sizes; ++j) {@\\
\mbox{}\verb@    @\hbox{$\langle$Trim dictionary[Pattern\_Size[j]] to have at most Number\_Of\_Tests words {\footnotesize 53c}$\rangle$}\verb@@\\
\mbox{}\verb@    Increment = (S1.size() - Pattern_Size[j]) / Number_Of_Tests;@\\
\mbox{}\verb@    cerr << Pattern_Size[j] << " " << flush;@\\
\mbox{}\verb@    const char* separator = "";@\\
\mbox{}\verb@    @\hbox{$\langle$Output header (C++) {\footnotesize 54b}$\rangle$}\verb@@\\
\mbox{}\verb@    @\hbox{$\langle$Run tests searching for selected subsequences (C++) {\footnotesize 54c}$\rangle$}\verb@@\\
\mbox{}\verb@    @\hbox{$\langle$Run tests searching for dictionary words (C++) {\footnotesize 54e}$\rangle$}\verb@@\\
\mbox{}\verb@  }@\\
\mbox{}\verb@}@\\
\mbox{}\verb@@\end{list}
\vspace{-2ex}
\end{flushleft}
\begin{flushleft} \small
\begin{minipage}{\linewidth} \label{__n:gensearch.w:part93}
$\langle$Read test parameters (C++) {\footnotesize 53a}$\rangle\equiv$
\vspace{-1ex}
\begin{list}{}{} \item
\mbox{}\verb@cout << "Input number of tests (for each pattern size): " << flush;@\\
\mbox{}\verb@cin >> Number_Of_Tests;@\\
\mbox{}\verb@cout << "Input number of pattern sizes: " << flush;@\\
\mbox{}\verb@cin >> Number_Of_Pattern_Sizes;@\\
\mbox{}\verb@cout << "Input pattern sizes: " << flush;@\\
\mbox{}\verb@vector<int> Pattern_Size(Number_Of_Pattern_Sizes);@\\
\mbox{}\verb@for (j = 0; j < Number_Of_Pattern_Sizes; ++j)@\\
\mbox{}\verb@  cin >> Pattern_Size[j];@\\
\mbox{}\verb@cout << "\nNumber of tests: " << Number_Of_Tests << endl;@\\
\mbox{}\verb@cout << "Pattern sizes: ";@\\
\mbox{}\verb@for (j = 0; j < Number_Of_Pattern_Sizes; ++j) @\\
\mbox{}\verb@  cout << Pattern_Size[j] << " ";@\\
\mbox{}\verb@cout << endl;@\\
\mbox{}\verb@@\end{list}
\vspace{-1ex}
\footnotesize\addtolength{\baselineskip}{-1ex}
\begin{list}{}{\setlength{\itemsep}{-\parsep}\setlength{\itemindent}{-\leftmargin}}
\item Used in parts 52, 54f, 57b, 59b, 63b, 65b.
\end{list}
\end{minipage}\\[4ex]
\end{flushleft}
\begin{flushleft} \small
\begin{minipage}{\linewidth} \label{__n:gensearch.w:part94}
$\langle$Read dictionary from file, placing words of size j in dictionary[j] {\footnotesize 53b}$\rangle\equiv$
\vspace{-1ex}
\begin{list}{}{} \item
\mbox{}\verb@ifstream dictfile(wordFileName);@\\
\mbox{}\verb@typedef istream_iterator<string> string_input;@\\
\mbox{}\verb@typedef map<int, vector<sequence>, less<int> > map_type;@\\
\mbox{}\verb@map_type dictionary;@\\
\mbox{}\verb@sequence S;@\\
\mbox{}\verb@string S0;@\\
\mbox{}\verb@string_input si(dictfile);@\\
\mbox{}\verb@while (si != string_input()) {@\\
\mbox{}\verb@  S0 = *si++;@\\
\mbox{}\verb@  sequence S(S0.begin(), S0.end());@\\
\mbox{}\verb@  dictionary[S.size()].push_back(S);@\\
\mbox{}\verb@}@\\
\mbox{}\verb@@\end{list}
\vspace{-1ex}
\footnotesize\addtolength{\baselineskip}{-1ex}
\begin{list}{}{\setlength{\itemsep}{-\parsep}\setlength{\itemindent}{-\leftmargin}}
\item Used in parts 52, 54f, 59b.
\end{list}
\end{minipage}\\[4ex]
\end{flushleft}
\begin{flushleft} \small
\begin{minipage}{\linewidth} \label{__n:gensearch.w:part95}
$\langle$Trim dictionary[Pattern\_Size[j]] to have at most Number\_Of\_Tests words {\footnotesize 53c}$\rangle\equiv$
\vspace{-1ex}
\begin{list}{}{} \item
\mbox{}\verb@vector<sequence>& diction = dictionary[Pattern_Size[j]];@\\
\mbox{}\verb@if (diction.size() > Number_Of_Tests) {@\\
\mbox{}\verb@  vector<sequence> temp;@\\
\mbox{}\verb@  int Skip_Amount = diction.size() / Number_Of_Tests;@\\
\mbox{}\verb@  for (unsigned int T = 0; T < Number_Of_Tests; ++T) {@\\
\mbox{}\verb@     temp.push_back(diction[T * Skip_Amount]);@\\
\mbox{}\verb@  }@\\
\mbox{}\verb@  diction = temp;@\\
\mbox{}\verb@}@\\
\mbox{}\verb@@\end{list}
\vspace{-1ex}
\footnotesize\addtolength{\baselineskip}{-1ex}
\begin{list}{}{\setlength{\itemsep}{-\parsep}\setlength{\itemindent}{-\leftmargin}}
\item Used in parts 52, 54f, 59b.
\end{list}
\end{minipage}\\[4ex]
\end{flushleft}
\begin{flushleft} \small
\begin{minipage}{\linewidth} \label{__n:gensearch.w:part96}
$\langle$Read character sequence from file (C++) {\footnotesize 54a}$\rangle\equiv$
\vspace{-1ex}
\begin{list}{}{} \item
\mbox{}\verb@ifstream ifs(textFileName);@\\
\mbox{}\verb@char C;@\\
\mbox{}\verb@while (ifs.get(C))@\\
\mbox{}\verb@  S1.push_back(C);@\\
\mbox{}\verb@cout << S1.size() << " characters read." << endl;@\\
\mbox{}\verb@@\end{list}
\vspace{-1ex}
\footnotesize\addtolength{\baselineskip}{-1ex}
\begin{list}{}{\setlength{\itemsep}{-\parsep}\setlength{\itemindent}{-\leftmargin}}
\item Used in parts 52, 54f, 59b.
\end{list}
\end{minipage}\\[4ex]
\end{flushleft}
\begin{flushleft} \small
\begin{minipage}{\linewidth} \label{__n:gensearch.w:part97}
$\langle$Output header (C++) {\footnotesize 54b}$\rangle\equiv$
\vspace{-1ex}
\begin{list}{}{} \item
\mbox{}\verb@cout << "\n\n-----------------------------------------------------------\n"@\\
\mbox{}\verb@     << "Searching for patterns of size " << Pattern_Size[j] @\\
\mbox{}\verb@     << "..." << endl;@\\
\mbox{}\verb@cout << "(" << Number_Of_Tests << " patterns from the text, "@\\
\mbox{}\verb@     << dictionary[Pattern_Size[j]].size() << "  from the dictionary)" << endl;@\\
\mbox{}\verb@@\end{list}
\vspace{-1ex}
\footnotesize\addtolength{\baselineskip}{-1ex}
\begin{list}{}{\setlength{\itemsep}{-\parsep}\setlength{\itemindent}{-\leftmargin}}
\item Used in parts 52, 54f, 57b, 59b, 63b, 65b.
\end{list}
\end{minipage}\\[4ex]
\end{flushleft}
\begin{flushleft} \small
\begin{minipage}{\linewidth} \label{__n:gensearch.w:part98}
$\langle$Run tests searching for selected subsequences (C++) {\footnotesize 54c}$\rangle\equiv$
\vspace{-1ex}
\begin{list}{}{} \item
\mbox{}\verb@F = 0;@\\
\mbox{}\verb@for (K = 1; K <= Number_Of_Tests; ++K) {@\\
\mbox{}\verb@  @\hbox{$\langle$Select sequence S2 to search for in S1 (C++) {\footnotesize 54d}$\rangle$}\verb@@\\
\mbox{}\verb@  @\hbox{$\langle$Run tests and report results (C++) {\footnotesize 51c}$\rangle$}\verb@@\\
\mbox{}\verb@}@\\
\mbox{}\verb@@\end{list}
\vspace{-1ex}
\footnotesize\addtolength{\baselineskip}{-1ex}
\begin{list}{}{\setlength{\itemsep}{-\parsep}\setlength{\itemindent}{-\leftmargin}}
\item Used in parts 52, 63b.
\end{list}
\end{minipage}\\[4ex]
\end{flushleft}
\begin{flushleft} \small
\begin{minipage}{\linewidth} \label{__n:gensearch.w:part99}
$\langle$Select sequence S2 to search for in S1 (C++) {\footnotesize 54d}$\rangle\equiv$
\vspace{-1ex}
\begin{list}{}{} \item
\mbox{}\verb@S2.erase(S2.begin(), S2.end());@\\
\mbox{}\verb@copy(S1.begin() + F, S1.begin() + F + Pattern_Size[j], back_inserter(S2));@\\
\mbox{}\verb@F += Increment;@\\
\mbox{}\verb@@\end{list}
\vspace{-1ex}
\footnotesize\addtolength{\baselineskip}{-1ex}
\begin{list}{}{\setlength{\itemsep}{-\parsep}\setlength{\itemindent}{-\leftmargin}}
\item Used in part 54c.
\end{list}
\end{minipage}\\[4ex]
\end{flushleft}
\begin{flushleft} \small
\begin{minipage}{\linewidth} \label{__n:gensearch.w:part100}
$\langle$Run tests searching for dictionary words (C++) {\footnotesize 54e}$\rangle\equiv$
\vspace{-1ex}
\begin{list}{}{} \item
\mbox{}\verb@for (K = 0; K < dictionary[Pattern_Size[j]].size(); ++K) {@\\
\mbox{}\verb@  S2 = dictionary[Pattern_Size[j]][K];@\\
\mbox{}\verb@  @\hbox{$\langle$Run tests and report results (C++) {\footnotesize 51c}$\rangle$}\verb@@\\
\mbox{}\verb@}@\\
\mbox{}\verb@@\end{list}
\vspace{-1ex}
\footnotesize\addtolength{\baselineskip}{-1ex}
\begin{list}{}{\setlength{\itemsep}{-\parsep}\setlength{\itemindent}{-\leftmargin}}
\item Used in part 52.
\end{list}
\end{minipage}\\[4ex]
\end{flushleft}
\subsection{Timed Tests}

Again, the following program for timing searches conducts the same
searches as in the Ada version, plus searches for words of the requested
pattern size selected from a given dictionary.

\begin{flushleft} \small \label{__n:gensearch.w:part101}
\verb@"time_long_search.cpp"@ {\footnotesize 54f }$\equiv$
\vspace{-1ex}
\begin{list}{}{} \item
\mbox{}\verb@@\hbox{$\langle$Include algorithms header with existing search function renamed {\footnotesize 48}$\rangle$}\verb@@\\
\mbox{}\verb@#include "new_search.h"@\\
\mbox{}\verb@#include "hume.hh"@\\
\mbox{}\verb@#include "DNA_search.h"@\\
\mbox{}\verb@#include <iterator>@\\
\mbox{}\verb@#include <deque>@\\
\mbox{}\verb@#include <vector>@\\
\mbox{}\verb@#include <map>@\\
\mbox{}\verb@#include <iostream>@\\
\mbox{}\verb@#include <fstream>@\\
\mbox{}\verb@#include <ctime>@\\
\mbox{}\verb@#include <string>@\\
\mbox{}\verb@using namespace std;@\\
\mbox{}\verb@@\\
\mbox{}\verb@typedef unsigned char data;@\\
\mbox{}\verb@typedef vector<data> sequence;@\\
\mbox{}\verb@sequence S1;@\\
\mbox{}\verb@int Base_Line;@\\
\mbox{}\verb@unsigned int Number_Of_Tests, Number_Of_Pattern_Sizes, Increment;@\\
\mbox{}\verb@double Base_Time = 0.0;@\\
\mbox{}\verb@@\hbox{$\langle$Define algorithm enumeration type, names, and selector function (C++) {\footnotesize 49b}$\rangle$}\verb@@\\
\mbox{}\verb@@\hbox{$\langle$Define run procedure (C++ forward) {\footnotesize 55}$\rangle$}\verb@@\\
\mbox{}\verb@@\\
\mbox{}\verb@int main()@\\
\mbox{}\verb@{ @\\
\mbox{}\verb@  int j;@\\
\mbox{}\verb@  @\hbox{$\langle$Read test parameters (C++) {\footnotesize 53a}$\rangle$}\verb@@\\
\mbox{}\verb@  @\hbox{$\langle$Read character sequence from file (C++) {\footnotesize 54a}$\rangle$}\verb@@\\
\mbox{}\verb@  @\hbox{$\langle$Read dictionary from file, placing words of size j in dictionary[j] {\footnotesize 53b}$\rangle$}\verb@@\\
\mbox{}\verb@  for (j = 0; j < Number_Of_Pattern_Sizes; ++j) {@\\
\mbox{}\verb@    @\hbox{$\langle$Trim dictionary[Pattern\_Size[j]] to have at most Number\_Of\_Tests words {\footnotesize 53c}$\rangle$}\verb@@\\
\mbox{}\verb@    Increment = (S1.size() - Pattern_Size[j]) / Number_Of_Tests;@\\
\mbox{}\verb@    @\hbox{$\langle$Output header (C++) {\footnotesize 54b}$\rangle$}\verb@@\\
\mbox{}\verb@    cerr << Pattern_Size[j] << " " << flush;@\\
\mbox{}\verb@    @\hbox{$\langle$Run and time tests searching for selected subsequences (C++) {\footnotesize 56b}$\rangle$}\verb@@\\
\mbox{}\verb@  }@\\
\mbox{}\verb@  cerr << endl;@\\
\mbox{}\verb@}@\\
\mbox{}\verb@@\end{list}
\vspace{-2ex}
\end{flushleft}
The following test procedure is programmed using forward iterator
operations only, so that it can be applied to a non-random access
container (e.g., list), assuming the designated algorithm works
with forward iterators.

\begin{flushleft} \small \label{__n:gensearch.w:part102}
$\langle$Define run procedure (C++ forward) {\footnotesize 55}$\rangle\equiv$
\vspace{-1ex}
\begin{list}{}{} \item
\mbox{}\verb@template <typename Container>@\\
\mbox{}\verb@void Run(int k, const Container& S1, @\\
\mbox{}\verb@         const vector<Container>& dictionary, int Pattern_Size)@\\
\mbox{}\verb@{@\\
\mbox{}\verb@  typename Container::const_iterator P;@\\
\mbox{}\verb@  int F = 0, d, K;@\\
\mbox{}\verb@  double Start_Time, Finish_Time, Time_Taken;@\\
\mbox{}\verb@  long Total_Search = 0;@\\
\mbox{}\verb@  Start_Time = clock();@\\
\mbox{}\verb@  Container S2;@\\
\mbox{}\verb@  for (K = 1; K <= Number_Of_Tests; ++K) {@\\
\mbox{}\verb@    typename Container::const_iterator u = S1.begin();@\\
\mbox{}\verb@    advance(u, F);@\\
\mbox{}\verb@    S2.erase(S2.begin(), S2.end());@\\
\mbox{}\verb@    for (int I = 0; I < Pattern_Size; ++I)@\\
\mbox{}\verb@      S2.push_back(*u++);@\\
\mbox{}\verb@    F += Increment;@\\
\mbox{}\verb@    @\hbox{$\langle$Run algorithm and record search distance {\footnotesize 56a}$\rangle$}\verb@@\\
\mbox{}\verb@  }@\\
\mbox{}\verb@  for (K = 0; K < dictionary.size(); ++K) {@\\
\mbox{}\verb@    S2 = dictionary[K];@\\
\mbox{}\verb@    @\hbox{$\langle$Run algorithm and record search distance {\footnotesize 56a}$\rangle$}\verb@@\\
\mbox{}\verb@  }@\\
\mbox{}\verb@  Finish_Time = clock();@\\
\mbox{}\verb@  @\hbox{$\langle$Output statistics (C++) {\footnotesize 56c}$\rangle$}\verb@@\\
\mbox{}\verb@}@\\
\mbox{}\verb@@\end{list}
\vspace{-1ex}
\footnotesize\addtolength{\baselineskip}{-1ex}
\begin{list}{}{\setlength{\itemsep}{-\parsep}\setlength{\itemindent}{-\leftmargin}}
\item Used in parts 54f, 57b, 65b.
\end{list}
\end{flushleft}
\begin{flushleft} \small
\begin{minipage}{\linewidth} \label{__n:gensearch.w:part103}
$\langle$Run algorithm and record search distance {\footnotesize 56a}$\rangle\equiv$
\vspace{-1ex}
\begin{list}{}{} \item
\mbox{}\verb@Algorithm(k, S1, S2, P);@\\
\mbox{}\verb@d = 0;@\\
\mbox{}\verb@distance(S1.begin(), P, d);@\\
\mbox{}\verb@Total_Search += d + Pattern_Size;@\\
\mbox{}\verb@@\end{list}
\vspace{-1ex}
\footnotesize\addtolength{\baselineskip}{-1ex}
\begin{list}{}{\setlength{\itemsep}{-\parsep}\setlength{\itemindent}{-\leftmargin}}
\item Used in parts 55, 62b.
\end{list}
\end{minipage}\\[4ex]
\end{flushleft}
\begin{flushleft} \small
\begin{minipage}{\linewidth} \label{__n:gensearch.w:part104}
$\langle$Run and time tests searching for selected subsequences (C++) {\footnotesize 56b}$\rangle\equiv$
\vspace{-1ex}
\begin{list}{}{} \item
\mbox{}\verb@Base_Time = 0.0;@\\
\mbox{}\verb@for (int k = 0; k < number_of_algorithms; ++k) {@\\
\mbox{}\verb@  if (k != 0) @\\
\mbox{}\verb@    cout << "Timing " << algorithm_names[k] << ":" << endl;@\\
\mbox{}\verb@  Run(k, S1, dictionary[Pattern_Size[j]], Pattern_Size[j]);@\\
\mbox{}\verb@}@\\
\mbox{}\verb@cout << endl;@\\
\mbox{}\verb@@\end{list}
\vspace{-1ex}
\footnotesize\addtolength{\baselineskip}{-1ex}
\begin{list}{}{\setlength{\itemsep}{-\parsep}\setlength{\itemindent}{-\leftmargin}}
\item Used in parts 54f, 57b, 65b.
\end{list}
\end{minipage}\\[4ex]
\end{flushleft}
\begin{flushleft} \small
\begin{minipage}{\linewidth} \label{__n:gensearch.w:part105}
$\langle$Output statistics (C++) {\footnotesize 56c}$\rangle\equiv$
\vspace{-1ex}
\begin{list}{}{} \item
\mbox{}\verb@Time_Taken = (Finish_Time - Start_Time)/CLOCKS_PER_SEC - Base_Time;@\\
\mbox{}\verb@if (k == 0) @\\
\mbox{}\verb@  Base_Time = Time_Taken;  @\\
\mbox{}\verb@else {@\\
\mbox{}\verb@  cout << "Total search length: " << Total_Search << " elements" << endl;@\\
\mbox{}\verb@  cout << "Time: " << Time_Taken << " seconds." << endl;@\\
\mbox{}\verb@  double Speed = Time_Taken == 0.0 ? 0.0 : @\\
\mbox{}\verb@    static_cast<double>(Total_Search) / 1000000 / Time_Taken;@\\
\mbox{}\verb@  cout << "Speed: " << Speed << " elements/microsecond." << endl << endl;@\\
\mbox{}\verb@}@\\
\mbox{}\verb@@\end{list}
\vspace{-1ex}
\footnotesize\addtolength{\baselineskip}{-1ex}
\begin{list}{}{\setlength{\itemsep}{-\parsep}\setlength{\itemindent}{-\leftmargin}}
\item Used in part 55.
\end{list}
\end{minipage}\\[4ex]
\end{flushleft}
\subsection{Timed Tests (Large Alphabet)}

Again, the following program for timing searches conducts the same
searches as in the Ada version, plus searches for words of the requested
pattern size selected from a given dictionary.

\begin{flushleft} \small \label{__n:gensearch.w:part106}
$\langle$Define algorithm enumeration type, names, and selector function (C++ large alphabet) {\footnotesize 57a}$\rangle\equiv$
\vspace{-1ex}
\begin{list}{}{} \item
\mbox{}\verb@enum algorithm_enumeration {@\\
\mbox{}\verb@     Dummy, SF, L, HAL, NHAL@\\
\mbox{}\verb@};@\\
\mbox{}\verb@const char* algorithm_names[] = {@\\
\mbox{}\verb@     "selection code", "SF", "L", "HAL", "NHAL"@\\
\mbox{}\verb@};@\\
\mbox{}\verb@@\\
\mbox{}\verb@const int number_of_algorithms = 5;@\\
\mbox{}\verb@@\\
\mbox{}\verb@template <typename Container, typename Container__const_iterator>@\\
\mbox{}\verb@inline void@\\
\mbox{}\verb@   Algorithm(int k, const Container& x, const Container& y, @\\
\mbox{}\verb@             Container__const_iterator& result)@\\
\mbox{}\verb@{@\\
\mbox{}\verb@  switch (algorithm_enumeration(k)) {@\\
\mbox{}\verb@  case Dummy: @\\
\mbox{}\verb@     // does nothing, used for timing overhead of test loop@\\
\mbox{}\verb@     result = x.begin(); return; @\\
\mbox{}\verb@  case SF: @\\
\mbox{}\verb@     result = stl_search(x.begin(), x.end(), y.begin(), y.end()); return;@\\
\mbox{}\verb@  case L: @\\
\mbox{}\verb@     result =  __search_L(x.begin(), x.end(), y.begin(), y.end() ); return;@\\
\mbox{}\verb@  case HAL: @\\
\mbox{}\verb@     result = search(x.begin(), x.end(), y.begin(), y.end() ); return;@\\
\mbox{}\verb@  case NHAL: @\\
\mbox{}\verb@     result = search_no_hashing(x.begin(), x.end(), y.begin(), y.end() ); return;@\\
\mbox{}\verb@  }@\\
\mbox{}\verb@  result = x.begin(); return;@\\
\mbox{}\verb@}@\\
\mbox{}\verb@@\end{list}
\vspace{-1ex}
\footnotesize\addtolength{\baselineskip}{-1ex}
\begin{list}{}{\setlength{\itemsep}{-\parsep}\setlength{\itemindent}{-\leftmargin}}
\item Used in part 57b.
\end{list}
\end{flushleft}
\begin{flushleft} \small \label{__n:gensearch.w:part107}
\verb@"experimental_search.cpp"@ {\footnotesize 57b }$\equiv$
\vspace{-1ex}
\begin{list}{}{} \item
\mbox{}\verb@@\hbox{$\langle$Include algorithms header with existing search function renamed {\footnotesize 48}$\rangle$}\verb@@\\
\mbox{}\verb@#include "new_search.h"@\\
\mbox{}\verb@#include "experimental_search.h"@\\
\mbox{}\verb@#include <iterator>@\\
\mbox{}\verb@#include <deque>@\\
\mbox{}\verb@#include <vector>@\\
\mbox{}\verb@#include <map>@\\
\mbox{}\verb@#include <iostream>@\\
\mbox{}\verb@#include <fstream>@\\
\mbox{}\verb@#include <ctime>@\\
\mbox{}\verb@using namespace std;@\\
\mbox{}\verb@@\\
\mbox{}\verb@typedef unsigned short data;@\\
\mbox{}\verb@typedef vector<data> sequence;@\\
\mbox{}\verb@@\\
\mbox{}\verb@sequence S1;@\\
\mbox{}\verb@@\\
\mbox{}\verb@int Base_Line, Number_Of_Tests, Number_Of_Pattern_Sizes, Increment;@\\
\mbox{}\verb@double Base_Time = 0.0;@\\
\mbox{}\verb@@\hbox{$\langle$Define algorithm enumeration type, names, and selector function (C++ large alphabet) {\footnotesize 57a}$\rangle$}\verb@@\\
\mbox{}\verb@@\hbox{$\langle$Define run procedure (C++ forward) {\footnotesize 55}$\rangle$}\verb@@\\
\mbox{}\verb@@\hbox{$\langle$Define RandomNumberGenerator class {\footnotesize 58a}$\rangle$}\verb@@\\
\mbox{}\verb@@\\
\mbox{}\verb@int main()@\\
\mbox{}\verb@{ @\\
\mbox{}\verb@  int j;@\\
\mbox{}\verb@  @\hbox{$\langle$Read test parameters (C++) {\footnotesize 53a}$\rangle$}\verb@@\\
\mbox{}\verb@  @\hbox{$\langle$Generate data sequence {\footnotesize 58b}$\rangle$}\verb@@\\
\mbox{}\verb@  @\hbox{$\langle$Generate dictionary {\footnotesize 59a}$\rangle$}\verb@@\\
\mbox{}\verb@  for (j = 0; j < Number_Of_Pattern_Sizes; ++j) {@\\
\mbox{}\verb@    Increment = (S1.size() - Pattern_Size[j]) / Number_Of_Tests;@\\
\mbox{}\verb@    @\hbox{$\langle$Output header (C++) {\footnotesize 54b}$\rangle$}\verb@@\\
\mbox{}\verb@    cerr << Pattern_Size[j] << " " << flush;@\\
\mbox{}\verb@    @\hbox{$\langle$Run and time tests searching for selected subsequences (C++) {\footnotesize 56b}$\rangle$}\verb@@\\
\mbox{}\verb@  }@\\
\mbox{}\verb@  cerr << endl;@\\
\mbox{}\verb@}@\\
\mbox{}\verb@@\end{list}
\vspace{-2ex}
\end{flushleft}
\begin{flushleft} \small
\begin{minipage}{\linewidth} \label{__n:gensearch.w:part108}
$\langle$Define RandomNumberGenerator class {\footnotesize 58a}$\rangle\equiv$
\vspace{-1ex}
\begin{list}{}{} \item
\mbox{}\verb@int random(int max_value) { return rand() % max_value; }@\\
\mbox{}\verb@@\\
\mbox{}\verb@template <int MAX_VALUE> struct RandomNumberGenerator {@\\
\mbox{}\verb@  int operator() () { return random(MAX_VALUE); }@\\
\mbox{}\verb@};@\\
\mbox{}\verb@@\end{list}
\vspace{-1ex}
\footnotesize\addtolength{\baselineskip}{-1ex}
\begin{list}{}{\setlength{\itemsep}{-\parsep}\setlength{\itemindent}{-\leftmargin}}
\item Used in part 57b.
\end{list}
\end{minipage}\\[4ex]
\end{flushleft}
\begin{flushleft} \small
\begin{minipage}{\linewidth} \label{__n:gensearch.w:part109}
$\langle$Generate data sequence {\footnotesize 58b}$\rangle\equiv$
\vspace{-1ex}
\begin{list}{}{} \item
\mbox{}\verb@generate_n(back_inserter(S1), 100000, RandomNumberGenerator<65535>());@\\
\mbox{}\verb@@\end{list}
\vspace{-1ex}
\footnotesize\addtolength{\baselineskip}{-1ex}
\begin{list}{}{\setlength{\itemsep}{-\parsep}\setlength{\itemindent}{-\leftmargin}}
\item Used in part 57b.
\end{list}
\end{minipage}\\[4ex]
\end{flushleft}
\begin{flushleft} \small
\begin{minipage}{\linewidth} \label{__n:gensearch.w:part110}
$\langle$Generate dictionary {\footnotesize 59a}$\rangle\equiv$
\vspace{-1ex}
\begin{list}{}{} \item
\mbox{}\verb@typedef map<int, vector<sequence >, less<int> > map_type;@\\
\mbox{}\verb@map_type dictionary;@\\
\mbox{}\verb@@\\
\mbox{}\verb@for(int i = 0; i < Number_Of_Pattern_Sizes; ++i) {@\\
\mbox{}\verb@  int pattern_size = Pattern_Size[i];@\\
\mbox{}\verb@@\\
\mbox{}\verb@  for(int j = 0; j < Number_Of_Tests; ++j) {@\\
\mbox{}\verb@    int position = random(S1.size() - pattern_size);@\\
\mbox{}\verb@    dictionary[pattern_size].push_back( sequence() );@\\
\mbox{}\verb@    copy(S1.begin() + position, S1.begin() + position + pattern_size, @\\
\mbox{}\verb@         back_inserter( dictionary[pattern_size].back() ) ) ;@\\
\mbox{}\verb@  }@\\
\mbox{}\verb@}@\\
\mbox{}\verb@@\end{list}
\vspace{-1ex}
\footnotesize\addtolength{\baselineskip}{-1ex}
\begin{list}{}{\setlength{\itemsep}{-\parsep}\setlength{\itemindent}{-\leftmargin}}
\item Used in part 57b.
\end{list}
\end{minipage}\\[4ex]
\end{flushleft}
\subsection{Counted Tests}

The following program runs the same searches as in the timing program,
but in addition to times it records and reports counts of many
different types of operations, including equality comparisons on data,
iterator operations, and ``distance operations,'' which are arithmetic
operations on integer results of iterator subtractions.  These counts
are obtained without modifying the source code of the algorithms at
all, by specializing their type parameters with classes whose
operations have counters built into them.

\begin{flushleft} \small \label{__n:gensearch.w:part111}
\verb@"count_long_search.cpp"@ {\footnotesize 59b }$\equiv$
\vspace{-1ex}
\begin{list}{}{} \item
\mbox{}\verb@@\hbox{$\langle$Include algorithms header with existing search function renamed {\footnotesize 48}$\rangle$}\verb@@\\
\mbox{}\verb@#define LIST_TEST@\\
\mbox{}\verb@#include "new_search.h"@\\
\mbox{}\verb@#include "hume.hh"@\\
\mbox{}\verb@#include <iterator>@\\
\mbox{}\verb@#include <vector>@\\
\mbox{}\verb@#include <map>@\\
\mbox{}\verb@#include <iostream>@\\
\mbox{}\verb@#include <fstream>@\\
\mbox{}\verb@#include <ctime>@\\
\mbox{}\verb@#include <string>@\\
\mbox{}\verb@using namespace std;@\\
\mbox{}\verb@@\\
\mbox{}\verb@@\hbox{$\langle$Define types needed for counting operations {\footnotesize 60a}$\rangle$}\verb@@\\
\mbox{}\verb@typedef vector<data> sequence;@\\
\mbox{}\verb@sequence S1;@\\
\mbox{}\verb@int Base_Line;@\\
\mbox{}\verb@unsigned int Number_Of_Tests, Number_Of_Pattern_Sizes, Increment;@\\
\mbox{}\verb@double Base_Time = 0.0;@\\
\mbox{}\verb@@\hbox{$\langle$Define algorithm enumeration type, names, and selector function (C++ counter) {\footnotesize 60b}$\rangle$}\verb@@\\
\mbox{}\verb@@\hbox{$\langle$Define run procedure (C++ counter) {\footnotesize 62b}$\rangle$}\verb@@\\
\mbox{}\verb@@\\
\mbox{}\verb@int main()@\\
\mbox{}\verb@{ @\\
\mbox{}\verb@  recorder<> stats[number_of_algorithms];@\\
\mbox{}\verb@  int j;@\\
\mbox{}\verb@  @\hbox{$\langle$Read test parameters (C++) {\footnotesize 53a}$\rangle$}\verb@@\\
\mbox{}\verb@  @\hbox{$\langle$Read character sequence from file (C++) {\footnotesize 54a}$\rangle$}\verb@@\\
\mbox{}\verb@  @\hbox{$\langle$Read dictionary from file, placing words of size j in dictionary[j] {\footnotesize 53b}$\rangle$}\verb@@\\
\mbox{}\verb@  for (j = 0; j < Number_Of_Pattern_Sizes; ++j) {@\\
\mbox{}\verb@    @\hbox{$\langle$Trim dictionary[Pattern\_Size[j]] to have at most Number\_Of\_Tests words {\footnotesize 53c}$\rangle$}\verb@@\\
\mbox{}\verb@    Increment = (S1.size() - Pattern_Size[j]) / Number_Of_Tests;@\\
\mbox{}\verb@    @\hbox{$\langle$Output header (C++) {\footnotesize 54b}$\rangle$}\verb@@\\
\mbox{}\verb@    cerr << Pattern_Size[j] << " " << flush;@\\
\mbox{}\verb@    @\hbox{$\langle$Run and time tests searching for selected subsequences (C++ counter) {\footnotesize 62a}$\rangle$}\verb@@\\
\mbox{}\verb@  }@\\
\mbox{}\verb@  cerr << endl;@\\
\mbox{}\verb@}@\\
\mbox{}\verb@@\end{list}
\vspace{-2ex}
\end{flushleft}
\begin{flushleft} \small \label{__n:gensearch.w:part112}
$\langle$Define types needed for counting operations {\footnotesize 60a}$\rangle\equiv$
\vspace{-1ex}
\begin{list}{}{} \item
\mbox{}\verb@#include <utility>@\\
\mbox{}\verb@using namespace std;@\\
\mbox{}\verb@namespace std { namespace rel_ops {} };@\\
\mbox{}\verb@using namespace std::rel_ops;@\\
\mbox{}\verb@#define DEFAULT_COUNTER_TYPE double@\\
\mbox{}\verb@#include "counters.h"@\\
\mbox{}\verb@@\\
\mbox{}\verb@typedef unsigned char basedata;@\\
\mbox{}\verb@@\\
\mbox{}\verb@typedef value_counter<basedata> data;@\\
\mbox{}\verb@typedef iterator_counter<vector<data>::const_iterator> citer;@\\
\mbox{}\verb@@\\
\mbox{}\verb@struct search_trait_for_counting {@\\
\mbox{}\verb@  enum {hash_range_max = 256};@\\
\mbox{}\verb@  enum {suffix_size = 1};@\\
\mbox{}\verb@  inline static unsigned int hash(const citer& i) {return (*i).base();}@\\
\mbox{}\verb@};@\\
\mbox{}\verb@@\\
\mbox{}\verb@@\end{list}
\vspace{-1ex}
\footnotesize\addtolength{\baselineskip}{-1ex}
\begin{list}{}{\setlength{\itemsep}{-\parsep}\setlength{\itemindent}{-\leftmargin}}
\item Used in part 59b.
\end{list}
\end{flushleft}
\begin{flushleft} \small \label{__n:gensearch.w:part113}
$\langle$Define algorithm enumeration type, names, and selector function (C++ counter) {\footnotesize 60b}$\rangle\equiv$
\vspace{-1ex}
\begin{list}{}{} \item
\mbox{}\verb@enum algorithm_enumeration {@\\
\mbox{}\verb@     Dummy, STL_search, L, HAL, ABM, TBM@\\
\mbox{}\verb@};@\\
\mbox{}\verb@const char* algorithm_names[] = {@\\
\mbox{}\verb@     "selection code", "SF", "L", "HAL", "ABM", "TBM"@\\
\mbox{}\verb@};@\\
\mbox{}\verb@#ifndef LIST_TEST@\\
\mbox{}\verb@const int number_of_algorithms = 6;@\\
\mbox{}\verb@#else@\\
\mbox{}\verb@const int number_of_algorithms = 3;@\\
\mbox{}\verb@#endif@\\
\mbox{}\verb@@\\
\mbox{}\verb@const char textFileName[] = "long.txt";@\\
\mbox{}\verb@const char wordFileName[] = "words.txt";@\\
\mbox{}\verb@@\\
\mbox{}\verb@template <typename Container, typename Container__const_iterator>@\\
\mbox{}\verb@void Algorithm(int k, const Container& x, const Container& y,@\\
\mbox{}\verb@               Container__const_iterator& result)@\\
\mbox{}\verb@{@\\
\mbox{}\verb@  switch (algorithm_enumeration(k)) {@\\
\mbox{}\verb@  case Dummy: @\\
\mbox{}\verb@     // does nothing, used for timing overhead of test loop@\\
\mbox{}\verb@     result = x.begin();  @\\
\mbox{}\verb@     return;@\\
\mbox{}\verb@  case STL_search: @\\
\mbox{}\verb@     result = stl_search(citer(x.begin()), citer(x.end()), @\\
\mbox{}\verb@                         citer(y.begin()), citer(y.end())).base();@\\
\mbox{}\verb@     return;@\\
\mbox{}\verb@  case L: @\\
\mbox{}\verb@     result = __search_L(citer(x.begin()), citer(x.end()), @\\
\mbox{}\verb@                         citer(y.begin()), citer(y.end())).base();@\\
\mbox{}\verb@     return;@\\
\mbox{}\verb@@\\
\mbox{}\verb@#ifndef LIST_TEST@\\
\mbox{}\verb@  case HAL: @\\
\mbox{}\verb@     result = search_hashed(citer(x.begin()), citer(x.end()), @\\
\mbox{}\verb@                            citer(y.begin()), citer(y.end()), @\\
\mbox{}\verb@                            static_cast<search_trait_for_counting*>(0)).base();@\\
\mbox{}\verb@     return;@\\
\mbox{}\verb@  case ABM:@\\
\mbox{}\verb@     fbmprep((const basedata*)y.begin(), y.size());@\\
\mbox{}\verb@     result = (typename Container::const_iterator)@\\
\mbox{}\verb@                fbmexec_cnt((const basedata*)x.begin(), x.size());@\\
\mbox{}\verb@//     data::accesses += ::pat.accs;@\\
\mbox{}\verb@//     data::equal_comparisons += ::pat.cmps;@\\
\mbox{}\verb@     return;@\\
\mbox{}\verb@  case TBM:@\\
\mbox{}\verb@     humprep((const basedata*)y.begin(), y.size());@\\
\mbox{}\verb@     result = (typename Container::const_iterator)@\\
\mbox{}\verb@                humexec_cnt((const basedata*)x.begin(), x.size());@\\
\mbox{}\verb@//     data::accesses += ::pat.accs;@\\
\mbox{}\verb@//     data::equal_comparisons += ::pat.cmps;@\\
\mbox{}\verb@     result = result;@\\
\mbox{}\verb@     return;@\\
\mbox{}\verb@#endif@\\
\mbox{}\verb@  }@\\
\mbox{}\verb@  result = x.begin();@\\
\mbox{}\verb@  return;@\\
\mbox{}\verb@}@\\
\mbox{}\verb@@\end{list}
\vspace{-1ex}
\footnotesize\addtolength{\baselineskip}{-1ex}
\begin{list}{}{\setlength{\itemsep}{-\parsep}\setlength{\itemindent}{-\leftmargin}}
\item Used in part 59b.
\end{list}
\end{flushleft}
\begin{flushleft} \small
\begin{minipage}{\linewidth} \label{__n:gensearch.w:part114}
$\langle$Run and time tests searching for selected subsequences (C++ counter) {\footnotesize 62a}$\rangle\equiv$
\vspace{-1ex}
\begin{list}{}{} \item
\mbox{}\verb@Base_Time = 0.0;@\\
\mbox{}\verb@for (int k = 0; k < number_of_algorithms; ++k) {@\\
\mbox{}\verb@  if (k != 0) @\\
\mbox{}\verb@    cout << "Timing " << algorithm_names[k] << ":" << endl;@\\
\mbox{}\verb@  Run(k, S1, dictionary[Pattern_Size[j]], Pattern_Size[j], stats);@\\
\mbox{}\verb@}@\\
\mbox{}\verb@cout << endl;@\\
\mbox{}\verb@@\end{list}
\vspace{-1ex}
\footnotesize\addtolength{\baselineskip}{-1ex}
\begin{list}{}{\setlength{\itemsep}{-\parsep}\setlength{\itemindent}{-\leftmargin}}
\item Used in part 59b.
\end{list}
\end{minipage}\\[4ex]
\end{flushleft}
\begin{flushleft} \small \label{__n:gensearch.w:part115}
$\langle$Define run procedure (C++ counter) {\footnotesize 62b}$\rangle\equiv$
\vspace{-1ex}
\begin{list}{}{} \item
\mbox{}\verb@template <typename Container>@\\
\mbox{}\verb@void Run(int k, const Container& S1, @\\
\mbox{}\verb@         const vector<Container>& dictionary, int Pattern_Size, recorder<>* stats)@\\
\mbox{}\verb@{@\\
\mbox{}\verb@  typename Container::const_iterator P;@\\
\mbox{}\verb@  int F = 0, K, d;@\\
\mbox{}\verb@  double Start_Time, Finish_Time, Time_Taken;@\\
\mbox{}\verb@  long Total_Search = 0;@\\
\mbox{}\verb@  stats[k].reset();@\\
\mbox{}\verb@  Start_Time = clock();@\\
\mbox{}\verb@  Container S2;@\\
\mbox{}\verb@  for (K = 1; K <= Number_Of_Tests; ++K) {@\\
\mbox{}\verb@    typename Container::const_iterator u = S1.begin();@\\
\mbox{}\verb@    advance(u, F);@\\
\mbox{}\verb@    S2.erase(S2.begin(), S2.end());@\\
\mbox{}\verb@    for (int I = 0; I < Pattern_Size; ++I)@\\
\mbox{}\verb@      S2.push_back(*u++);@\\
\mbox{}\verb@    F += Increment;@\\
\mbox{}\verb@    @\hbox{$\langle$Run algorithm and record search distance {\footnotesize 56a}$\rangle$}\verb@@\\
\mbox{}\verb@  }@\\
\mbox{}\verb@  for (K = 0; K < dictionary.size(); ++K) {@\\
\mbox{}\verb@    S2 = dictionary[K];@\\
\mbox{}\verb@    @\hbox{$\langle$Run algorithm and record search distance {\footnotesize 56a}$\rangle$}\verb@@\\
\mbox{}\verb@  }@\\
\mbox{}\verb@  stats[k].record();@\\
\mbox{}\verb@  Finish_Time = clock();@\\
\mbox{}\verb@  @\hbox{$\langle$Output statistics (C++ counter) {\footnotesize 63a}$\rangle$}\verb@@\\
\mbox{}\verb@}@\\
\mbox{}\verb@@\end{list}
\vspace{-1ex}
\footnotesize\addtolength{\baselineskip}{-1ex}
\begin{list}{}{\setlength{\itemsep}{-\parsep}\setlength{\itemindent}{-\leftmargin}}
\item Used in part 59b.
\end{list}
\end{flushleft}
\begin{flushleft} \small
\begin{minipage}{\linewidth} \label{__n:gensearch.w:part116}
$\langle$Output statistics (C++ counter) {\footnotesize 63a}$\rangle\equiv$
\vspace{-1ex}
\begin{list}{}{} \item
\mbox{}\verb@Time_Taken = (Finish_Time - Start_Time)/CLOCKS_PER_SEC - Base_Time;@\\
\mbox{}\verb@if (k == 0) @\\
\mbox{}\verb@  Base_Time = Time_Taken;  @\\
\mbox{}\verb@else {@\\
\mbox{}\verb@  operator_type group0[] = { LESS_THAN, EQUALITY };@\\
\mbox{}\verb@  operator_type group1[] = { DEFAULT_CTOR, COPY_CTOR, OTHER_CTOR, ASSIGNMENT };@\\
\mbox{}\verb@  operator_type group2[] = { @\\
\mbox{}\verb@    CONVERSION, POST_INCREMENT, PRE_INCREMENT, @\\
\mbox{}\verb@    POST_DECREMENT, PRE_DECREMENT, @\\
\mbox{}\verb@    UNARY_PLUS, UNARY_MINUS, PLUS, MINUS, TIMES, DIVIDE, MODULO,@\\
\mbox{}\verb@    LEFT_SHIFT, RIGHT_SHIFT,@\\
\mbox{}\verb@    PLUS_ASSIGN, MINUS_ASSIGN, @\\
\mbox{}\verb@    TIMES_ASSIGN, DIVIDE_ASSIGN, MODULO_ASSIGN,@\\
\mbox{}\verb@    LEFT_SHIFT_ASSIGN, RIGHT_SHIFT_ASSIGN,@\\
\mbox{}\verb@    DEREFERENCE, MEMBER, SUBSCRIPT, FUNCTION_CALL,@\\
\mbox{}\verb@    NEGATE, AND, AND_ASSIGN, OR, OR_ASSIGN, XOR, XOR_ASSIGN,@\\
\mbox{}\verb@  };@\\
\mbox{}\verb@  scale_recorder_visitor<group_recorder_visitor<> > visitor(.001);@\\
\mbox{}\verb@  visitor.add("Comps", group0, group0 + sizeof(group0) / sizeof(group0[0]));@\\
\mbox{}\verb@  visitor.add("Asmts", group1, group1 + sizeof(group1) / sizeof(group1[0]));@\\
\mbox{}\verb@  visitor.add("Other", group2, group2 + sizeof(group2) / sizeof(group2[0]));@\\
\mbox{}\verb@  operator_type group3[] = { BASE, DTOR, GENERATION, };@\\
\mbox{}\verb@  visitor.hide(group3, group3 + sizeof(group3) / sizeof(group3[0]));@\\
\mbox{}\verb@@\\
\mbox{}\verb@  stats[k].report(cout, visitor);@\\
\mbox{}\verb@  cout << "Total search length: " << Total_Search << " elements" << endl;@\\
\mbox{}\verb@  cout << "Time: " << Time_Taken << " seconds." << endl;@\\
\mbox{}\verb@  double Speed = Time_Taken == 0.0 ? 0.0 : @\\
\mbox{}\verb@    (double)Total_Search / 1000000 / Time_Taken;@\\
\mbox{}\verb@  cout << "Speed: " << Speed << " elements/microsecond." << endl << endl;@\\
\mbox{}\verb@}@\\
\mbox{}\verb@@\end{list}
\vspace{-1ex}
\footnotesize\addtolength{\baselineskip}{-1ex}
\begin{list}{}{\setlength{\itemsep}{-\parsep}\setlength{\itemindent}{-\leftmargin}}
\item Used in part 62b.
\end{list}
\end{minipage}\\[4ex]
\end{flushleft}
\subsection{Application to Matching Sequences of Words}

\subsubsection{Large Tests}

This \Cpp\ program specializes the generic search functions to work
with sequences of words (character strings).  It reads a text file
in as a sequence of words, and for each of a specified
set of pattern sizes, it searches for word sequences of that size selected
from evenly spaced positions in the target sequence.  These searches
are the counterpart of the first kind of searches done in the previous
programs on character sequences; the dictionary word searches of
the previous programs are omitted here.

\begin{flushleft} \small \label{__n:gensearch.w:part117}
\verb@"test_word_search.cpp"@ {\footnotesize 63b }$\equiv$
\vspace{-1ex}
\begin{list}{}{} \item
\mbox{}\verb@@\hbox{$\langle$Include algorithms header with existing search function renamed {\footnotesize 48}$\rangle$}\verb@@\\
\mbox{}\verb@#include "new_search.h"@\\
\mbox{}\verb@#include <iterator>@\\
\mbox{}\verb@#include <vector>@\\
\mbox{}\verb@#include <map>@\\
\mbox{}\verb@#include <iostream>@\\
\mbox{}\verb@#include <fstream>@\\
\mbox{}\verb@#include <string>@\\
\mbox{}\verb@using namespace std;@\\
\mbox{}\verb@@\\
\mbox{}\verb@typedef string data;@\\
\mbox{}\verb@typedef vector<data> sequence;@\\
\mbox{}\verb@@\\
\mbox{}\verb@sequence S1, S2;@\\
\mbox{}\verb@int Base_Line, Number_Of_Tests, Number_Of_Pattern_Sizes, Increment;@\\
\mbox{}\verb@@\hbox{$\langle$Define search trait for word searches {\footnotesize 64a}$\rangle$}\verb@@\\
\mbox{}\verb@@\hbox{$\langle$Define algorithm enumeration type, names, and selector function (C++ word) {\footnotesize 64b}$\rangle$}\verb@@\\
\mbox{}\verb@@\hbox{$\langle$Define Report procedure (C++) {\footnotesize 51d}$\rangle$}\verb@@\\
\mbox{}\verb@int main() @\\
\mbox{}\verb@{  @\\
\mbox{}\verb@  int F, K, j;@\\
\mbox{}\verb@  @\hbox{$\langle$Read test parameters (C++) {\footnotesize 53a}$\rangle$}\verb@@\\
\mbox{}\verb@  typedef map<int, vector<sequence >, less<int> > map_type;@\\
\mbox{}\verb@  map_type dictionary;@\\
\mbox{}\verb@  @\hbox{$\langle$Read word sequence from file (C++) {\footnotesize 65a}$\rangle$}\verb@@\\
\mbox{}\verb@  cout << S1.size() << " words read." << endl;@\\
\mbox{}\verb@  const char* separator = " ";@\\
\mbox{}\verb@  for (j = 0; j < Number_Of_Pattern_Sizes; ++j) {@\\
\mbox{}\verb@    Increment = (S1.size() - Pattern_Size[j]) / Number_Of_Tests;@\\
\mbox{}\verb@    @\hbox{$\langle$Output header (C++) {\footnotesize 54b}$\rangle$}\verb@@\\
\mbox{}\verb@    @\hbox{$\langle$Run tests searching for selected subsequences (C++) {\footnotesize 54c}$\rangle$}\verb@@\\
\mbox{}\verb@  }@\\
\mbox{}\verb@}@\\
\mbox{}\verb@@\end{list}
\vspace{-2ex}
\end{flushleft}
For a hash function the program uses a mapping of a word to its first
character.  Although this would not be a good hash function in hashed
associative table lookup, it works satisfactorily here because there
is less need for uniformity of hash value distribution.

\begin{flushleft} \small
\begin{minipage}{\linewidth} \label{__n:gensearch.w:part118}
$\langle$Define search trait for word searches {\footnotesize 64a}$\rangle\equiv$
\vspace{-1ex}
\begin{list}{}{} \item
\mbox{}\verb@struct search_word_trait {@\\
\mbox{}\verb@  typedef vector<string>::const_iterator RAI;@\\
\mbox{}\verb@  enum {hash_range_max = 256};@\\
\mbox{}\verb@  enum {suffix_size = 1};@\\
\mbox{}\verb@  inline static unsigned int hash(RAI i) {@\\
\mbox{}\verb@    return (*i)[0];@\\
\mbox{}\verb@  }@\\
\mbox{}\verb@};@\\
\mbox{}\verb@@\end{list}
\vspace{-1ex}
\footnotesize\addtolength{\baselineskip}{-1ex}
\begin{list}{}{\setlength{\itemsep}{-\parsep}\setlength{\itemindent}{-\leftmargin}}
\item Used in parts 63b, 65b.
\end{list}
\end{minipage}\\[4ex]
\end{flushleft}
\begin{flushleft} \small \label{__n:gensearch.w:part119}
$\langle$Define algorithm enumeration type, names, and selector function (C++ word) {\footnotesize 64b}$\rangle\equiv$
\vspace{-1ex}
\begin{list}{}{} \item
\mbox{}\verb@enum algorithm_enumeration {@\\
\mbox{}\verb@     Dummy, STL_search, L, HAL@\\
\mbox{}\verb@};@\\
\mbox{}\verb@const char* algorithm_names[] = {@\\
\mbox{}\verb@     "selection code", "SF", "L", "HAL"@\\
\mbox{}\verb@};@\\
\mbox{}\verb@#ifndef LIST_TEST@\\
\mbox{}\verb@const int number_of_algorithms = 4;@\\
\mbox{}\verb@#else@\\
\mbox{}\verb@const int number_of_algorithms = 3;@\\
\mbox{}\verb@#endif@\\
\mbox{}\verb@@\\
\mbox{}\verb@template <typename Container, typename Container__const_iterator>@\\
\mbox{}\verb@inline void@\\
\mbox{}\verb@   Algorithm(int k, const Container& x, const Container& y, @\\
\mbox{}\verb@             Container__const_iterator& result)@\\
\mbox{}\verb@{@\\
\mbox{}\verb@  switch (algorithm_enumeration(k)) {@\\
\mbox{}\verb@  case Dummy: @\\
\mbox{}\verb@     // does nothing, used for timing overhead of test loop@\\
\mbox{}\verb@     result = x.begin(); return;@\\
\mbox{}\verb@  case STL_search: @\\
\mbox{}\verb@     result = stl_search(x.begin(), x.end(), y.begin(), y.end()); return;@\\
\mbox{}\verb@  case L: @\\
\mbox{}\verb@     result = __search_L(x.begin(), x.end(), y.begin(), y.end()); return;@\\
\mbox{}\verb@#ifndef LIST_TEST@\\
\mbox{}\verb@  case HAL: @\\
\mbox{}\verb@     result = search_hashed(x.begin(), x.end(), y.begin(), y.end(), @\\
\mbox{}\verb@        (search_word_trait*)0); return;@\\
\mbox{}\verb@#endif@\\
\mbox{}\verb@  }@\\
\mbox{}\verb@  result = x.begin(); return;@\\
\mbox{}\verb@}@\\
\mbox{}\verb@@\end{list}
\vspace{-1ex}
\footnotesize\addtolength{\baselineskip}{-1ex}
\begin{list}{}{\setlength{\itemsep}{-\parsep}\setlength{\itemindent}{-\leftmargin}}
\item Used in parts 63b, 65b.
\end{list}
\end{flushleft}
\begin{flushleft} \small
\begin{minipage}{\linewidth} \label{__n:gensearch.w:part120}
$\langle$Read word sequence from file (C++) {\footnotesize 65a}$\rangle\equiv$
\vspace{-1ex}
\begin{list}{}{} \item
\mbox{}\verb@ifstream ifs("long.txt");@\\
\mbox{}\verb@typedef istream_iterator<string> string_input;@\\
\mbox{}\verb@copy(string_input(ifs), string_input(), back_inserter(S1));@\\
\mbox{}\verb@@\end{list}
\vspace{-1ex}
\footnotesize\addtolength{\baselineskip}{-1ex}
\begin{list}{}{\setlength{\itemsep}{-\parsep}\setlength{\itemindent}{-\leftmargin}}
\item Used in parts 63b, 65b.
\end{list}
\end{minipage}\\[4ex]
\end{flushleft}
\subsubsection{Timed Tests}

We also omit the dictionary searches in the following program which
times searches for selected subsequences, in this case by defining a
map from ints to empty dictionaries (in order to reuse some of the previous
code).

\begin{flushleft} \small \label{__n:gensearch.w:part121}
\verb@"time_word_search.cpp"@ {\footnotesize 65b }$\equiv$
\vspace{-1ex}
\begin{list}{}{} \item
\mbox{}\verb@@\hbox{$\langle$Include algorithms header with existing search function renamed {\footnotesize 48}$\rangle$}\verb@@\\
\mbox{}\verb@#include "new_search.h"@\\
\mbox{}\verb@#include <iterator>@\\
\mbox{}\verb@#include <vector>@\\
\mbox{}\verb@#include <map>@\\
\mbox{}\verb@#include <iostream>@\\
\mbox{}\verb@#include <fstream>@\\
\mbox{}\verb@#include <string>@\\
\mbox{}\verb@#include <ctime>@\\
\mbox{}\verb@//#include <list>@\\
\mbox{}\verb@//#define LIST_TEST@\\
\mbox{}\verb@using namespace std;@\\
\mbox{}\verb@@\\
\mbox{}\verb@typedef string data;@\\
\mbox{}\verb@typedef vector<data> sequence;@\\
\mbox{}\verb@@\\
\mbox{}\verb@sequence S1, S2;@\\
\mbox{}\verb@int Base_Line, Number_Of_Tests, Number_Of_Pattern_Sizes, Increment;@\\
\mbox{}\verb@double Base_Time = 0.0;@\\
\mbox{}\verb@@\hbox{$\langle$Define search trait for word searches {\footnotesize 64a}$\rangle$}\verb@@\\
\mbox{}\verb@@\hbox{$\langle$Define algorithm enumeration type, names, and selector function (C++ word) {\footnotesize 64b}$\rangle$}\verb@@\\
\mbox{}\verb@@\hbox{$\langle$Define run procedure (C++ forward) {\footnotesize 55}$\rangle$}\verb@@\\
\mbox{}\verb@int main() @\\
\mbox{}\verb@{  @\\
\mbox{}\verb@  int j;@\\
\mbox{}\verb@  @\hbox{$\langle$Read test parameters (C++) {\footnotesize 53a}$\rangle$}\verb@@\\
\mbox{}\verb@  typedef map<int, vector<sequence >, less<int> > map_type;@\\
\mbox{}\verb@  map_type dictionary;@\\
\mbox{}\verb@  @\hbox{$\langle$Read word sequence from file (C++) {\footnotesize 65a}$\rangle$}\verb@@\\
\mbox{}\verb@  cout << S1.size() << " words read." << endl;@\\
\mbox{}\verb@  for (j = 0; j < Number_Of_Pattern_Sizes; ++j) {@\\
\mbox{}\verb@    Increment = (S1.size() - Pattern_Size[j]) / Number_Of_Tests;@\\
\mbox{}\verb@    @\hbox{$\langle$Output header (C++) {\footnotesize 54b}$\rangle$}\verb@@\\
\mbox{}\verb@    @\hbox{$\langle$Run and time tests searching for selected subsequences (C++) {\footnotesize 56b}$\rangle$}\verb@@\\
\mbox{}\verb@  }@\\
\mbox{}\verb@}@\\
\mbox{}\verb@@\end{list}
\vspace{-2ex}
\end{flushleft}
\section{Index of Part Names}

{\small\begin{list}{}{\setlength{\itemsep}{-\parsep}\setlength{\itemindent}{-\leftmargin}}
\item $\langle$Accelerated Linear algorithm, no hashing (C++) {\footnotesize 46a}$\rangle$ {\footnotesize Referenced in part 44c.}
\item $\langle$Accelerated Linear algorithm, preliminary version {\footnotesize 5c}$\rangle$ {\footnotesize Referenced in part 29c.}
\item $\langle$Accelerated Linear algorithm {\footnotesize 7b}$\rangle$ {\footnotesize Referenced in part 27c.}
\item $\langle$Additional algorithms {\footnotesize 29c}$\rangle$ {\footnotesize Referenced in parts 31b, 35a, 36e.
}
\item $\langle$Algorithm L, optimized linear pattern search (C++) {\footnotesize 21a}$\rangle$ {\footnotesize Referenced in part 19b.}
\item $\langle$Algorithm L, optimized linear pattern search {\footnotesize 3a}$\rangle$ {\footnotesize Referenced in part 27c.}
\item $\langle$Algorithm subprogram declarations {\footnotesize 27b}$\rangle$ {\footnotesize Referenced in parts 31b, 35a, 36e.
}
\item $\langle$Basic KMP {\footnotesize 2}$\rangle$ {\footnotesize Referenced in part 27c.}
\item $\langle$Bidirectional iterator case {\footnotesize 40b}$\rangle$ {\footnotesize Referenced in part 38b.}
\item $\langle$Check for unexpected end of file (C++) {\footnotesize 51a}$\rangle$ {\footnotesize Referenced in part 50.}
\item $\langle$Check for unexpected end of file {\footnotesize 33a}$\rangle$ {\footnotesize Referenced in part 32.}
\item $\langle$Compute and return position of match {\footnotesize 22b}$\rangle$ {\footnotesize Referenced in part 22a.}
\item $\langle$Compute next table (C++ forward) {\footnotesize 20a}$\rangle$ {\footnotesize Referenced in part 19b.}
\item $\langle$Compute next table (C++) {\footnotesize 44a}$\rangle$ {\footnotesize Referenced in parts 42a, 46a.
}
\item $\langle$Compute next table {\footnotesize 31a}$\rangle$ {\footnotesize Referenced in parts 5c, 7b, 12b, 27c.
}
\item $\langle$Compute skip table and mismatch shift using the hash function (C++) {\footnotesize 43c}$\rangle$ {\footnotesize Referenced in part 42a.}
\item $\langle$Compute skip table and mismatch shift using the hash function {\footnotesize 12a}$\rangle$ {\footnotesize Referenced in part 12b.}
\item $\langle$Compute skip table and mismatch shift, no hashing (C++) {\footnotesize 46b}$\rangle$ {\footnotesize Referenced in part 46a.}
\item $\langle$Compute skip table and mismatch shift {\footnotesize 6a}$\rangle$ {\footnotesize Referenced in parts 5c, 7b.
}
\item $\langle$Data declarations {\footnotesize 35b}$\rangle$ {\footnotesize Referenced in parts 35a, 36e.
}
\item $\langle$Define DNA search traits {\footnotesize 47b}$\rangle$ {\footnotesize Referenced in part 47a.}
\item $\langle$Define RandomNumberGenerator class {\footnotesize 58a}$\rangle$ {\footnotesize Referenced in part 57b.}
\item $\langle$Define Report procedure (C++) {\footnotesize 51d}$\rangle$ {\footnotesize Referenced in parts 49a, 52, 63b.
}
\item $\langle$Define Report procedure {\footnotesize 33e}$\rangle$ {\footnotesize Referenced in parts 31b, 35a.
}
\item $\langle$Define algorithm enumeration type, names, and selector function (C++ counter) {\footnotesize 60b}$\rangle$ {\footnotesize Referenced in part 59b.}
\item $\langle$Define algorithm enumeration type, names, and selector function (C++ large alphabet) {\footnotesize 57a}$\rangle$ {\footnotesize Referenced in part 57b.}
\item $\langle$Define algorithm enumeration type, names, and selector function (C++ word) {\footnotesize 64b}$\rangle$ {\footnotesize Referenced in parts 63b, 65b.
}
\item $\langle$Define algorithm enumeration type, names, and selector function (C++) {\footnotesize 49b}$\rangle$ {\footnotesize Referenced in parts 49a, 52, 54f.
}
\item $\langle$Define algorithm enumeration type, names, and selector function {\footnotesize 31e}$\rangle$ {\footnotesize Referenced in parts 31b, 35a, 36e.
}
\item $\langle$Define procedure to compute next table (C++ forward) {\footnotesize 20b}$\rangle$ {\footnotesize Referenced in part 38b.}
\item $\langle$Define procedure to compute next table (C++) {\footnotesize 43d}$\rangle$ {\footnotesize Referenced in part 38b.}
\item $\langle$Define procedure to compute next table {\footnotesize 30}$\rangle$ {\footnotesize Referenced in part 27b.}
\item $\langle$Define procedure to output sequence {\footnotesize 33c}$\rangle$ {\footnotesize Referenced in part 31b.}
\item $\langle$Define procedure to read string into sequence (C++) {\footnotesize 51b}$\rangle$ {\footnotesize Referenced in part 49a.}
\item $\langle$Define procedure to read string into sequence {\footnotesize 33b}$\rangle$ {\footnotesize Referenced in part 31b.}
\item $\langle$Define run procedure (C++ counter) {\footnotesize 62b}$\rangle$ {\footnotesize Referenced in part 59b.}
\item $\langle$Define run procedure (C++ forward) {\footnotesize 55}$\rangle$ {\footnotesize Referenced in parts 54f, 57b, 65b.
}
\item $\langle$Define run procedure {\footnotesize 37b}$\rangle$ {\footnotesize Referenced in part 36e.}
\item $\langle$Define search trait for word searches {\footnotesize 64a}$\rangle$ {\footnotesize Referenced in parts 63b, 65b.
}
\item $\langle$Define types needed for counting operations {\footnotesize 60a}$\rangle$ {\footnotesize Referenced in part 59b.}
\item $\langle$Experimental search function with skip loop without hashing {\footnotesize 44c}$\rangle$ {\footnotesize Referenced in part 44b.}
\item $\langle$Forward iterator case {\footnotesize 19b}$\rangle$ {\footnotesize Referenced in part 38b.}
\item $\langle$Generate data sequence {\footnotesize 58b}$\rangle$ {\footnotesize Referenced in part 57b.}
\item $\langle$Generate dictionary {\footnotesize 59a}$\rangle$ {\footnotesize Referenced in part 57b.}
\item $\langle$Generic search trait {\footnotesize 39b}$\rangle$ {\footnotesize Referenced in part 39a.}
\item $\langle$HAL declaration {\footnotesize 29b}$\rangle$ {\footnotesize Referenced in part 27b.}
\item $\langle$HAL with random access iterators, no trait passed {\footnotesize 41a}$\rangle$ {\footnotesize Referenced in part 38b.}
\item $\langle$Handle pattern size = 1 as a special case (C++) {\footnotesize 21b}$\rangle$ {\footnotesize Referenced in parts 21a, 42a, 46a.
}
\item $\langle$Handle pattern size = 1 as a special case {\footnotesize 3b}$\rangle$ {\footnotesize Referenced in parts 3a, 5c, 7b, 12b, 27c.
}
\item $\langle$Hashed Accelerated Linear algorithm (C++) {\footnotesize 42a}$\rangle$ {\footnotesize Referenced in part 41b.}
\item $\langle$Hashed Accelerated Linear algorithm {\footnotesize 12b}$\rangle$ {\footnotesize Referenced in part 29b.}
\item $\langle$Include algorithms header with existing search function renamed {\footnotesize 48}$\rangle$ {\footnotesize Referenced in parts 49a, 52, 54f, 57b, 59b, 63b, 65b.
}
\item $\langle$Non-hashed algorithms {\footnotesize 27c}$\rangle$ {\footnotesize Referenced in part 27b.}
\item $\langle$Output S2 {\footnotesize 34a}$\rangle$ {\footnotesize Referenced in part 33e.}
\item $\langle$Output header (C++) {\footnotesize 54b}$\rangle$ {\footnotesize Referenced in parts 52, 54f, 57b, 59b, 63b, 65b.
}
\item $\langle$Output statistics (C++ counter) {\footnotesize 63a}$\rangle$ {\footnotesize Referenced in part 62b.}
\item $\langle$Output statistics (C++) {\footnotesize 56c}$\rangle$ {\footnotesize Referenced in part 55.}
\item $\langle$Output statistics {\footnotesize 38a}$\rangle$ {\footnotesize Referenced in part 37b.}
\item $\langle$Read character sequence from file (C++) {\footnotesize 54a}$\rangle$ {\footnotesize Referenced in parts 52, 54f, 59b.
}
\item $\langle$Read character sequence from file {\footnotesize 36a}$\rangle$ {\footnotesize Referenced in parts 35a, 36e.
}
\item $\langle$Read dictionary from file, placing words of size j in dictionary[j] {\footnotesize 53b}$\rangle$ {\footnotesize Referenced in parts 52, 54f, 59b.
}
\item $\langle$Read test parameters (C++) {\footnotesize 53a}$\rangle$ {\footnotesize Referenced in parts 52, 54f, 57b, 59b, 63b, 65b.
}
\item $\langle$Read test parameters {\footnotesize 35d}$\rangle$ {\footnotesize Referenced in parts 35a, 36e.
}
\item $\langle$Read test sequences from file (C++) {\footnotesize 50}$\rangle$ {\footnotesize Referenced in part 49a.}
\item $\langle$Read test sequences from file {\footnotesize 32}$\rangle$ {\footnotesize Referenced in part 31b.}
\item $\langle$Read word sequence from file (C++) {\footnotesize 65a}$\rangle$ {\footnotesize Referenced in parts 63b, 65b.
}
\item $\langle$Recover from a mismatch using the next table (C++ forward) {\footnotesize 22a}$\rangle$ {\footnotesize Referenced in part 21a.}
\item $\langle$Recover from a mismatch using the next table (C++) {\footnotesize 43b}$\rangle$ {\footnotesize Referenced in part 42c.}
\item $\langle$Recover from a mismatch using the next table, with k translated {\footnotesize 8c}$\rangle$ {\footnotesize Referenced in part 8a.}
\item $\langle$Recover from a mismatch using the next table {\footnotesize 4b}$\rangle$ {\footnotesize Referenced in parts 3a, 5c.
}
\item $\langle$Run algorithm and record search distance {\footnotesize 56a}$\rangle$ {\footnotesize Referenced in parts 55, 62b.
}
\item $\langle$Run and time tests searching for selected subsequences (C++ counter) {\footnotesize 62a}$\rangle$ {\footnotesize Referenced in part 59b.}
\item $\langle$Run and time tests searching for selected subsequences (C++) {\footnotesize 56b}$\rangle$ {\footnotesize Referenced in parts 54f, 57b, 65b.
}
\item $\langle$Run and time tests searching for selected subsequences {\footnotesize 37a}$\rangle$ {\footnotesize Referenced in part 36e.}
\item $\langle$Run tests and report results (C++) {\footnotesize 51c}$\rangle$ {\footnotesize Referenced in parts 49a, 54ce.
}
\item $\langle$Run tests and report results {\footnotesize 33d}$\rangle$ {\footnotesize Referenced in part 31b.}
\item $\langle$Run tests searching for dictionary words (C++) {\footnotesize 54e}$\rangle$ {\footnotesize Referenced in part 52.}
\item $\langle$Run tests searching for selected subsequences (C++) {\footnotesize 54c}$\rangle$ {\footnotesize Referenced in parts 52, 63b.
}
\item $\langle$Run tests searching for selected subsequences {\footnotesize 36b}$\rangle$ {\footnotesize Referenced in part 35a.}
\item $\langle$Run tests {\footnotesize 36d}$\rangle$ {\footnotesize Referenced in part 36b.}
\item $\langle$Scan the text for a possible match (C++) {\footnotesize 21c}$\rangle$ {\footnotesize Referenced in part 21a.}
\item $\langle$Scan the text for a possible match {\footnotesize 3c}$\rangle$ {\footnotesize Referenced in parts 3a, 27c.
}
\item $\langle$Scan the text using a single-test skip loop with hashing (C++) {\footnotesize 42b}$\rangle$ {\footnotesize Referenced in part 42a.}
\item $\langle$Scan the text using a single-test skip loop with hashing {\footnotesize 11}$\rangle$ {\footnotesize Referenced in part 12b.}
\item $\langle$Scan the text using a single-test skip loop, no hashing (C++) {\footnotesize 46c}$\rangle$ {\footnotesize Referenced in part 46a.}
\item $\langle$Scan the text using a single-test skip loop, with k translated {\footnotesize 7a}$\rangle$ {\footnotesize Referenced in part 7b.}
\item $\langle$Scan the text using a single-test skip loop {\footnotesize 6b}$\rangle$ {\footnotesize Not referenced.}
\item $\langle$Scan the text using the skip loop {\footnotesize 5a}$\rangle$ {\footnotesize Referenced in part 5c.}
\item $\langle$Search traits for character sequences {\footnotesize 40a}$\rangle$ {\footnotesize Referenced in part 39a.}
\item $\langle$Select sequence S2 to search for in S1 (C++) {\footnotesize 54d}$\rangle$ {\footnotesize Referenced in part 54c.}
\item $\langle$Select sequence S2 to search for in S1 {\footnotesize 36c}$\rangle$ {\footnotesize Referenced in parts 36b, 37b.
}
\item $\langle$Sequence declarations {\footnotesize 27a}$\rangle$ {\footnotesize Referenced in parts 31b, 35a, 36e.
}
\item $\langle$Set file long.txt as input file {\footnotesize 35c}$\rangle$ {\footnotesize Referenced in parts 35a, 36e.
}
\item $\langle$Set file small.txt as input file {\footnotesize 31d}$\rangle$ {\footnotesize Referenced in part 31b.}
\item $\langle$Simple hash function declarations {\footnotesize 29a}$\rangle$ {\footnotesize Referenced in part 27b.}
\item $\langle$Trim dictionary[Pattern\_Size[j]] to have at most Number\_Of\_Tests words {\footnotesize 53c}$\rangle$ {\footnotesize Referenced in parts 52, 54f, 59b.
}
\item $\langle$User level search function with trait argument {\footnotesize 41b}$\rangle$ {\footnotesize Referenced in part 38b.}
\item $\langle$User level search function {\footnotesize 19a}$\rangle$ {\footnotesize Referenced in part 38b.}
\item $\langle$Variable declarations {\footnotesize 31c}$\rangle$ {\footnotesize Referenced in part 31b.}
\item $\langle$Verify match or recover from mismatch (C++) {\footnotesize 42c}$\rangle$ {\footnotesize Referenced in parts 42a, 46a.
}
\item $\langle$Verify match or recover from mismatch {\footnotesize 8a}$\rangle$ {\footnotesize Referenced in parts 7b, 12b.
}
\item $\langle$Verify the match for positions 1 through pattern\_size - 1 (C++) {\footnotesize 43a}$\rangle$ {\footnotesize Referenced in part 42c.}
\item $\langle$Verify the match for positions a + 1 through m - 1, with k translated {\footnotesize 8b}$\rangle$ {\footnotesize Referenced in part 8a.}
\item $\langle$Verify the match for positions a through m - 2 {\footnotesize 5b}$\rangle$ {\footnotesize Referenced in part 5c.}
\item $\langle$Verify whether a match is possible at the position found (C++) {\footnotesize 21d}$\rangle$ {\footnotesize Referenced in part 21a.}
\item $\langle$Verify whether a match is possible at the position found {\footnotesize 4a}$\rangle$ {\footnotesize Referenced in parts 3a, 27c.
}
\end{list}}


\clearpage

\begin{thebibliography}{MMM83}

\bibitem[BM77]{BM} R.~Boyer and S.~Moore. A fast string matching algorithm.
{\em CACM, 20(1977),762--772.}

\bibitem[Briggs]{Briggs} P.~Briggs, \emph{Nuweb, a simple literate
programming tool}, Version 0.87, 1989.

\bibitem[Cole96]{Cole} R.~Cole. Tight bounds on the complexity of the
Boyer-Moore string matching algorithm, \emph{SIAM Journal on
Computing} 5 (1994): 1075--1091.

\bibitem[CGG90]{FourThirds} L.~Colussi, Z.~Galil, R.~Giancarlo. On the
Exact Complexity of String Matching. {\em Proceedings of the Thirty First
Annual IEEE Symposium on the Foundations of Computer Science}, 
1990, 135--143.

\bibitem[DNAsource]{DNAsource} H.S.~Bilofsky, C.~Burks,
The GenBank(r) genetic sequence data bank.
{\em Nucl. Acids Res.} 16 (1988), 1861--1864.

\bibitem[Ga79]{Galil} Z.~Galil. On Improving the worst case running time of
the Boyer-Moore string matching algorithm.
{\em CACM} 22 (1979), 505--508.

\bibitem[GS83]{ConstantSpace} Z.~Galil, J.~Seiferas. Time space
optimal string matching.  {\em JCSS} 26 (1983), 280--294.

\bibitem[DraftCPP]{DraftCPP} Accredited Standards Committee X3
(American National Standards Institute), Information Processing
Systems, \emph{Working paper for draft proposed international
standard for information systems---programming language \Cpp}.
Doc No. X3J16/95-0185, WG21/N0785.[[Check for most recent version.]]

\bibitem[GO77]{Guibas} L.J.~Guibas, A.M.~Odlyzko, A new proof of the
linearity of the Boyer-Moore string searching algorithm.  {\em
Proc. 18th Ann. IEEE Symp. Foundations of Comp. Sci., 1977, 189--195}

\bibitem[Horspool88]{Horspool} R.N.~Horspool. Practical fast searching
in strings {\em Soft.-Prac. and Exp., 10 (March 1980), 501--506}

\bibitem[Hume88]{Hume} A.~Hume. A tale of two greps.
{\em Soft.-Prac. and Exp.} 18 (November 1988), 1063--1072.

\bibitem[HS91]{HumeSunday} A.~Hume, S.~Sunday. Fast string searching.
{\em Soft.-Prac. and Exp.} 21 (November 1991), 1221--1248.

\bibitem[Knuth84]{Knuth:literate} D.E.~Knuth, Literate programming.
{\em Computer Journal} 27 (1984), 97--111.

\bibitem[KMP77]{KMP} D.E.~Knuth, J.~Morris, V. Pratt. Fast pattern matching
in strings. {\em SIAM Journal on Computing} 6 (1977), 323--350.

\bibitem[Mu96]{MusserCounting} D.R.~Musser. \emph{Measuring Computing Times
 and Operation Counts}, http://www.cs.rpi.edu/musser/gp/timing.html.

\bibitem[MS96]{STLBook} D.R.~Musser, A.~Saini. \emph{STL Tutorial and
Reference Guide: \Cpp\ Programming with Standard Template Library}.
Addison-Wesley, Reading, MA, 1996.

\bibitem[SGI96]{STLSGI} Silicon Graphics Standard Template Library
Programming Guide, online guide, http://www.sgi.com/Technology/STL/.

\bibitem[Sm82]{Comp3} G.V.~Smit. A comparison of three string matching
algorithms.  {\em Soft.-Prac. and Exp.} 12, 1 (Jan 1982), 57--66.

\bibitem[StepanovLee]{StepanovLee} A.A.~Stepanov, M.~Lee, 
\emph{The Standard Template Library}, Tech.\ Report HPL-94-34, 
April 1994, revised October 31, 1995.

\bibitem[Su90]{Sunday} D.M.~Sunday. A very fast substring search
algorithm.  {\em CACM} 33 (August 1990), 132--142.

\end{thebibliography}
\end{document}